\documentclass[fleqn,usenatbib]{mnras}

\usepackage{savesym}
\usepackage{amsmath}
\savesymbol{iint}
\savesymbol{iiint}
\savesymbol{iiiint}
\usepackage{txfonts}
\restoresymbol{TXF}{iint}
\restoresymbol{TXF}{iiint}
\restoresymbol{TXF}{iiiint}

\usepackage[T1]{fontenc}
\usepackage{ae,aecompl}

\usepackage{graphicx}	
\usepackage{amssymb}	
\usepackage{soul}
\usepackage{array}
\usepackage{mathptmx}
\usepackage{mathtools}
\usepackage[usenames]{color}
\usepackage{pdflscape}
\usepackage{hyperref}
\usepackage[all]{hypcap}
\hypersetup{colorlinks=true,citecolor=blue,linkcolor=purple,filecolor=black,runcolor=black,breaklinks=true}
\usepackage{etoolbox}
\makeatletter
\patchcmd\@combinedblfloats{\box\@outputbox}{\unvbox\@outputbox}{}{%
   \errmessage{\noexpand\@combinedblfloats could not be patched}%
}%
 \makeatother
\DeclareMathAlphabet{\mathcal}{OMS}{cmsy}{m}{n}
\SetMathAlphabet{\mathcal}{bold}{OMS}{cmsy}{b}{n}

\newcommand{\comment}[1]{}
\definecolor{purple}{RGB}{160,32,240}

\newcommand{\Msun}{\;\mathrm{M}_{\odot}}

\newcommand{\vmax}{v_\mathrm{max}}

\newcommand{\vmp}{v_\mathrm{Mpeak}}

\newcommand{\sfrsf}{SFR_\mathrm{SF}}

\title[UniverseMachine Predictions for JWST]{The Universe at $z>10$: Predictions for JWST from the \textsc{UniverseMachine} DR1}

\author[P. Behroozi et al.]{Peter Behroozi,$^{1}$\thanks{E-mail: behroozi@arizona.edu} Charlie Conroy,$^2$ Risa H.\ Wechsler,$^{3,4}$ Andrew Hearin,$^5$\newauthor Christina C.\ Williams,$^{1,6}$ Benjamin P.\ Moster,$^7$ L. Y. Aaron\ Yung,$^{8,9}$ Rachel S.\ Somerville,$^{9,8}$\newauthor Stefan Gottl\"ober,$^{10}$ Gustavo Yepes,$^{11,12}$ Ryan Endsley$^{1}$
\\
$^{1}$ Department of Astronomy and Steward Observatory, University of Arizona, Tucson, AZ 85721, USA\\
$^{2}$ Department of Astronomy, Harvard University, Cambridge, MA 02138, USA\\
$^{3}$ Kavli Institute for Particle Astrophysics and Cosmology and Department of Physics, Stanford University, Stanford, CA 94305, USA\\
$^{4}$ Department of Particle Physics and Astrophysics,  SLAC National  Accelerator Laboratory, Stanford, CA 94305, USA\\
$^{5}$ High-Energy Physics Division, Argonne National Laboratory, Argonne, IL 60439, USA\\
$^{6}$ NSF Fellow\\
$^{7}$ Universit{\"a}ts-Sternwarte, Ludwig-Maximilians-Universit{\"a}t M{\"u}nchen, Scheinerstr. 1, 81679 M{\"u}nchen, Germany\\
$^{8}$ Department of Physics and Astronomy, Rutgers University, 136 Frelinghuysen Road, Piscataway, NJ 08854, USA\\
$^{9}$ Center for Computational Astrophysics, Flatiron Institute, 162 5th Ave, New York, NY 10010, USA\\
$^{10}$Leibniz-Institut f\"ur Astrophysik, An der Sternwarte 16, 14482 Potsdam, Germany\\
$^{11}$Departamento de F\'{\i}sica Te\'orica, Universidad Aut\'onoma de Madrid, Cantoblanco E-28049, Madrid, Spain \\
$^{12}$Centro de Investigaci\'on Avanzada en F\'{\i}sica Fundamental,  Facultad de Ciencias, Universidad Aut\'onoma de Madrid, E-28049 Madrid, Spain \\
}

\pubyear{2020}

\begin{document}
\label{firstpage}
\pagerange{\pageref{firstpage}--\pageref{lastpage}}
\maketitle

\begin{abstract}
The \textit{James Webb Space Telescope} (\textit{JWST}) is expected to observe  galaxies at $z>10$ that are presently inaccessible.  Here, we use a self-consistent empirical model, the \textsc{UniverseMachine}, to generate mock galaxy catalogues and lightcones over the redshift range $z=0-15$.  These data include realistic galaxy properties (stellar masses, star formation rates, and UV luminosities), galaxy--halo relationships, and galaxy--galaxy clustering.  Mock observables are also provided for different model parameters spanning observational uncertainties at $z<10$.  We predict that Cycle 1 \textit{JWST} surveys will very likely detect galaxies with $M_*>10^7\Msun$ and/or $M_{1500}<-17$ out to at least $z\sim 13.5$.  Number density uncertainties at $z>12$ expand dramatically, so efforts to detect $z>12$ galaxies will provide the most valuable constraints on galaxy formation models.  
The faint-end slopes of the stellar mass/luminosity functions at a given mass/luminosity threshold steepen as redshift increases.  
This is because observable galaxies are hosted by haloes in the exponentially falling regime of the halo mass function at high redshifts.  Hence, these faint-end slopes are robustly predicted to become shallower below current observable limits ($M_\ast < 10^7\Msun$ or $M_\mathrm{1500}>-17$).  For reionization models, extrapolating luminosity functions with a constant faint-end slope from $M_{1500}=-17$ down to $M_{1500}=-12$ gives the most reasonable upper limit for the total UV luminosity and cosmic star formation rate up to $z\sim 12$.  We compare to three other empirical models and one semi-analytic model, showing that the range of predicted observables from our approach encompasses predictions from other techniques.   Public catalogues and lightcones for common fields are available online.
\end{abstract}




\section{Introduction}

\textit{JWST} will provide our first clear view of galaxy formation at $z>10$, offering the potential to study exotic physical systems (e.g., population III stars and direct-collapse black holes; see \citealt{Bromm11} for a review) and extreme conditions (e.g., low metallicities, high densities, high merger rates, and high accretion rates; see \citealt{Bromm11} and \citealt{Stark16} for reviews).  At the same time, \textit{JWST} will test whether key trends for galaxies at $z=4-10$ persist at higher redshifts.  Examples include steeper faint-end slopes for luminosity and mass functions \citep{Bouwens15,Finkelstein15,Song15}, rapid fall-off in observed cosmic star formation rates at $z>9$ \citep{Oesch14,Oesch18}, and increasing stellar mass--halo mass ratios (e.g., \citealt{BehrooziHighZ,Moster17,BWHC19}; cf.\ \citealt{RP17}).

Predictions for what \textit{JWST} will see have become increasingly common as its launch approaches (e.g., \citealt{Mason15,BehrooziHighZ,Tacchella18,Williams18,Yung19,Lagos19,Park20,Endsley20,Vogelsberger20,Hainline20,Griffin20, Kauffmann20}), as these predictions are essential for proposal planning.  In this paper, we provide public catalogues and lightcones containing a range of \textit{JWST} predictions generated by the \textsc{UniverseMachine} \citep{BWHC19}.  The \textsc{UniverseMachine} is an empirical model that self-consistently parametrizes galaxy star formation rates as a function of their host dark matter halo masses, mass accretion rates, and redshifts.  As with other empirical models \citep{Mutch13,Becker15,Cohn16,RP16b,Moster17}, this parametrization is applied to the merger trees of simulated dark matter haloes, thereby tracing the growth of model galaxies over time.  Empirical models have the unique ability to extract galaxy--halo relationships from observational constraints without making explicit assumptions about the particular physical processes driving galaxy growth \citep{Behroozi19}.  At the same time, empirical models can map the range of plausible galaxy formation scenarios consistent with all observations simultaneously.  See \cite{Somerville15} and \cite{Wechsler18} for reviews of this and other approaches to modelling galaxy formation.

The \textsc{UniverseMachine} is well-suited to high-redshift predictions for several reasons.  First, the model was calibrated directly to $z>4$ UV luminosity functions and IR Excess--UV (IRX--UV) relationships from the \textit{Atacama Large Millimeter Array} (\textit{ALMA}), instead of less-certain stellar mass functions.  Second, the constraining data included new and existing clustering measurements over $z=0-1$ \citep{Coil17,BWHC19}, and the mock catalog of the resulting best-fit model matches pair counts and clustering of observed galaxies to at least $z\sim 5$ \citep{Pandya19,Endsley20}.  Finally, the model agrees with clustering-derived stellar mass--halo mass relationships out to $z\sim 7$ \citep{Harikane16,Harikane18,Ishikawa17}.  These qualities allow the \textsc{UniverseMachine} to generate realistic galaxy properties (including UV luminosities, stellar masses, and star formation rates), realistic galaxy--halo relationships, and realistic galaxy clustering at high redshifts.

High-redshift galaxy evolution can be empirically predicted by smooth interpolation between two boundary conditions: 1) the Universe had no stars when it began, and 2) modelled galaxies at observable redshifts must match actual observations.  Combining these boundary conditions with Lambda Cold Dark Matter ($\Lambda$CDM) simulations on halo growth is surprisingly powerful. For example, \cite{BehrooziHighZ} showed that $z=0$ observations of the galaxy stellar mass function and specific star formation rates could successfully predict evolution of the stellar mass--halo mass relation to $z=3$, and similarly that $z\le 4$ constraints could successfully predict the stellar mass--halo mass relation to $z=8$.  To aid confidence that the range of our predictions encompasses a wide variety of other possible techniques, we compare our predictions both to other empirical models \citep{BehrooziHighZ,Moster17,Williams18} and to a semi-analytical model (the Santa Cruz model; \citealt{Somerville15b,Yung19,Yung19b}).

In this paper, Section \ref{s:methods} details the dark matter simulation that we use, provides an overview of the \textsc{UniverseMachine}, and discusses observational constraints at $z<10$.  Section \ref{s:results} describes results from the generated mock catalogues, including mass/luminosity functions and cosmic star formation rates.  We discuss these results in Section \ref{s:discussion} and conclude in Section \ref{s:conclusions}.  Appendix \ref{a:full} summarizes key equations for the \textsc{UniverseMachine}, Appendix \ref{a:res_tests} contains resolution tests, and Appendix \ref{a:cosmo} describes the effects of cosmology uncertainties. Throughout this paper, we adopt a flat, $\Lambda$CDM cosmology ($h=0.68$, $\Omega_M=0.307$, $\Omega_\Lambda=0.693$, $n_s=0.96$, $\sigma_8 = 0.823$) consistent with \textit{Planck} constraints \citep{Planck18}.  Halo masses use the virial spherical overdensity definition of \cite{mvir_conv}.  Stellar masses assume a \cite{Chabrier03} initial mass function (IMF), a \cite{bc-03} stellar population synthesis model, and a \cite{calzetti-00} dust law.  The adopted galaxy--halo modelling uses the \textsc{UniverseMachine} Data Release 1 (DR1) code release \citep{BWHC19}.  Except where otherwise specified, galaxy formation is assumed to be inefficient in haloes with $M_h < 10^8\Msun$ due to the atomic cooling limit \citep{OShea15,Xu16}.

\section{Methods}

\label{s:methods}

The original \textsc{UniverseMachine} analysis \citep{BWHC19} used a large-volume dark matter simulation (\textit{Bolshoi-Planck}; \citealt{Klypin14,RP16b}).  However, the halo mass resolution of \textit{Bolshoi-Planck} ($\sim 10^{10}\Msun$) is not sufficient to resolve most star formation above $z=10$, which likely occurs in $10^9 - 10^{10} \Msun$ haloes \citep{BehrooziHighZ}.  In this section, we describe the higher-resolution simulation used in this paper (\S \ref{s:dm_sim}), the \textsc{UniverseMachine} code (\S \ref{s:um}), our resolution tests (\S \ref{s:resolution}), and the lightcone generation process (\S \ref{s:lightcones}).

\subsection{Dark Matter Simulation}

\label{s:dm_sim}

Throughout this work, we use the public \textit{Very Small MultiDark-Planck} (\textit{VSMDPL}) simulation,\footnote{\url{https://www.cosmosim.org/cms/simulations/vsmdpl/}} which follows a periodic comoving cube of side length $160h^{-1}$ Mpc from $z=150$ to $z=0$ with 3840$^3$ particles.  This gives \textit{VSMDPL} both very high particle mass resolution (9.1$\times 10^6 \Msun$) and force resolution (1 $h^{-1}$ kpc at $z<1$, 2 $h^{-1}$ comoving kpc at $z>1$) resolution; the simulation thereby resolves $10^9 \Msun$ haloes with $>100$ particles.  As described in \S \ref{s:resolution}, this gives sufficient resolution to capture almost all star formation in haloes at $z\le15$.  The \textit{VSMDPL} box was run with the \textsc{GADGET-2} code \citep{Springel05}, with a flat, $\Lambda$CDM cosmology ($h=0.68$, $\Omega_M=0.307$, $\Omega_\Lambda=0.693$, $n_s=0.96$, $\sigma_8 = 0.823$).  Haloes were identified at 151 snapshots from $z=25$ to $z=0$ using the \textsc{Rockstar} phase-space halo finder \citep{Rockstar}.  Merger trees were constructed using the \textsc{Consistent Trees} code \citep{BehrooziTree}.

\subsection{The \textsc{UniverseMachine}}

\label{s:um}

\subsubsection{Overview}

The \textsc{UniverseMachine} is an empirical model that links galaxy star formation rates to properties of their host haloes \citep{BWHC19}.  Specifically, the model parametrizes the probability distribution of galaxy star formation rates (SFRs) as a function of host halo mass ($M_h$), mass accretion rate ($\dot{M}_h$), and redshift, $z$, i.e., $P(SFR|M_h, \dot{M}_h, z)$.  The \textsc{UniverseMachine} uses a guess in this parameter space to assign an SFR to each halo at every redshift of the simulation; each galaxy's SFR is integrated along the merger tree of its host halo to obtain a stellar mass and UV luminosity. This results in an entire mock universe populated with galaxy properties.  The \textsc{UniverseMachine} then compares statistics of this mock universe to real observations to obtain a likelihood for the original guess.  This likelihood is fed to a Markov Chain Monte Carlo algorithm, which repeatedly generates new guesses in parameter space (and new mock catalogues) until the chain converges to the posterior distribution of the model parameters.  The full 44-dimensional parametrization is designed to be flexible so that the model can approximate the true probability distribution of galaxy SFRs in haloes; see \cite{Behroozi19}.  Almost all these parameters control behaviour at low redshifts ($z\le 1$), where observable constraints are tightest.

\begin{figure}
\centering
    \vspace{-4ex}
    \hspace{-10ex}\includegraphics[width=1.1\columnwidth]{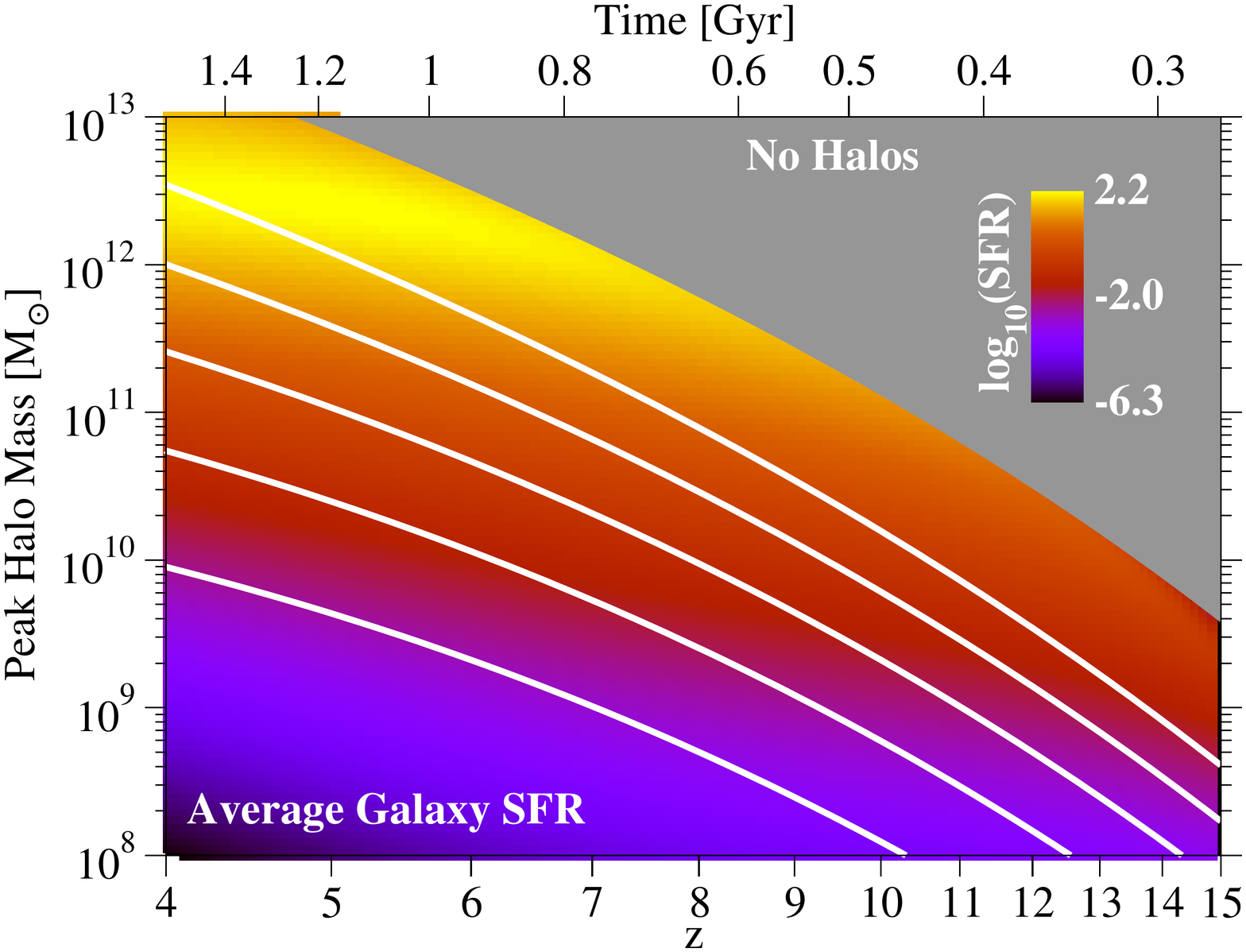}\\[-6ex]
    \caption{Average galaxy star formation rates (M$_\odot$ yr$^{-1}$) in the best-fit model of the \textsc{UniverseMachine} as a function of redshift and peak halo mass.  \textit{White lines} show typical halo growth histories for haloes of mass $10^{11}$, $10^{12}$, $10^{13}$, $10^{14}$, and $10^{15}\Msun$ at $z=0$.  The \textit{grey shaded region} shows haloes below the number densities expected to be observable with \textit{JWST}.}
    \label{fig:halo_sfrs}
    \vspace{-2ex}
    \hspace{-10ex}\includegraphics[width=1.1\columnwidth]{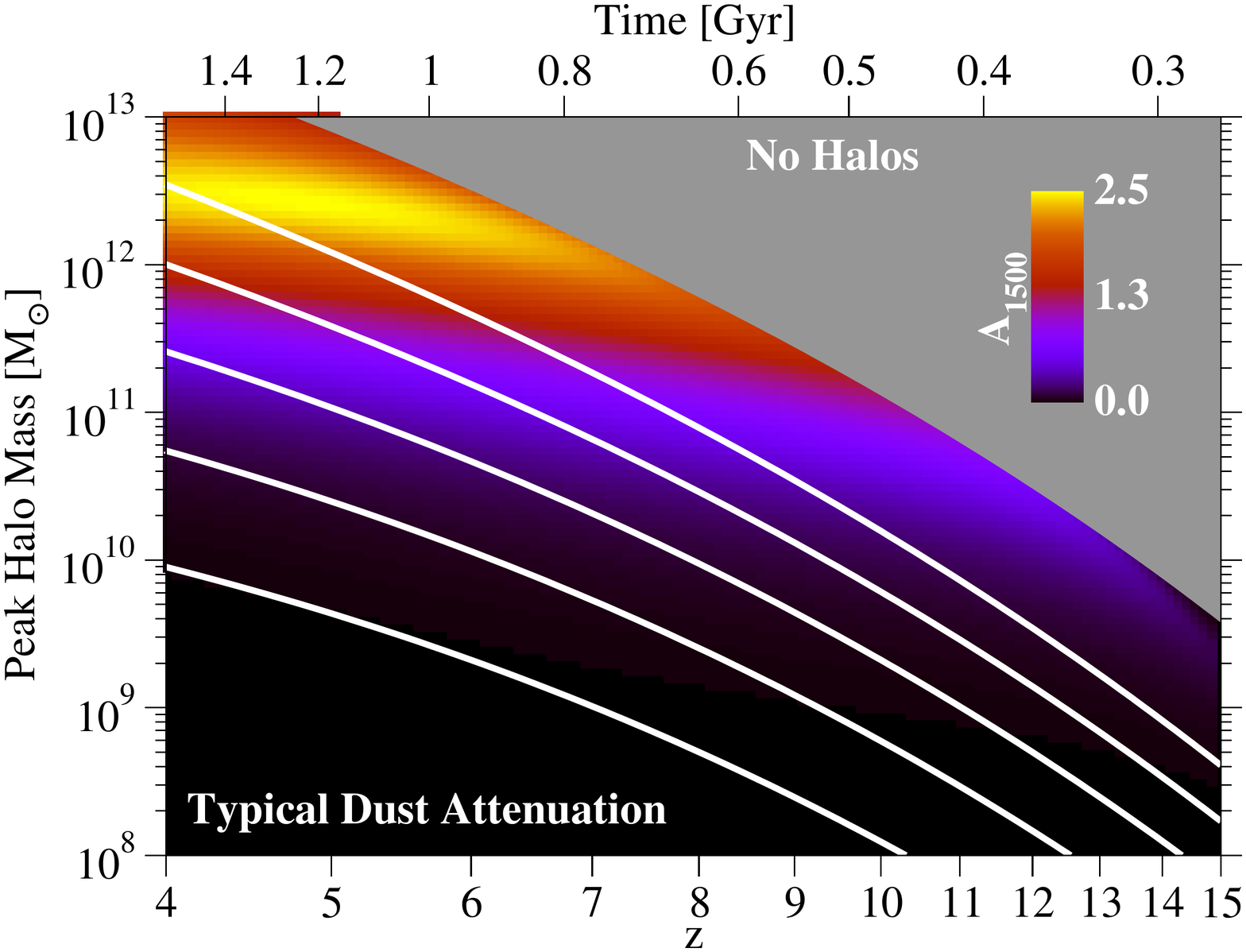}\\[-6ex]
    \caption{Typical dust attenuation ($A_{1500}$, magnitudes) in the best-fit model of the \textsc{UniverseMachine}  as a function of redshift and peak halo mass.  \textit{White lines} show typical halo growth histories as in Fig.\ \ref{fig:halo_sfrs}.  The \textit{grey shaded region} shows haloes below the number densities expected to be observable with \textit{JWST}.}
    \label{fig:halo_dust}
    \vspace{-2ex}
    \hspace{-10ex}\includegraphics[width=1.1\columnwidth]{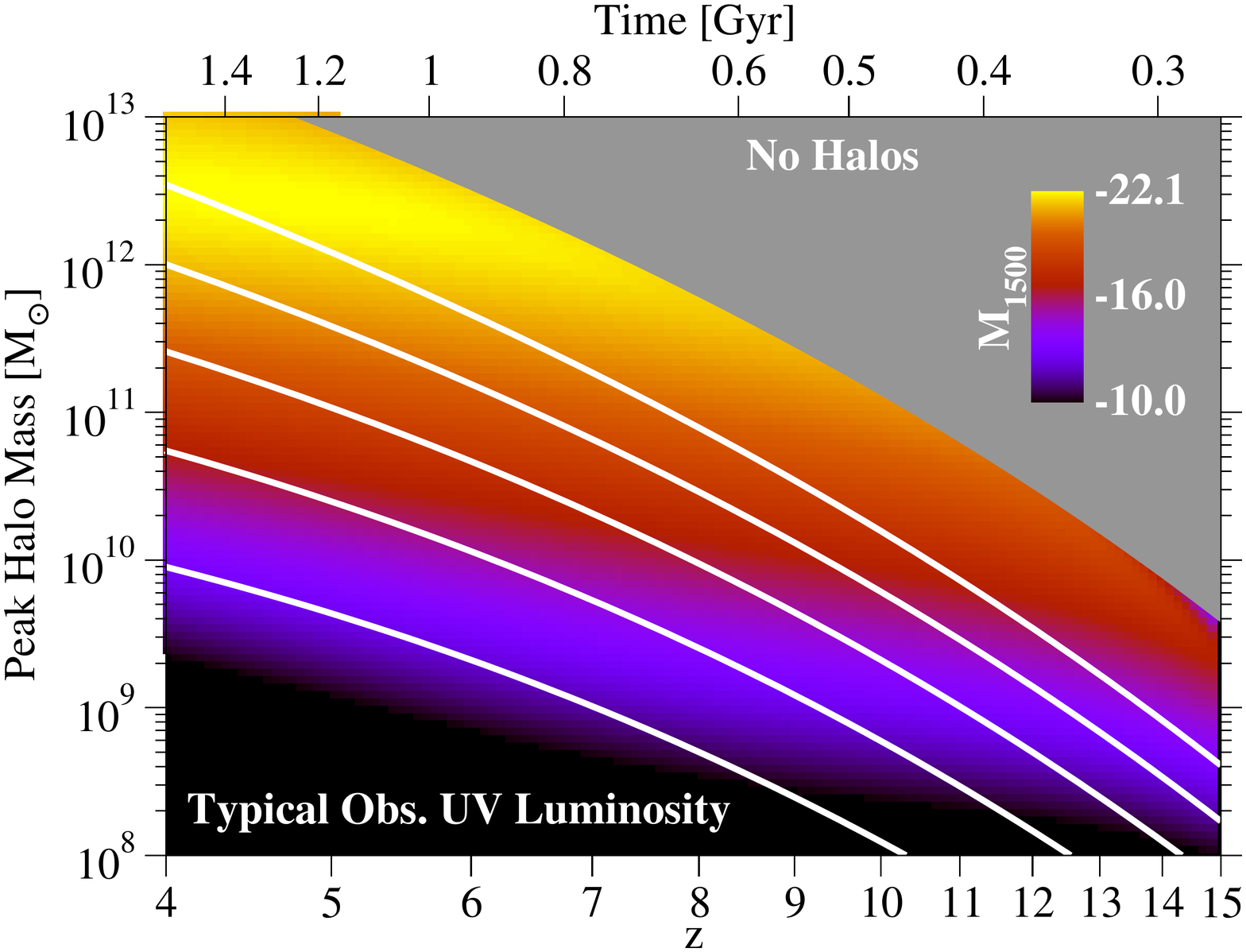}\\[-6ex]
    \caption{Typical observed (dust-attenuated) absolute UV luminosity  ($M_{1500}$, AB magnitudes) in the best-fit model of the \textsc{UniverseMachine} as a function of redshift and peak halo mass. \textit{White lines} show typical halo growth histories as in Fig.\ \ref{fig:halo_sfrs}.  The \textit{grey shaded region} shows haloes below the number densities expected to be observable with \textit{JWST}.}
    \label{fig:halo_uv}
\end{figure}

\begin{figure*}
 \centering
    \vspace{-11ex}
    \hspace{-14ex}\includegraphics[width=1.15\columnwidth]{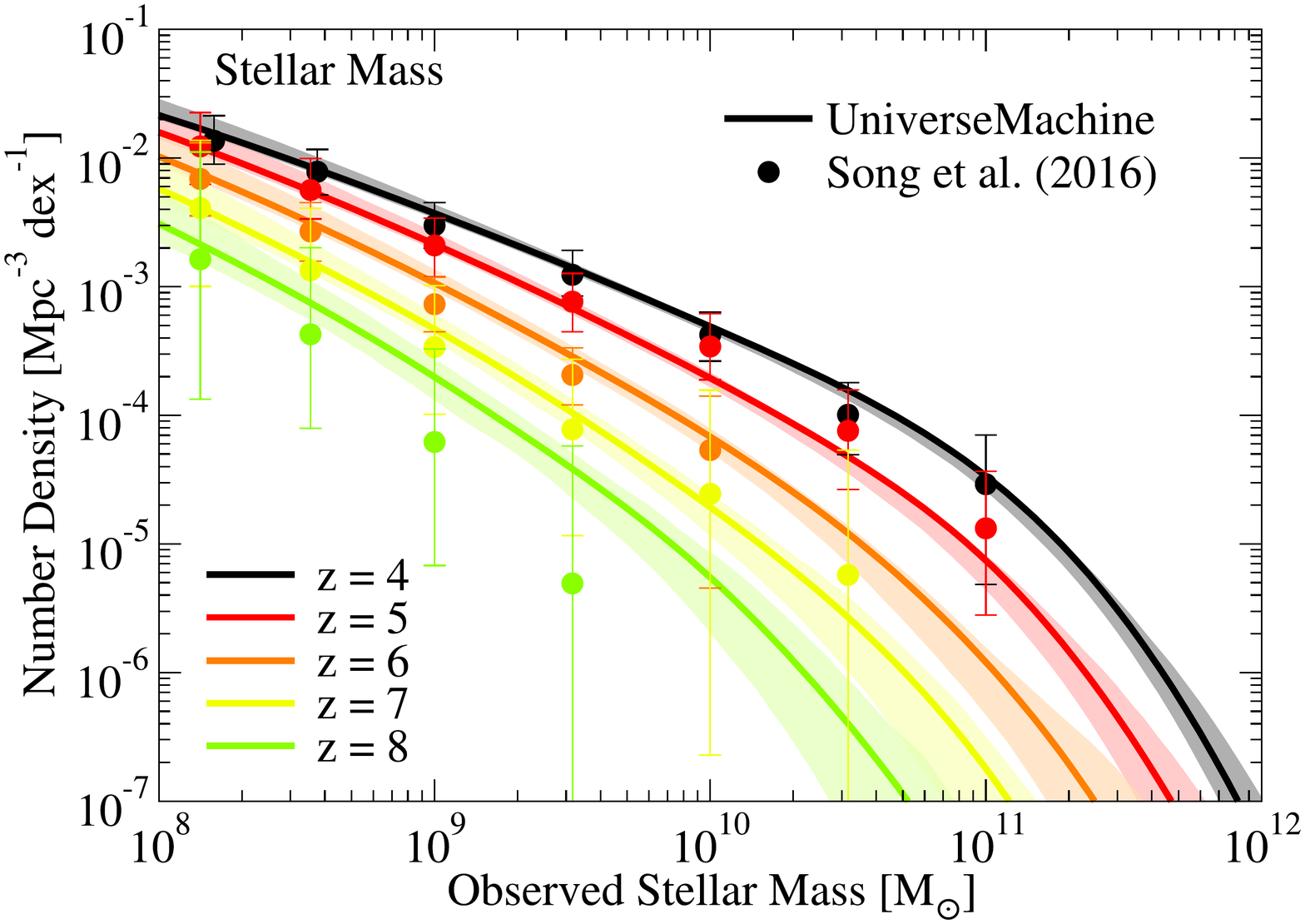}    \hspace{-7ex}\includegraphics[width=1.15\columnwidth]{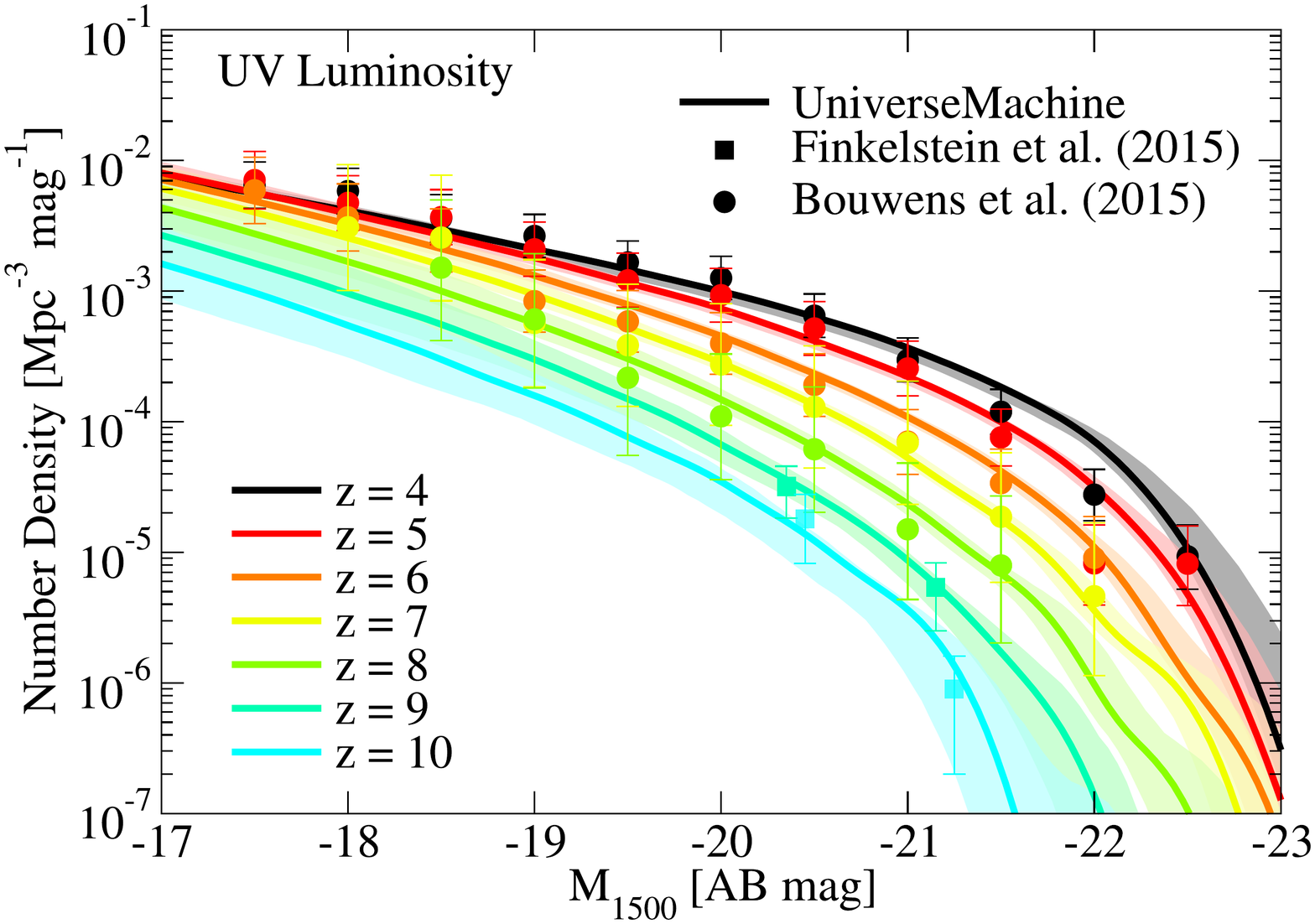}\hspace{-15ex}\\[-4ex]
    \caption{\textbf{Left} Panel: comparison between observed stellar mass functions \citep{Song15} and those generated by the \textsc{UniverseMachine} on the \textit{VSMDPL} high-resolution simulation.  These data were \textit{not} used as constraints to the \textsc{UniverseMachine}.  \textbf{Right} Panel: comparison between observed UV luminosity functions \citep{Finkelstein15,Bouwens15} and those generated by the \textsc{UniverseMachine} on \textit{VSMDPL}.  These data were used to constrain the \textsc{UniverseMachine} on the \textit{Bolshoi-Planck} simulation; as above, the \textsc{UniverseMachine} applied to \textit{VSMDPL} generates equivalent results.  For both panels, \textit{shaded regions} correspond to the $16-84^\mathrm{th}$ percentile confidence interval.}
    \label{fig:existing_data}
\end{figure*}

\subsubsection{Observational Constraints}

Full observational constraints are described in Appendix C of \citealt{BWHC19}.  The observational constraints used at $z>4$ include UV luminosity functions from $z=4-10$ \citep{Finkelstein15,Bouwens15b}, IRX--UV relationships from $z=4-7$ \citep{Bouwens16}, UV--SM relationships from the \textsc{Sedition} code \citep{BWHC19} applied to stacked SEDs from \cite{Song15} for $z=4-8$, galaxy specific star formation rates \citep{McLure11,Labbe12,Smit14,Salmon15}, and total cosmic star formation rates from UV galaxy surveys \citep{Yoshida06,Cucciati11,vdBurg10,Finkelstein15} and gamma-ray bursts \citep{Kistler13}.  See Section \ref{s:resolution} for comparisons between the \textsc{UniverseMachine} and observed stellar mass and luminosity functions at $z\ge 4$.

\subsubsection{Effective Behaviour at High Redshifts}

\label{s:effective}

Although we use the full \textsc{UniverseMachine} parametrization for the results in this paper (see Appendix \ref{a:full} for key equations and \citealt{BWHC19} for full details), its behaviour reduces to a much simpler effective model at high redshifts, which we describe here to guide intuition.

Fig.\ \ref{fig:halo_sfrs} shows average galaxy star formation rates for haloes at $z>4$ in the best-fit \textsc{UniverseMachine} model.  At $z>6$, the lack of massive and/or quenched haloes results in a simple power-law form for average star formation rates (SFRs):
\begin{eqnarray}
SFR(M_h, z) & \approx & 10^{\alpha(z)} \cdot M_h^{\beta(z)} \label{e:sfr1}\\
\alpha(z) & \approx  & \alpha_0 + \alpha_z \cdot z \label{e:sfr2}\\
\beta(z) & \approx  & \beta_0 + \beta_z \cdot z \label{e:sfr3},
\end{eqnarray}
where $M_h$ is the peak halo mass (i.e., maximum mass attained over the halo's assembly history).  The variation of SFR with the halo mass assembly rate ($\dot{M}_h$) is $\sim$0.3 dex, which is much smaller than the corresponding variation of SFR with either $M_h$ or redshift (both several dex; see Fig.\ \ref{fig:halo_sfrs}), so $\dot{M}_h$ does not appear in Eq.\ \ref{e:sfr1}. The effective values of $\alpha_0$ and $\beta_0$ are constrained principally by stellar mass functions at $z\le 4$, and those of $\alpha_z$ and $\beta_z$ are effectively constrained to match the evolution of UV luminosity functions over $z=4-10$.

The \textsc{UniverseMachine} integrates the SFR of each galaxy along the merger tree of its dark matter halo to obtain galaxy stellar mass and luminosity.  UV luminosities ($M_{1500}$) are calculated using the Flexible Stellar Population Synthesis code (FSPS) v3.0 \citep{Conroy09,Conroy10}.  A correction of $-$0.06 mag is applied to match luminosities produced by the \cite{bc-03} SPS model, i.e., the model assumed for all the stellar mass function constraints used in the \textsc{UniverseMachine}.  Dust at $z>4$ is modelled as a net attenuation:
\begin{eqnarray}
A_\mathrm{1500} & = & 2.5 \log_{10}(1+10^{0.4 \alpha_\mathrm{dust}(M_\mathrm{dust}(z)-M_\mathrm{1500,intrinsic})}) \label{e:dust1}\\
M_\mathrm{dust}(z) & = & M_\mathrm{dust,0} + M_\mathrm{dust,z}\cdot z. \label{e:dust2}
\end{eqnarray}
Here, the observed UV luminosity ($M_{1500}$) is $M_\mathrm{1500,intrinsic}+A_\mathrm{1500}$, where $M_\mathrm{1500,intrinsic}$ is the unattenuated UV luminosity from FSPS.  The free parameters $\alpha_\mathrm{dust}$, $M_\mathrm{dust,0}$, and $M_\mathrm{dust,z}$ are constrained to match the IRX--UV relationship obtained from \textit{ALMA} observations in \cite{Bouwens16} for $z=4-7$.  

At high redshift ($z>6$), stars are assumed to form with low metallicities; i.e., $\log_{10}(Z/Z_\odot)=-1.5$.  This is consistent with the extrapolation of redshift trends in \cite{Maiolino08}.  Typical 1500\AA{} UV luminosities for $z>6$ galaxies change by $<5\%$ over a metallicity range of $\log_{10}(Z/Z_\odot)=-1$ to $\log_{10}(Z/Z_\odot)=-2$ \citep{Madau14}.  Higher metallicities of $Z=Z_\odot$ would result in 20\% lower UV luminosities \citep{Madau14}.

Fig.\ \ref{fig:halo_dust} shows the typical dust attenuation as a function of redshift and halo mass in the best-fit \textsc{UniverseMachine} model (i.e., Eqs.\ \ref{e:dust1}--\ref{e:dust2} evaluated for the average galaxy SFRs in Fig.\ \ref{fig:halo_sfrs}).  The main effect is a modest ($1-2$ mag) average attenuation for highly star-forming galaxies ($M_\ast\sim 100 \Msun$ yr$^{-1}$).  Because the number densities of these galaxies drop significantly at higher redshifts, typical star-forming galaxies have lower dust attenuation at higher redshifts.  Fig.\ \ref{fig:halo_uv} shows the resulting typical rest-frame UV luminosities, assuming the median SFR--UV relation in the \textsc{UniverseMachine} (Eq.\ \ref{e:kappa_fuv}) and the attenuation in Fig.\ \ref{fig:halo_dust}.  Because SFRs increase with redshift at fixed halo mass (Fig.\ \ref{fig:halo_sfrs}), a fixed UV luminosity threshold will correspond to less-massive haloes at higher redshifts.

The 7 effective parameters in Eqs.\ \ref{e:sfr1}--\ref{e:dust2} (4 for SFR, 3 for dust) dominate the evolution of galaxy SFR and UV luminosity in the \textsc{UniverseMachine} at $z>6$.  Eqs.\ \ref{e:sfr1}--\ref{e:sfr3} are very general, in that they could approximate the emergent behaviour of a wide range of physical models.  Our adopted dust model is general enough to capture overall trends in dust with luminosity and redshift.

\subsubsection{Applying the \textsc{UniverseMachine} to VSMDPL}

The \textsc{UniverseMachine} is designed to be resolution-independent.  Fig.\ \ref{fig:existing_data} shows this qualitatively, in the sense that the best-fit \textsc{UniverseMachine} model from \textit{Bolshoi-Planck} applied to \textit{VSMDPL} (which has $\sim 20\times$ higher mass resolution) still gives very good fits to observed data.  More rigorous tests of resolution-independence are performed in Appendix \ref{a:res_tests}.    These demonstrate that it is not necessary to refit the entire model when a new simulation is used.  This is fortunate, as it would take approximately 20 million CPU hours to do so for \textit{VSMDPL}.  Instead, we randomly sample 1000 galaxy models from the \textsc{UniverseMachine} DR1 posterior distribution and apply each of these to the \textit{VSMDPL} simulation.  Measuring the posterior distribution of high-redshift observables from these models took 34,000 CPU hours.

\begin{figure*}
    \centering
    \vspace{-12ex}
    \hspace{-14ex}\includegraphics[width=1.2\columnwidth]{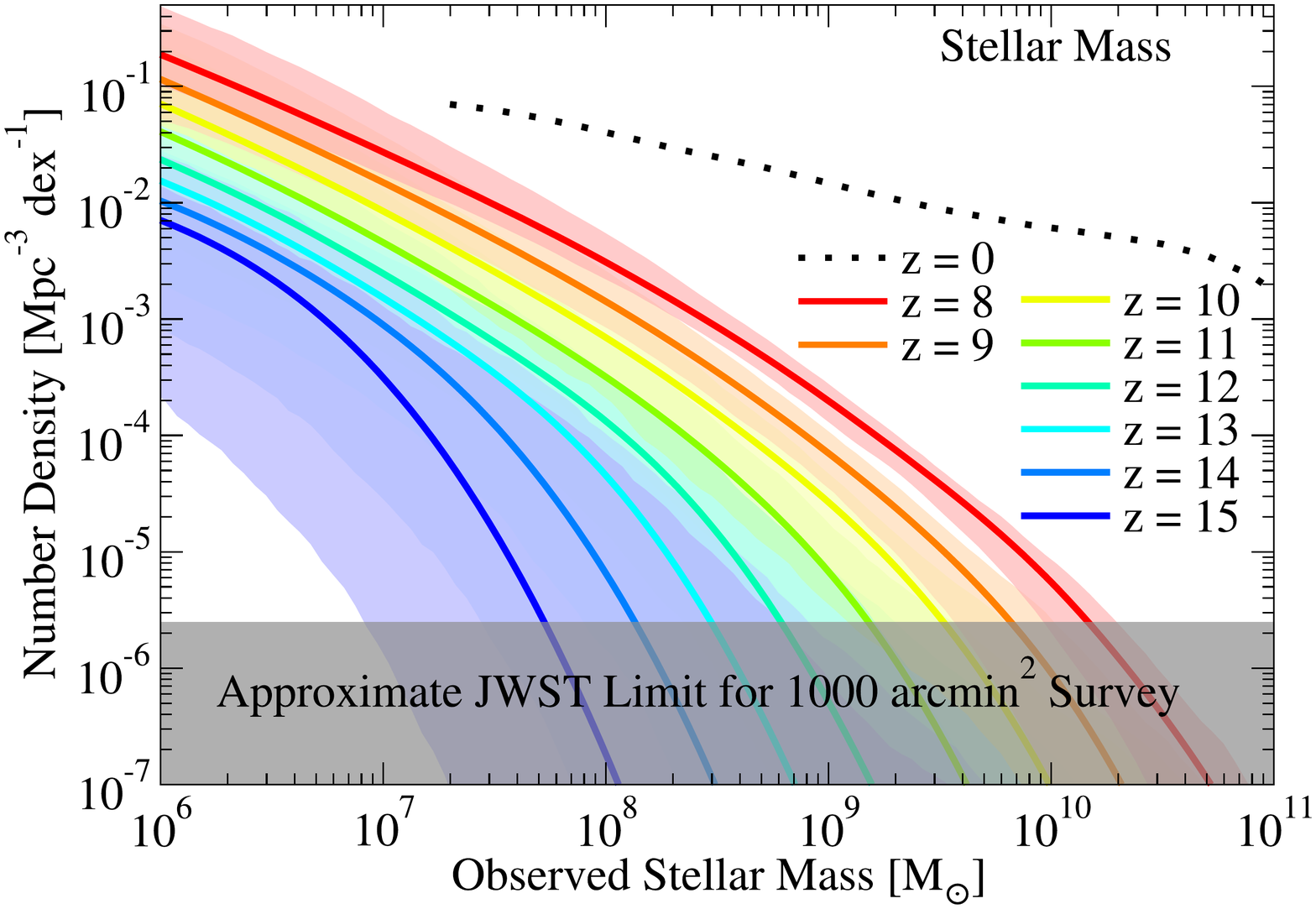}    \hspace{-7ex}\includegraphics[width=1.2\columnwidth]{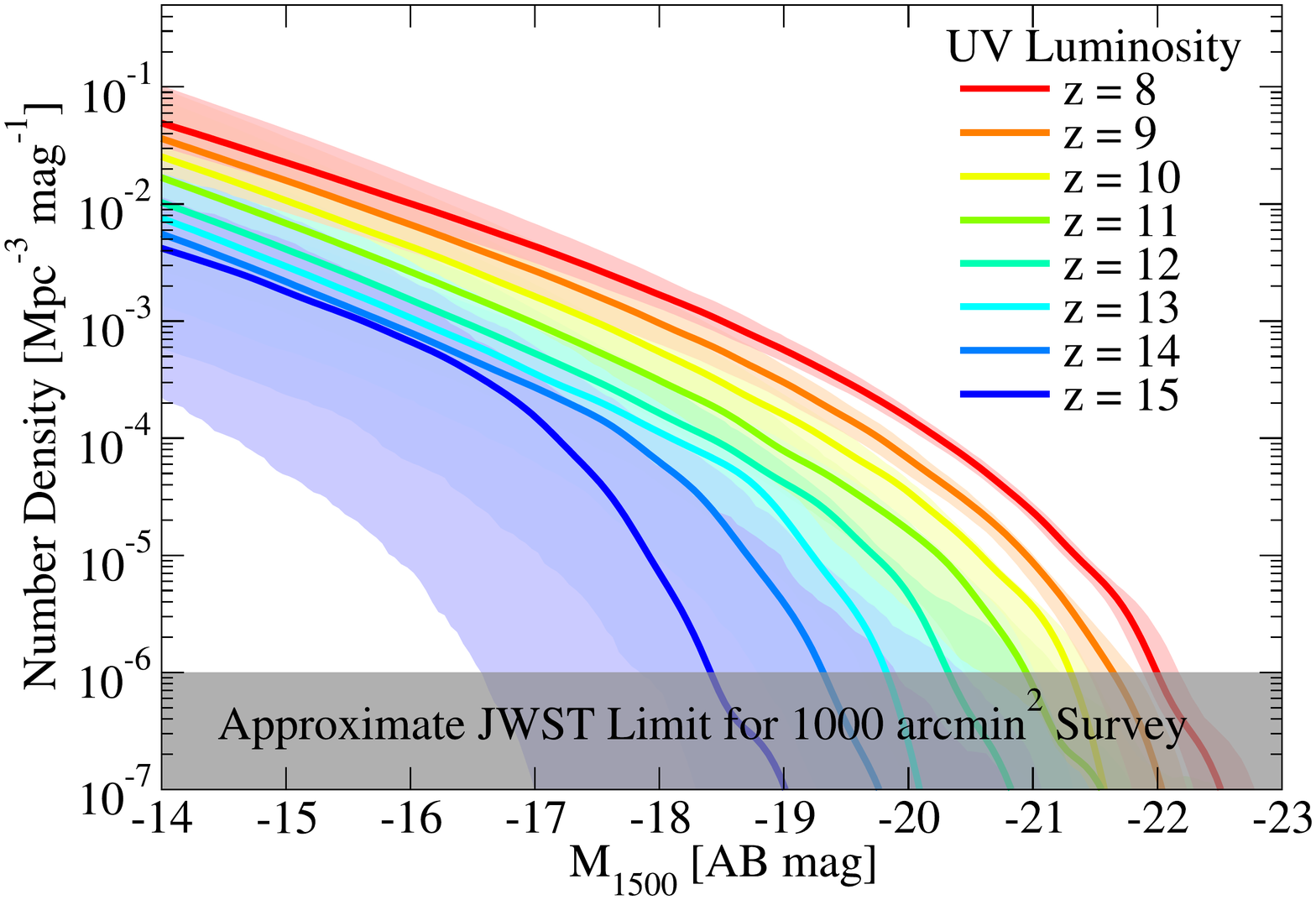} \hspace{-15ex}\\[-5ex]
    \caption{\textbf{Left} Panel: Predicted stellar mass functions for $z=8$ to $z=15$ from the \textsc{UniverseMachine}.  The \textit{dotted line} shows the stellar mass function at $z\sim 0$ (\citealt{Moustakas13}, from the Sloan Digital Sky Survey).  \textbf{Right} Panel: Predicted UV luminosity functions at 1500\AA{} for $z=8$ to $z=15$ from the \textsc{UniverseMachine}.  For both panels, \textit{shaded regions} correspond to the $16-84^\mathrm{th}$ percentile confidence interval.}
    \label{fig:nds}
    \vspace{-5.5ex}
    \hspace{-14ex}   \includegraphics[width=1.2\columnwidth]{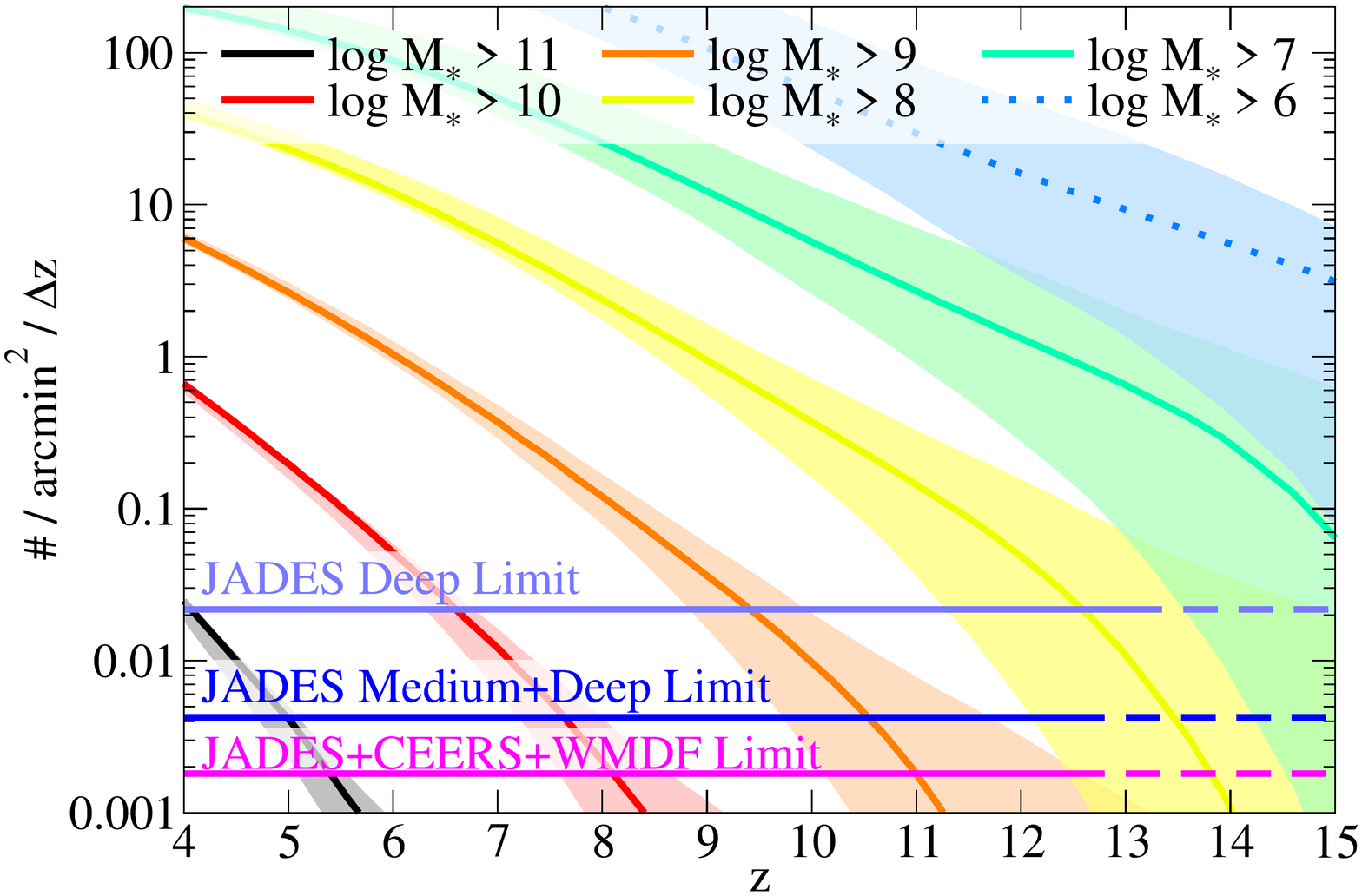}\hspace{-7ex}\includegraphics[width=1.2\columnwidth]{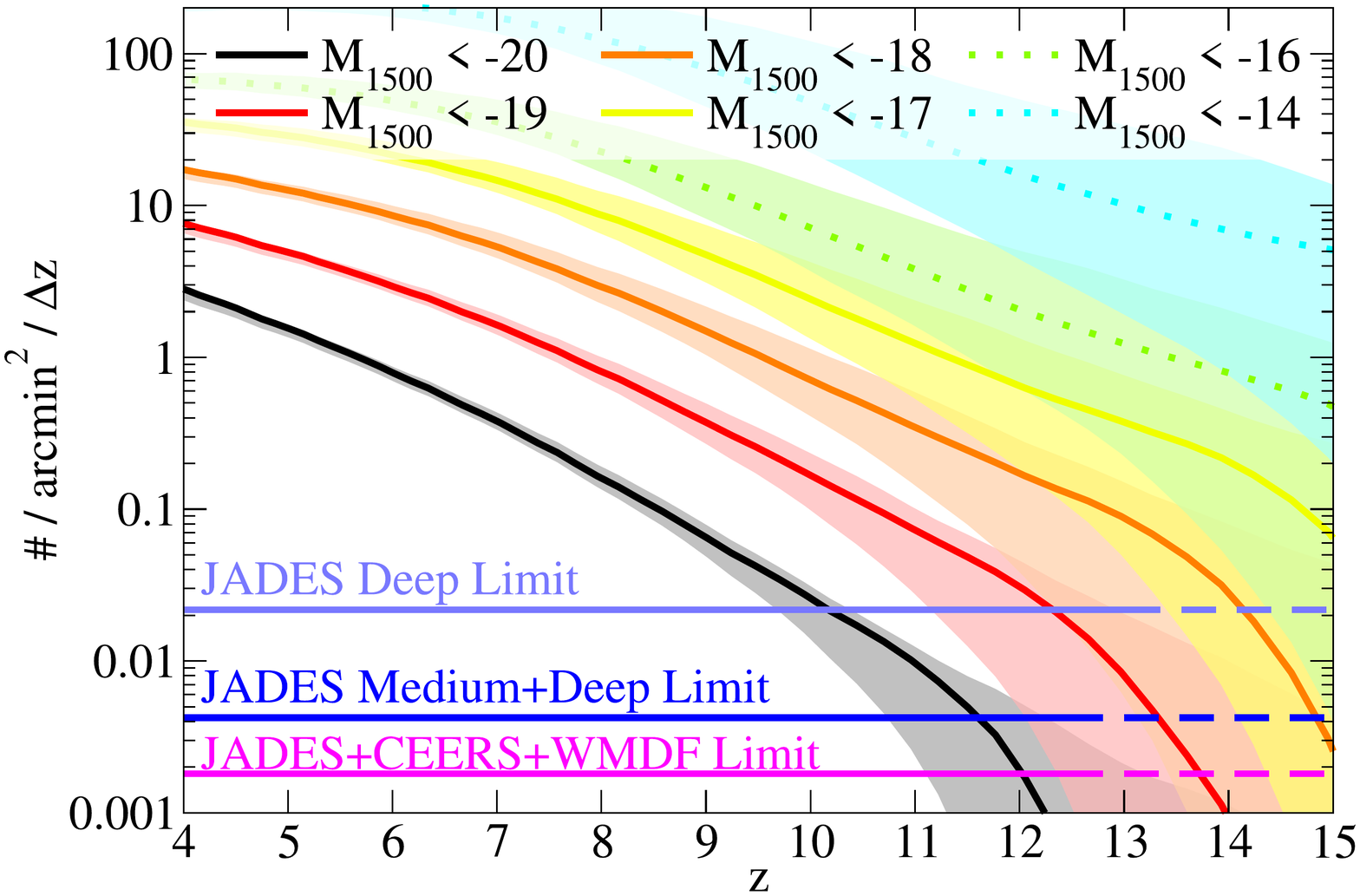}\hspace{-15ex}\\[-6.5ex]
    \caption{\textbf{Left} Panel: Cumulative  number densities for galaxies above specified stellar mass thresholds ($\log_{10} \Msun$) in the \textsc{UniverseMachine}.  Number densities are expressed as counts per unit angular area per unit redshift.  \textit{Horizontal lines} indicate the lowest number densities accessible for planned surveys (see Table \ref{t:obs_summary} for survey areas).  \textit{Solid} horizontal lines indicate redshifts at which the survey is likely ($>$85\% confidence) to detect galaxies given point-source limits in Table \ref{t:obs_summary}, and \textit{dashed} horizontal lines indicate lower confidence levels.  \textit{Dotted} lines indicate stellar masses below detection limits of current surveys.
    \textbf{Right} Panel: Cumulative number densities for galaxies brighter than specified UV luminosity thresholds (M$_{1500}$ AB).  Line styles have the same meaning as in the left panel.  For both panels, \textit{shaded regions} correspond to the $16-84^\mathrm{th}$ percentile confidence intervals.}
    \label{fig:cumul}
\end{figure*}

\begin{figure}
\vspace{-11ex}
\hspace{-10ex}
\includegraphics[width=1.15\columnwidth]{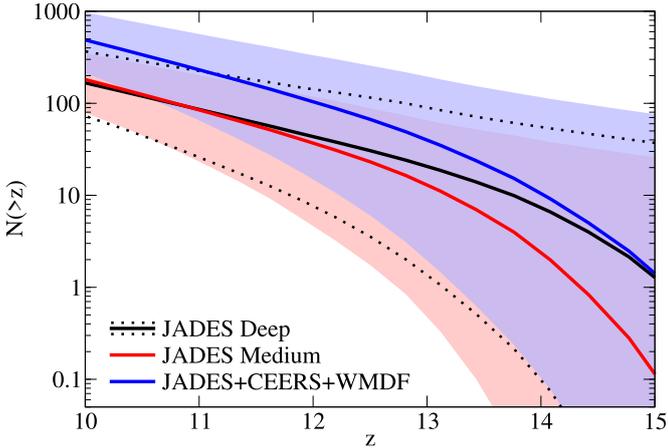}\\[-8ex]
    \caption{Mean expected galaxy number counts above a given redshift threshold for planned and combined \textit{JWST} Cycle 1 surveys (Table \ref{t:obs_summary}).  At redshifts $z>12$, JADES Deep is expected to provide the majority of detected objects.  \textit{Shaded regions} and \textit{dotted lines} (for JADES Deep) correspond to the $16-84^\mathrm{th}$ percentile confidence intervals.}
    \label{fig:survey_counts}
\end{figure}

\begin{figure*}
    \centering
    \vspace{-11ex}
    \hspace{-14ex}   \includegraphics[width=1.2\columnwidth]{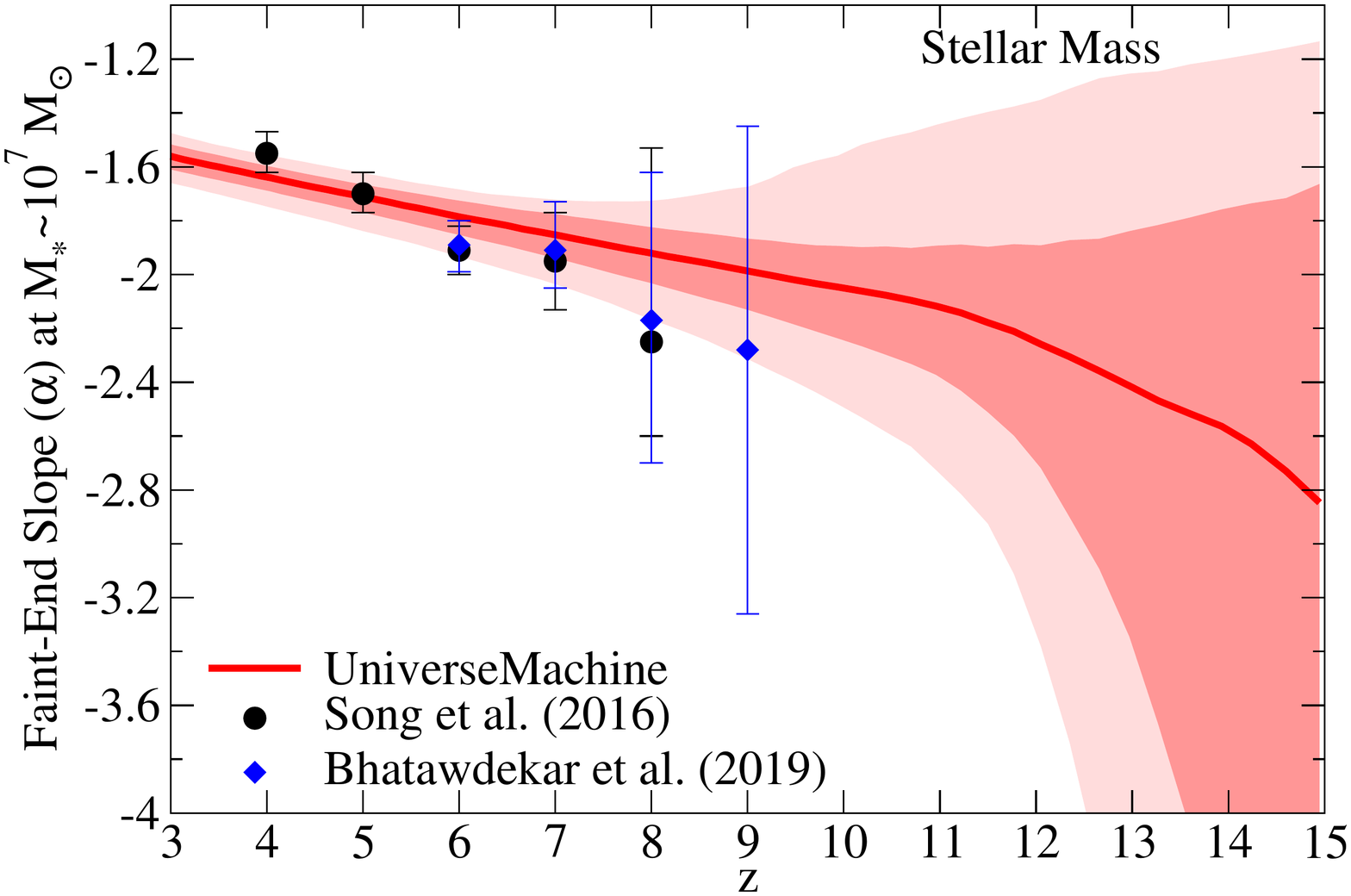}\hspace{-7ex}\includegraphics[width=1.2\columnwidth]{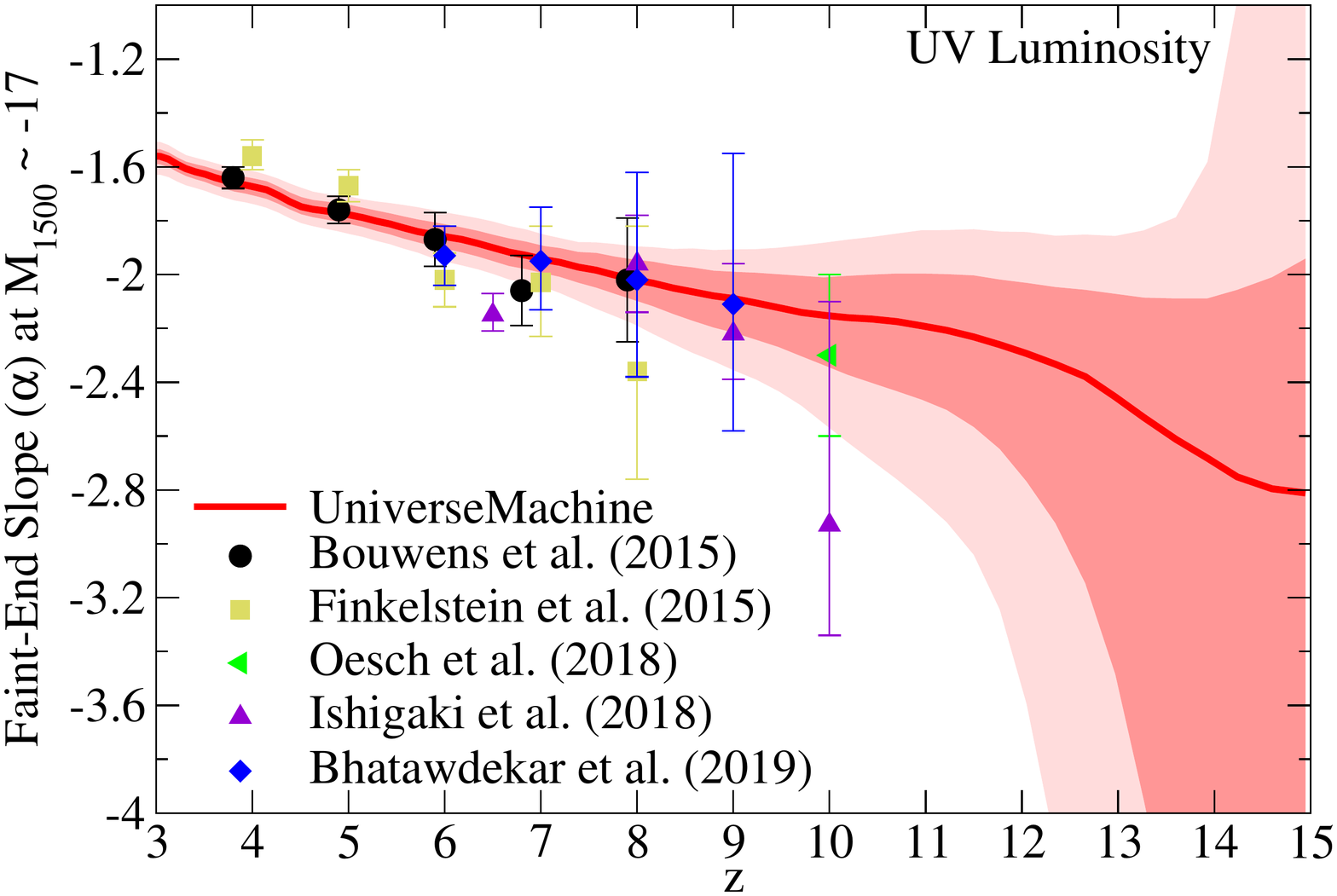}\hspace{-15ex}\\[-6.5ex]
    \caption{\textbf{Left} Panel: Constrained ($z< 8$) and predicted ($z \ge 8$) faint-end slopes of the stellar mass function as measured at $M_\ast = 10^{7}\Msun$ in the \textsc{UniverseMachine}.  Observationally measured slopes from \citet{Song15} and \citet{Bhatawdekar19} are shown for comparison.  \textbf{Right} Panel: Constrained ($z\le 8$) and predicted ($z > 8$) faint-end slopes of the stellar mass function as measured at $M_{1500} = -17$ in the \textsc{UniverseMachine}.  Observationally measured slopes from \citet{Bouwens15,Finkelstein15,Oesch18,Ishigaki18} and \citet{Bhatawdekar19} are shown for comparison.  For both panels, \textit{bold shaded regions} correspond to the $16-84^\mathrm{th}$ percentile confidence interval, and \textit{light shaded regions} correspond to the $3-97^\mathrm{th}$ percentile confidence interval.}
    \label{fig:slopes}
    \vspace{-5.5ex}
    \hspace{-14ex}\includegraphics[width=1.2\columnwidth]{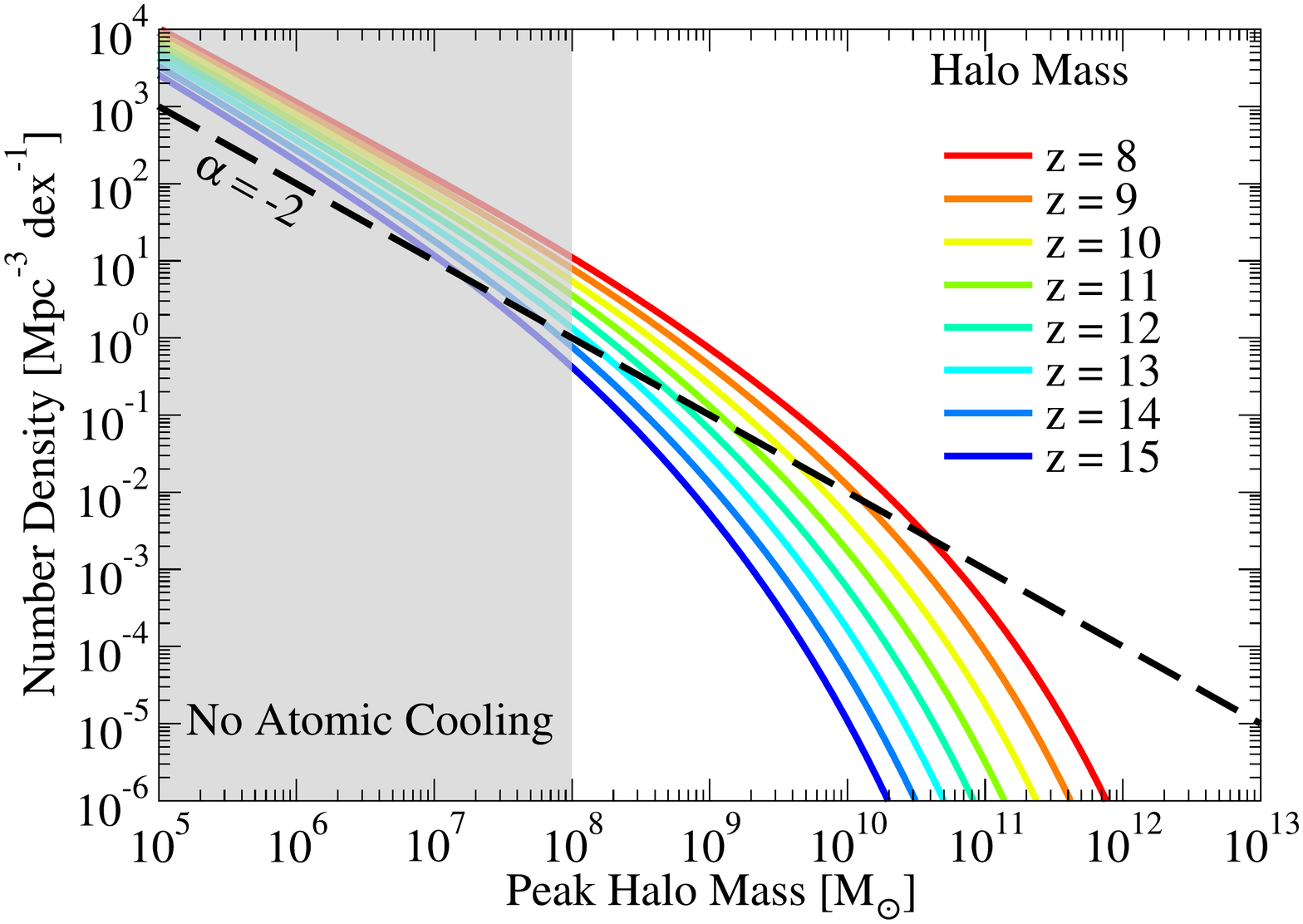}    \hspace{-7ex}\includegraphics[width=1.2\columnwidth]{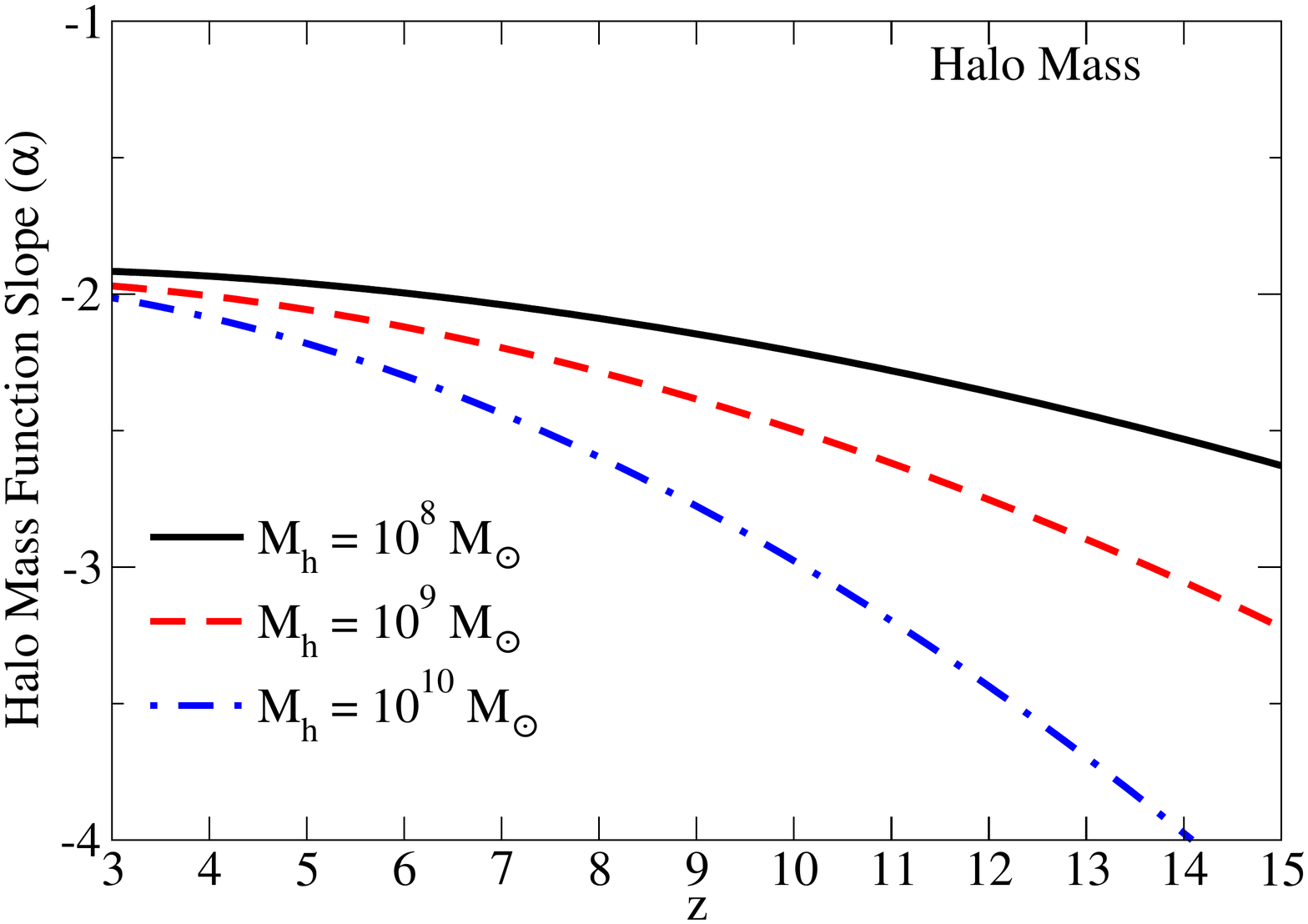}\hspace{-15ex}\\[-4ex]
    \caption{\textbf{Left} Panel: Halo mass functions for $z=8$ to $15$ for the \textit{Planck} cosmology \citep{Planck18}.  The \textit{dashed line} shows a constant power-law slope of $-2$.  Because the typical collapse mass at these redshifts is less than $10^5\Msun$, haloes above the atomic cooling limit ($\sim 10^8\Msun$) are all found in the exponentially falling regime of the halo mass function.
    \textbf{Right} Panel: Power-law slope of the halo mass function ($\frac{d\log N}{d\log M_h}-1$), evaluated at several halo masses.  For the $10^{10}\Msun$ haloes expected to host $10^7\Msun$ galaxies (Fig.\ \ref{fig:smhm}), the halo mass function slope is much steeper than $-2$ already by $z\sim 6$.  The slope also strongly depends on halo mass.  Extrapolating luminosity/mass functions to faint galaxies assuming fixed faint-end slopes hence requires care, as shown in Figures\ \ref{fig:extrap} and\ \ref{fig:csfr_completeness}.  See Section \ref{s:csfr} for a simple rule of thumb to minimize faint-end extrapolation biases.}
    \label{fig:hmfs}
    \centering
\end{figure*}
\begin{figure*}
\centering
    \vspace{-11ex}
    \hspace{-14ex}\includegraphics[width=1.15\columnwidth]{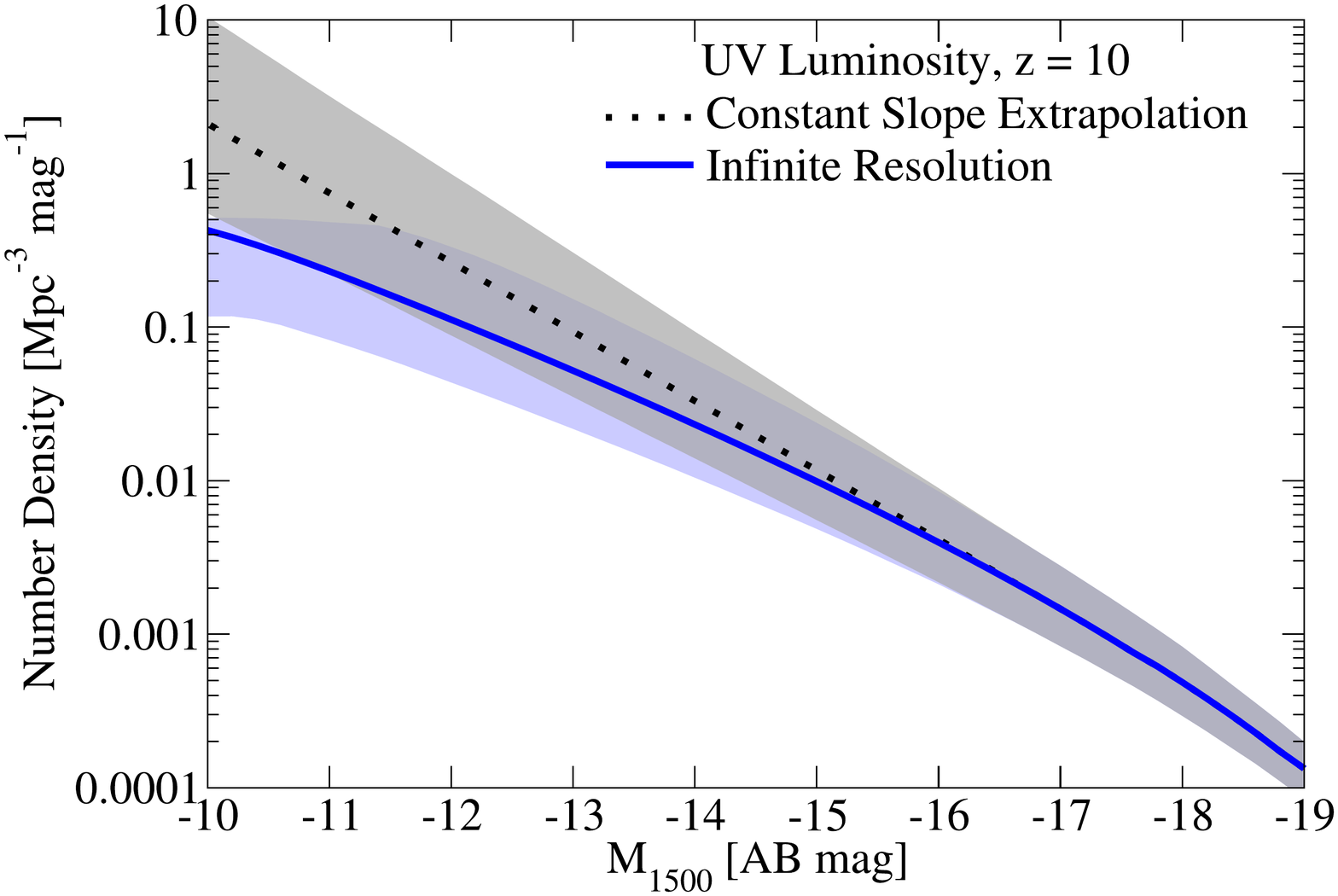}    \hspace{-5ex}\includegraphics[width=1.15\columnwidth]{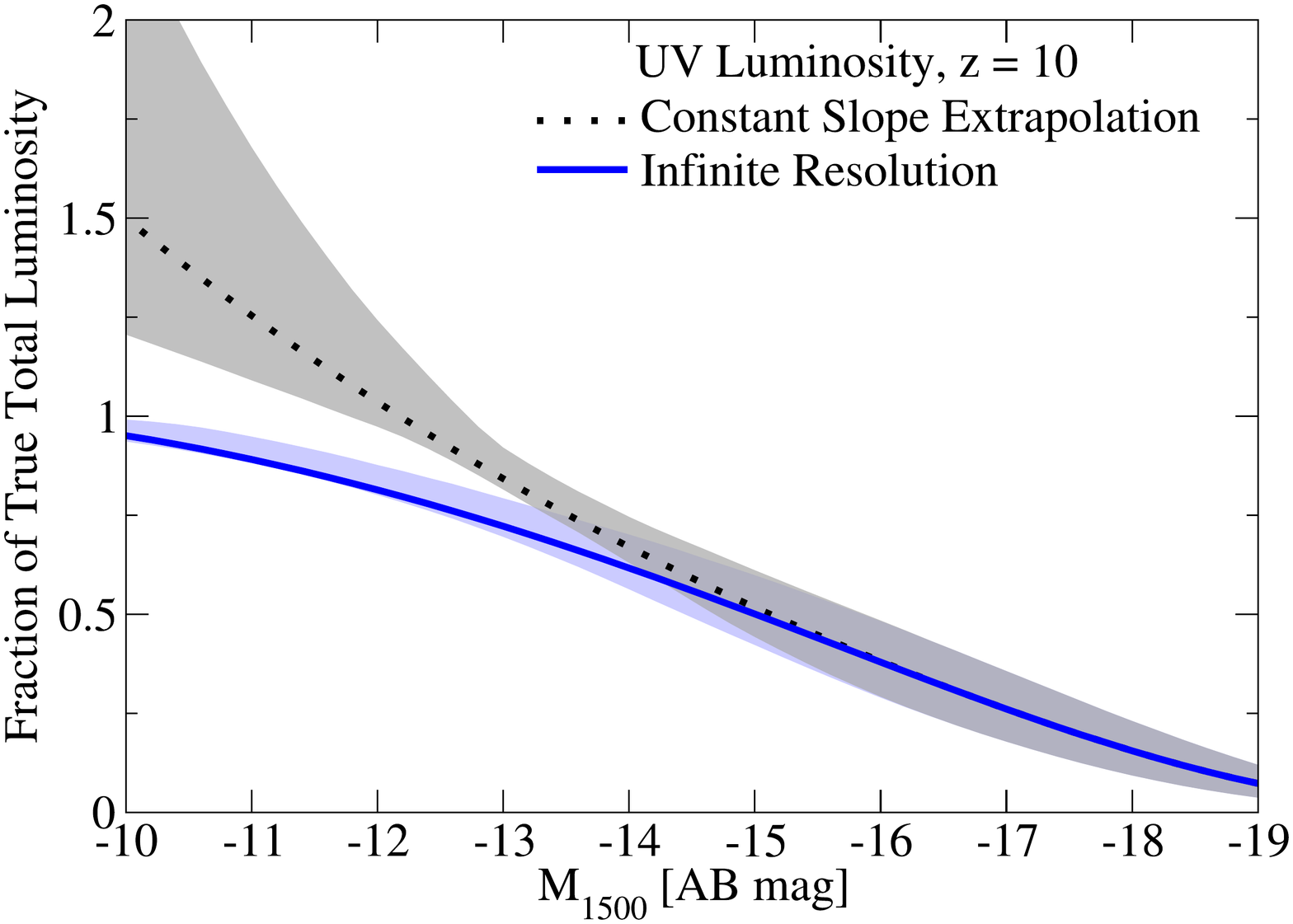}\hspace{-15ex}\\[-4ex]
    \caption{Assuming a constant faint-end slope for UV luminosity functions does not take into account the changing slope of the halo mass function, and therefore overestimates the number densities of faint galaxies.
    \textbf{Left} Panel: UV luminosity functions at $z=10$ resulting from applying the \textsc{UniverseMachine} to an infinite-resolution simulation (see Section \ref{s:resolution} and Appendix \ref{a:res_tests} for details and additional redshifts). The faint-end UV luminosity function slope becomes shallower for fainter galaxies due to the changing slope of the halo mass function (Fig.\ \ref{fig:hmfs}), and flattens entirely near the lowest halo mass for efficient star formation ($\sim 10^8 \Msun$; \citealt{OShea15,Xu16}).  For this extrapolation, the slope of the SFR--halo mass relation is held fixed.  Any additional physics that reduces star formation efficiency in low-mass haloes would result in even lower number densities of faint galaxies. 
    \textbf{Right} Panel: Fraction of total UV luminosity above a given magnitude threshold.  Assuming a constant UV luminosity function slope fainter than $M_{1500}=-17$ results in an overestimate of the total luminosity if integrated below $M_{1500}=-12$ (Section \ref{s:csfr}).  In both panels, error bars indicate the $16-84^\mathrm{th}$ percentile confidence interval.}
    \label{fig:extrap}
    \end{figure*}
    \begin{figure}
    \vspace{-8ex}
    \hspace{-10ex}\includegraphics[width=1.15\columnwidth]{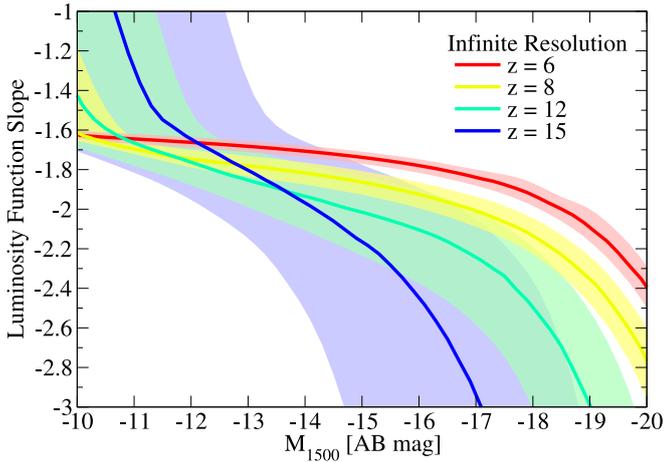}\\[-6ex]
    \caption{Luminosity function slopes ($\frac{\mathrm{d}\,\log N}{\mathrm{d}\,\log L}-1$) for the \textsc{UniverseMachine} applied to an infinite-resolution simulation, assuming no change in the form of the SFR--halo mass relation (Appendix \ref{a:res_tests}).  At $z>8$, there is \textit{never} an asymptotic faint-end slope.  This arises because the slope of the halo mass function (Fig.\ \ref{fig:hmfs}) continues increasing down to the assumed halo mass limit for forming stars ($M_h = 10^{8}\Msun$).  Below this limit, the luminosity function is assumed to turn over, resulting in a rapidly rising power-law slope at an $M_\mathrm{1500}$ of $-10$ to $-12$, depending on redshift.}
        \label{fig:extrap_slopes}
\end{figure}
\begin{figure*}
    \centering
    \vspace{-8ex}
    \hspace{-14ex}\includegraphics[width=1.15\columnwidth]{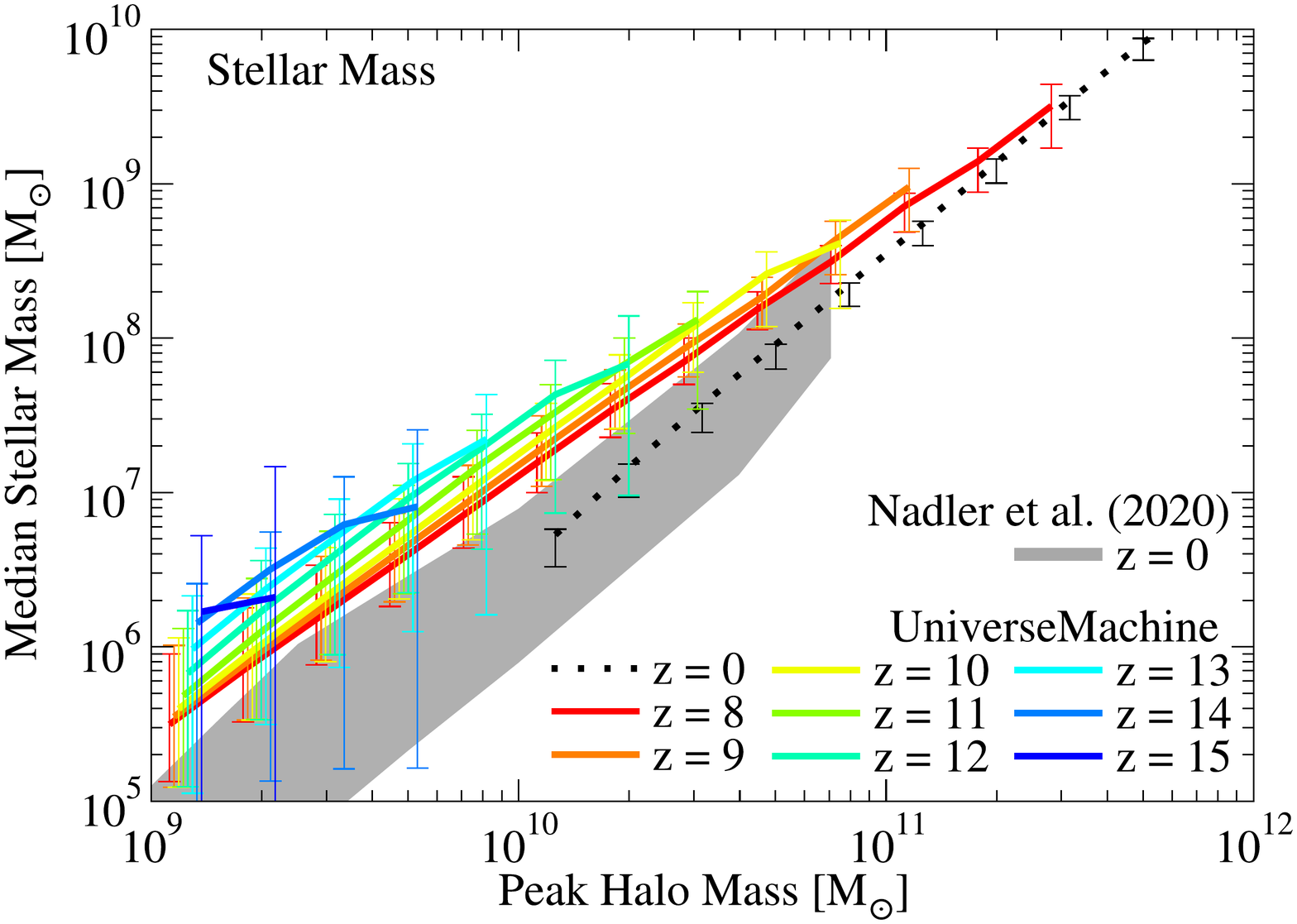}    \hspace{-7ex}\includegraphics[width=1.15\columnwidth]{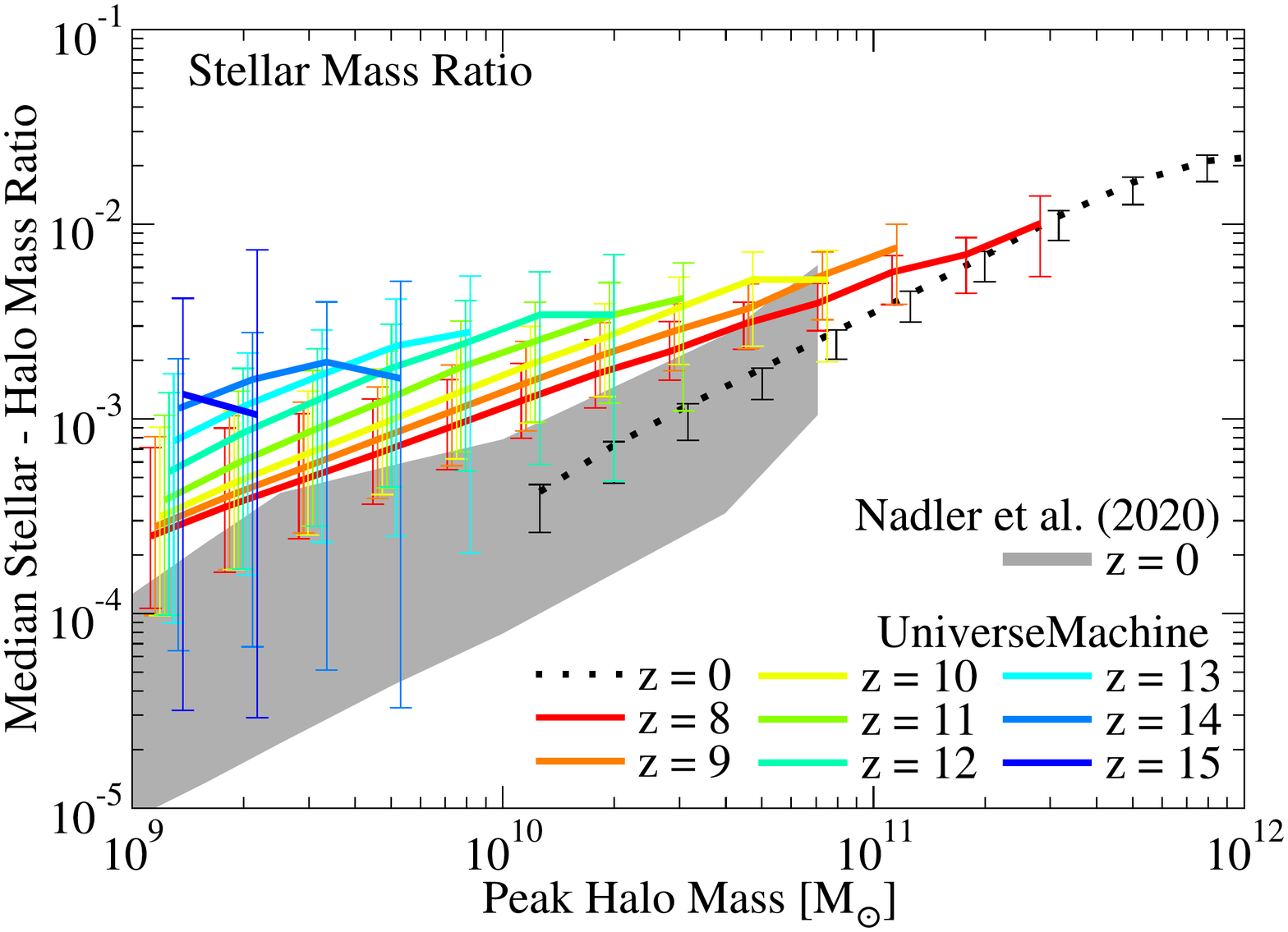}\hspace{-15ex}\\[-4ex]
    \caption{\textbf{Left} Panel: Median stellar masses as a function of halo mass from $z=0$ to $z=15$ from the \textsc{UniverseMachine}.  The grey shaded region shows the $z=0$ relation from \citet{Nadler20}.
    \textbf{Right} Panel: Same, except expressed as the ratio of stellar mass to halo mass.  In \textit{Planck} cosmologies, a ratio of $\sim 0.16$ would imply that 100\% of available baryons were converted to stars.  In both panels, error bars and shaded regions indicate the $16-84^\mathrm{th}$ percentile confidence interval.  Error bars have been offset by up to 0.05 dex in halo mass to increase clarity.}
    \label{fig:smhm}
    \vspace{-5ex}
    \hspace{-14ex}\includegraphics[width=1.15\columnwidth]{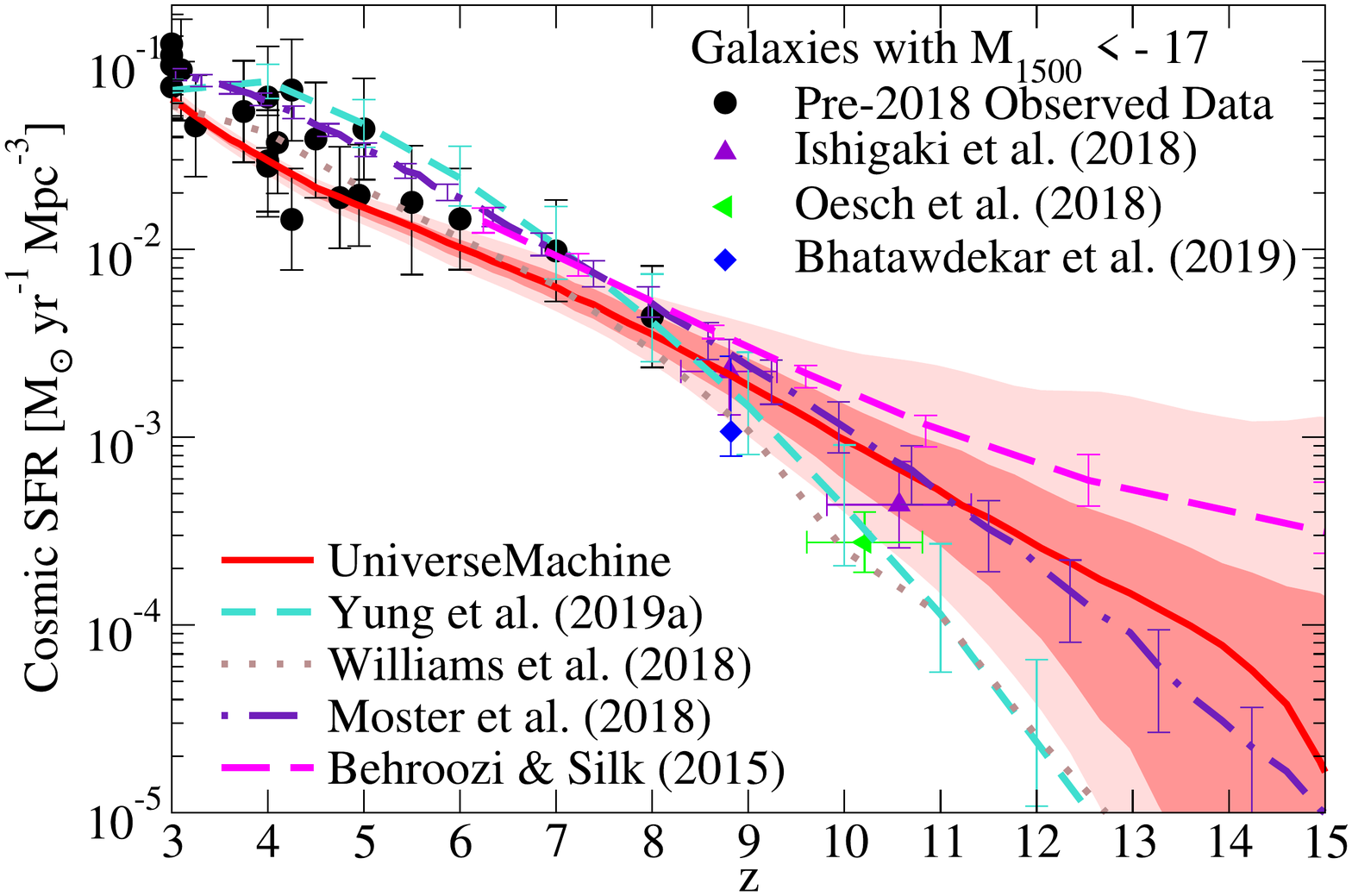}\hspace{-7ex}\includegraphics[width=1.15\columnwidth]{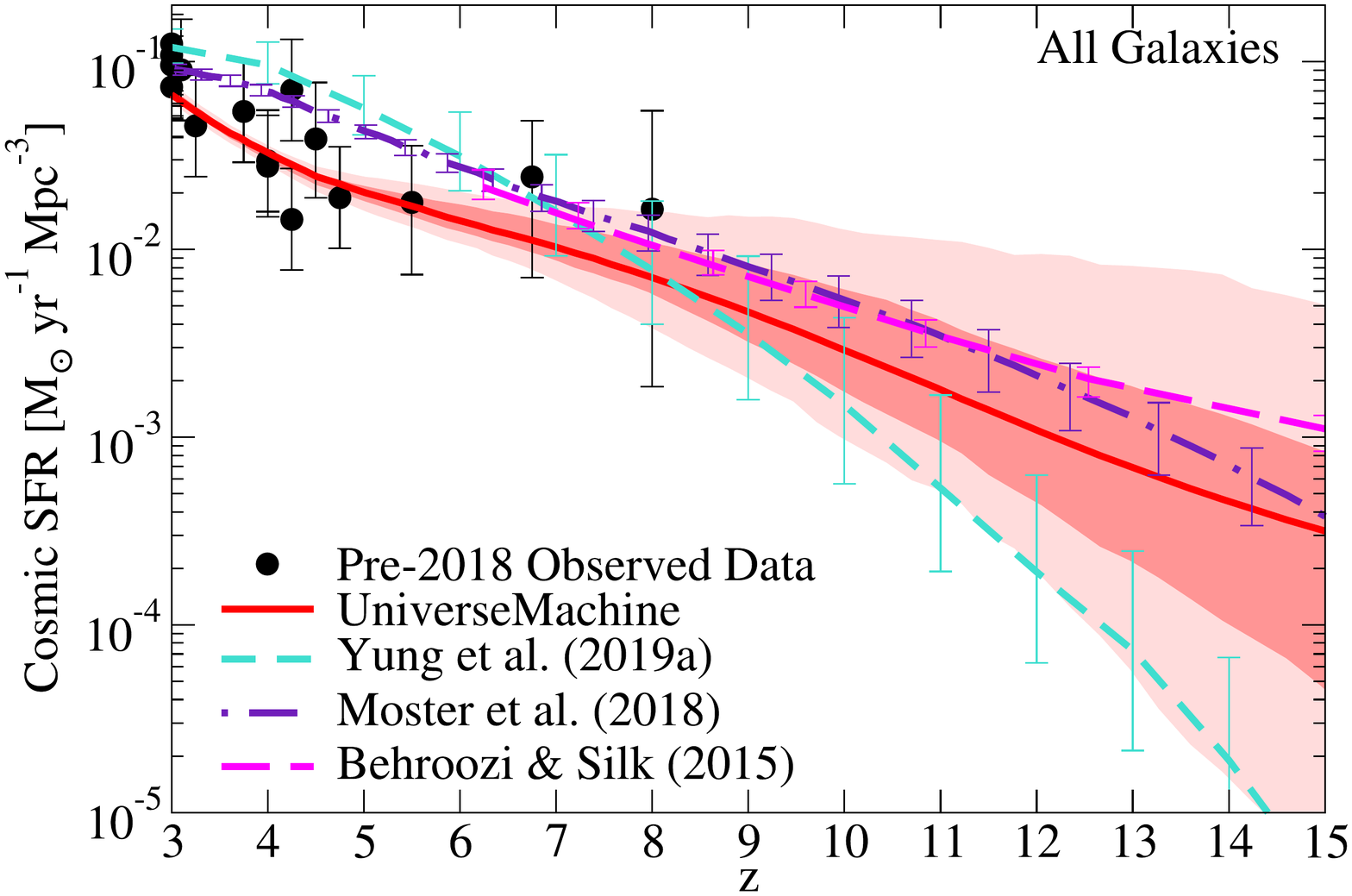}\\[-6ex]
    \caption{\textbf{Left} Panel: \underline{Observed} ($M_\mathrm{1500}<-17$) cosmic star formation rates from the \textsc{UniverseMachine}, with predictions extending to $z=15$.  \textit{Black points} are from multiple observations \citep{Yoshida06,vdBurg10,Cucciati11,Finkelstein15}.  \textbf{Right} Panel: \underline{Total} cosmic star formation rates from the \textsc{UniverseMachine}.  \textit{Black points} are from multiple observations \citep{Yoshida06,vdBurg10,Cucciati11,Kistler13}.  In both panels, bold and light shaded regions correspond to the $16-84^\mathrm{th}$ and $3-97^\mathrm{th}$ percentile confidence intervals, respectively.}
    \label{fig:csfrs}
    \vspace{-5ex}
    \hspace{-14ex}\includegraphics[width=1.15\columnwidth]{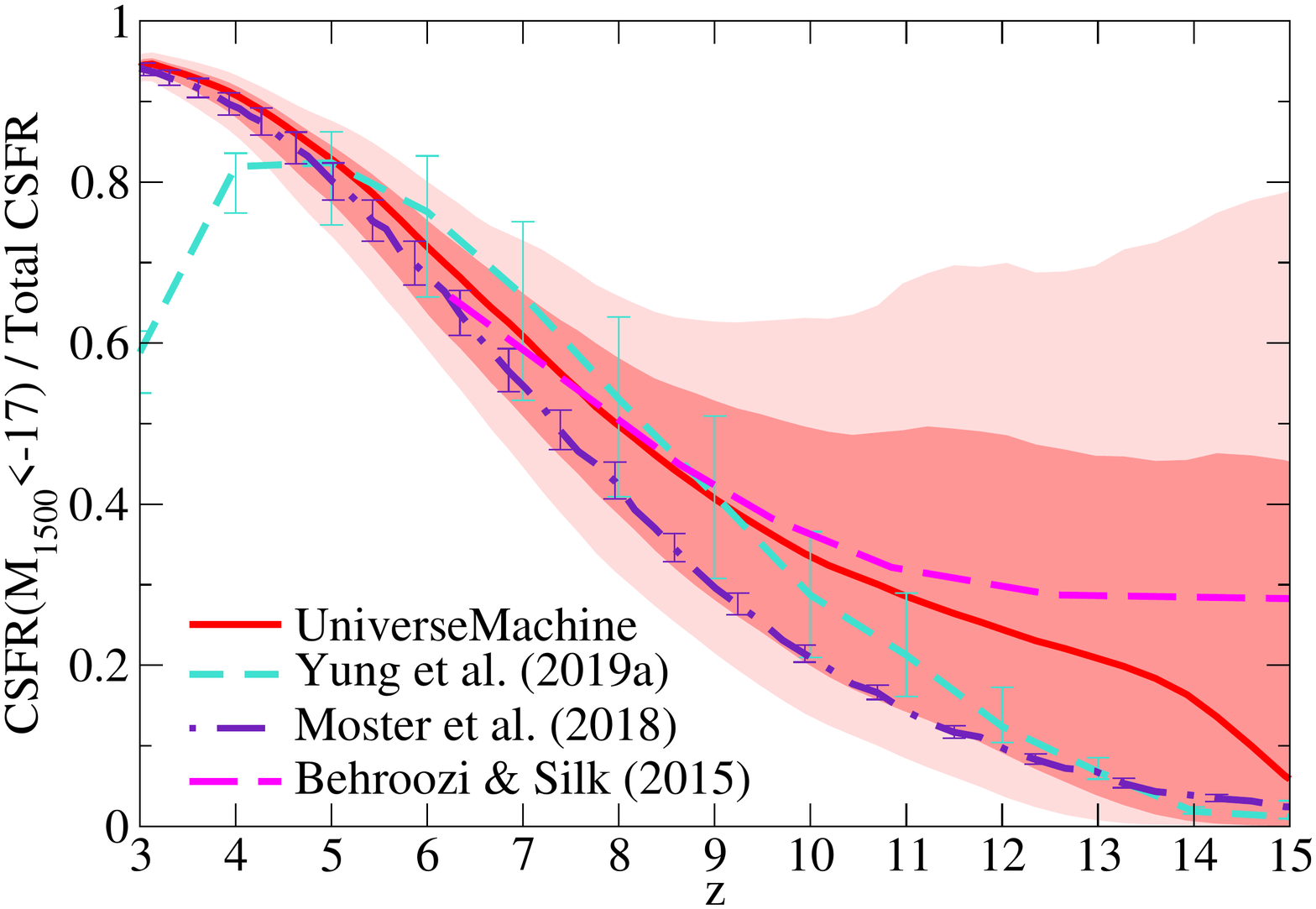}\hspace{-7ex}\includegraphics[width=1.15\columnwidth]{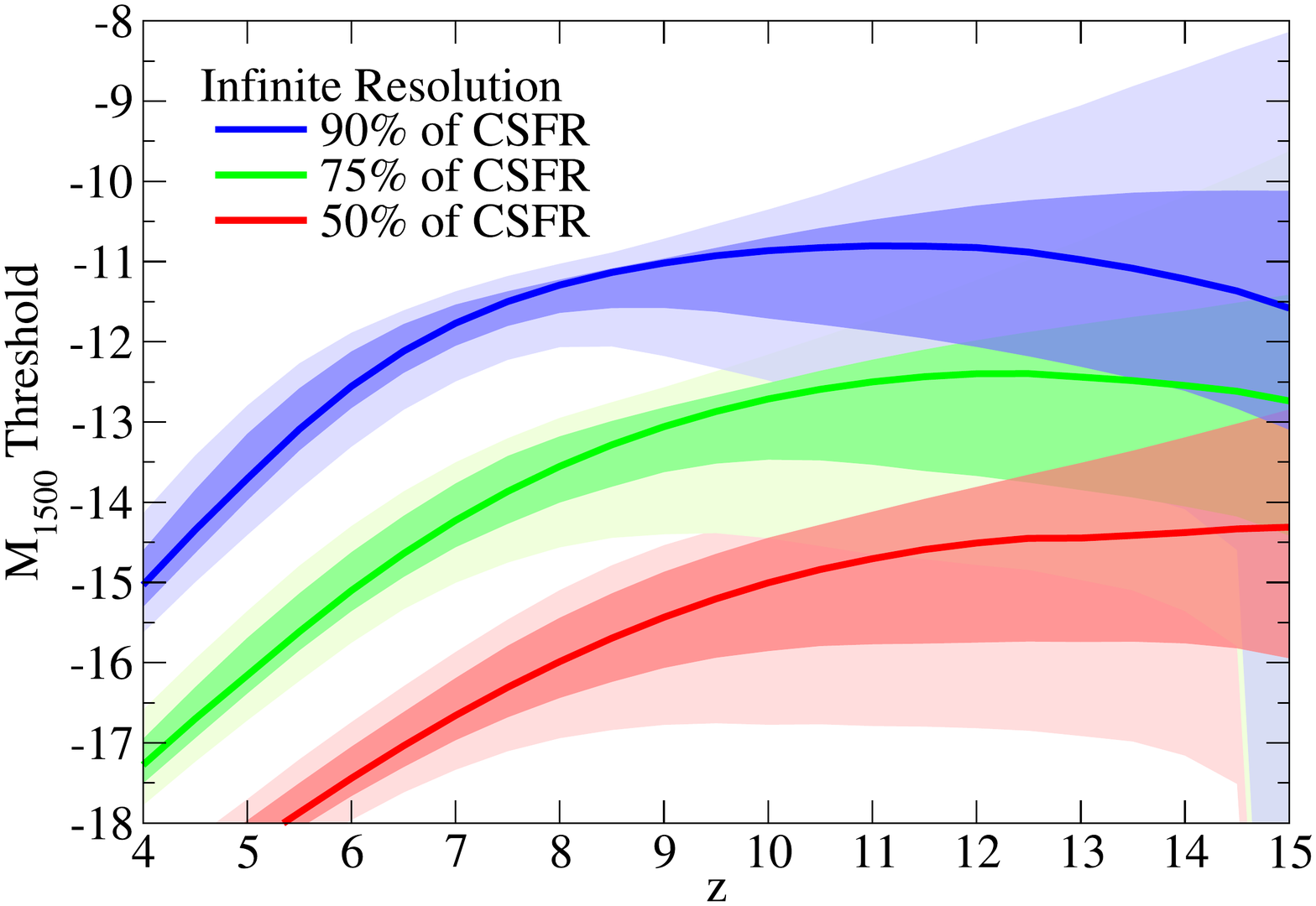}\\[-6ex]
    \caption{\textbf{Left} Panel: Predicted ratios between observed ($M_\mathrm{1500}<-17$) and total cosmic star formation rates from the \textsc{UniverseMachine}.  \textbf{Right} Panel: Luminosity thresholds below which 50\%, 75\%, and 90\% of all star formation occurs, for the \textsc{UniverseMachine} applied to an infinite-resolution simulation (Section \ref{s:resolution}, Appendix \ref{a:res_tests}). In both panels, bold and light shaded regions correspond to the $16-84^\mathrm{th}$ and $3-97^\mathrm{th}$ percentile confidence intervals, respectively.}
    \label{fig:csfr_completeness}
\end{figure*}

\begin{table*}
\caption{Planned extragalactic \textit{JWST} blank field surveys.}
\vspace{-4ex}
\begin{tabular}{lcclrclrcl}
    \phantom{ }& & & & \multicolumn{3}{c}{Mean Predicted} & \multicolumn{3}{c}{Mean Predicted}\\
    Survey Name & Area ($\square'$) & Depth & Reference & \multicolumn{3}{c}{$z>10$ Galaxies} & \multicolumn{3}{c}{$z>12$ Galaxies}\\
\hline
    JADES Deep & 46 & $\sim$-17 & \cite{Rieke19} & 72 & $-$ & 364 & 7.5 & $-$ & 142\\
    JADES Medium & 190 & $\sim$-18 & \cite{Rieke19} & 78 & $-$ & 350 & 4 & $-$ & 118\\
    CEERS & 97 & $\sim$-18.5 & \cite{Finkelstein17} & 18 & $-$ & 79 & 0.8 & $-$ & 22 \\
    WMDF & 220 & $\sim$-18.5 & \cite{Windhorst17} & 42 & $-$ & 179 & 1.7 & $-$ & 50\\
    \hline
    Total & 553 & $>$-18.5 & & \phantom{xyz}210 & $-$ & 972\phantom{xyz} & 14 & $-$ & 332\\
\end{tabular}
\parbox{1.7\columnwidth}{\textbf{Notes.} JADES: \textit{JWST} Advanced Deep Extragalactic Survey.  CEERS: Cosmic Evolution Early Release Science.  WMDF: Webb Medium-Deep Fields.  Depths are $M_\mathrm{1500}$ (AB) for point sources at the expected limiting redshift of the survey.  These were estimated from the 5-$\sigma$ point source depths with the F200W filter (2$\mu$m), assuming a flat UV spectrum.  Counts at $z>10$ and $z>12$ represent the $16-84^\mathrm{th}$ percentile range for the average number of galaxies per field.}  
\label{t:obs_summary}
\end{table*}

\begin{figure}
\vspace{-11ex}
\hspace{-10ex}\includegraphics[width=1.15\columnwidth]{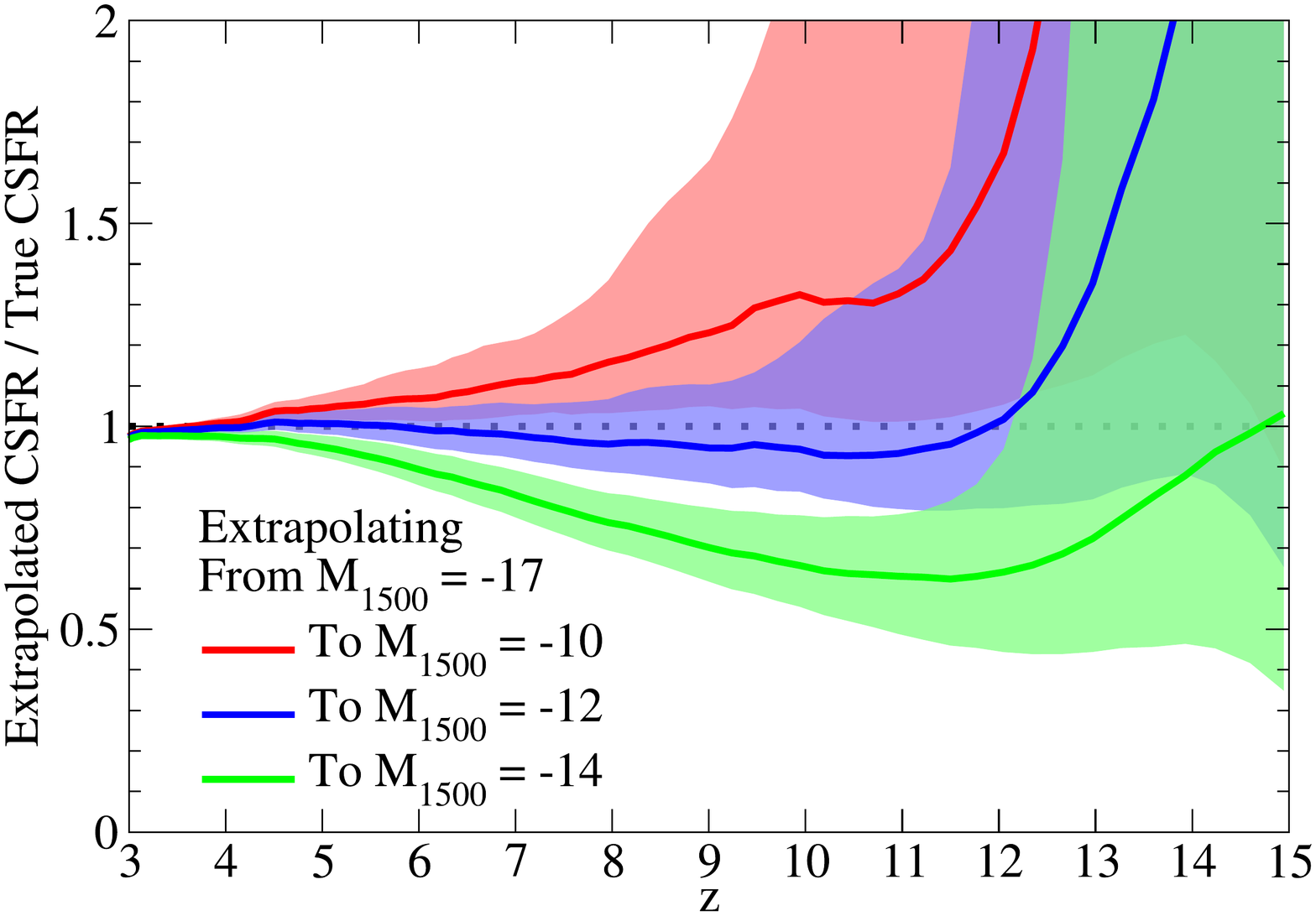}\\[-6ex]
\caption{Commonly, the total CSFR is estimated from extrapolating the observed luminosity function with a constant slope to faint magnitudes.  This figure shows the ratio between extrapolated and true total cosmic star formation rates from the \textsc{UniverseMachine}, with predictions extending to $z=15$.  Extrapolations assume a constant UV luminosity function slope from $M_{1500}=-17$ down to $M_{1500} = -10$, $-12$, and $-14$, respectively.  Extrapolations down to $M_{1500} = -12$ are closest to the true \textsc{UniverseMachine} total regardless of parameter set until $z\sim 12$.  At higher redshifts, the luminosity function at $M_{1500}<-17$ may contain too little information to make extrapolation feasible.  Hence, extrapolating to $M_{1500} = -12$ gives a reasonable upper limit to the total CSFR (see text).  Shaded regions correspond to the $16-84^\mathrm{th}$ percentile confidence intervals.}
\label{fig:csfr_integrate}
\vspace{-5ex} \hspace{-10ex}\includegraphics[width=1.15\columnwidth]{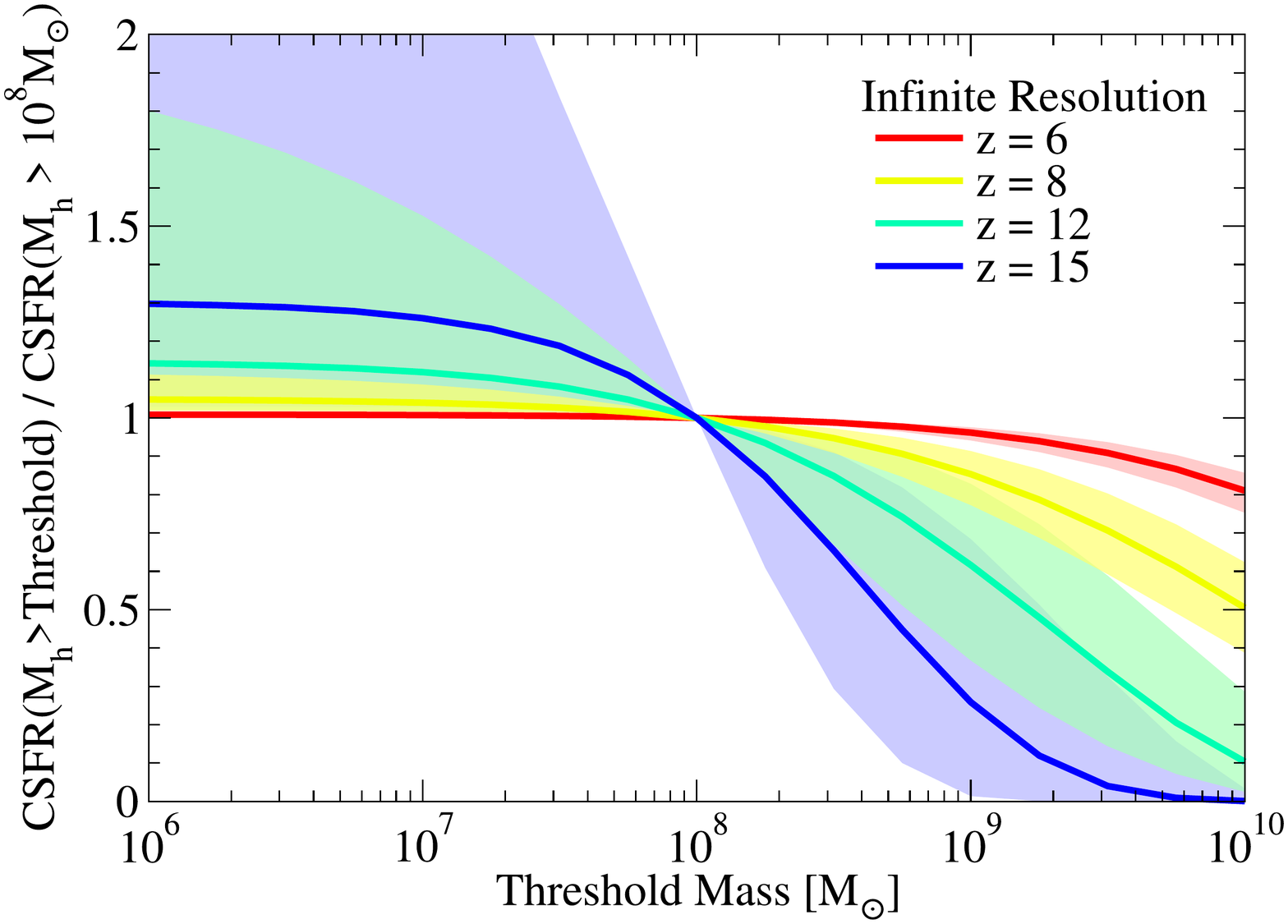}\\[-6ex]
\caption{Effect of the threshold halo mass for star formation on the total cosmic star formation rate, for the \textsc{UniverseMachine} applied to an infinite-resolution simulation (Section \ref{s:resolution}, Appendix \ref{a:res_tests}). Faint-end slopes for luminosity functions become shallower than $-2$ (Fig.\ \ref{fig:extrap_slopes}), so the threshold halo mass has relatively little effect at $z<12$ if it is below $10^{8.5}\Msun$.}
    \label{fig:threshold_mass}
\end{figure}

\subsection{Resolution and Cosmology Tests}

\label{s:resolution}

As explained in \S \ref{s:effective}, SFRs in the \textsc{UniverseMachine} at $z>4$ are mainly a function of halo mass and $z$.  Hence, given only halo number densities as a function of mass and redshift, we can still generate the cosmic distribution of SFRs:
\begin{equation}
    \phi(SFR, z) = \int_0^\infty P(SFR|M_h,z) \phi(M_h, z)dM_h, \label{e:infinite}
\end{equation}
where $P(SFR|M_h,z)$ is the distribution of SFRs as a function of peak halo mass ($M_h$) and $z$, and $\phi(M_h,z)$ is the halo mass function.  Similarly, we can generate UV luminosity functions by applying a UV--SFR relation (Eq.\ \ref{e:kappa_fuv}) as well as Eqs.\ \ref{e:dust1}--\ref{e:dust2} to the SFR distribution in Eq.\ \ref{e:infinite}.

As a result, we can generate several key comparisons (e.g., the evolution of the total CSFR, as well as the UV luminosity function) from halo mass functions alone.  Because halo mass function fits do not have intrinsic resolution limits, we can closely approximate the behaviour of the \textsc{UniverseMachine} on an infinite-resolution simulation.  In all plots in this paper, the label ``\textbf{Infinite Resolution}'' refers to applying Eqs.\ \ref{e:dust1}--\ref{e:infinite} and \ref{e:kappa_fuv} to the halo mass function in \citealt{tinker-umf}, as modified to include subhaloes by \citealt{BWC13}.  Full details of this approach, including the fitting of the UV--SFR relation used, are presented in Appendix \ref{a:res_tests}.

Even with infinite resolution, haloes with virial temperatures below the atomic cooling limit ($T_\mathrm{cool} \sim 8000$K) are not expected to form stars efficiently.  For the \textit{Planck} cosmology adopted here, we find that this halo mass limit is fit well by:
\begin{equation}
    M_\mathrm{thresh,vir}(a) = \frac{10^{9.76}\Msun \left(\frac{T_\mathrm{cool}}{8000\mathrm{K}}\right)^\frac{3}{2}}{(0.704a)^{-1.585} + (0.704a)^{-1}}.
\end{equation}
Over the redshift range considered here ($z=10$ to $15$), this varies by $<0.3$ dex, so we adopt for simplicity a fixed threshold mass of $10^{8}\Msun$, below which we assume that star formation does not occur.  This limit is similar to that found in cosmological simulations \citep{OShea15,Xu16}.  Although some residual star formation occurs in lower-mass haloes due to metal pollution and other reasons \citep{Smith15b,Aykutalp19,Nadler20}, this is not a significant contribution to either the observable luminosity function or the total CSFR at $z=10$ to $15$.

A natural application of the ``Infinite Resolution'' approximation is to verify that the \textit{VSMDPL} simulation resolves most star formation at $z<15$.  Details of these tests are provided in Appendix \ref{a:res_tests}.  Assuming that galaxy formation becomes rapidly more inefficient for haloes below the adopted atomic cooling limit ($M_h = 10^8\Msun$), \textit{VSMDPL} is $>90$\% complete for all star formation to at least $z=13$, and still 80\% complete by $z=15$.  We also use the ``Infinite Resolution'' approximation to investigate faint-end slopes of luminosity functions down to $M_{1500}=-10$ and the effects of a different threshold mass for star formation in haloes (Sections \ref{s:faint} and \ref{s:csfr}, as well as Appendix \ref{a:res_tests}).

Beyond resolution tests, we also evaluate cosmology uncertainties in Appendix \ref{a:cosmo}.  We find that \textit{Planck} cosmology uncertainties at $z>10$ are subdominant ($<0.2$ dex) to uncertainties in galaxy formation ($\gtrsim 0.4$ dex), even considering existing tensions in $h$.

\subsection{Lightcones}

\label{s:lightcones}

Lightcones are generated from galaxy catalogues using the \texttt{lightcone} tool provided with the \textsc{UniverseMachine} DR1.  For each lightcone realization, this tool chooses a random origin and viewing angle within the simulation, and includes all galaxies that fall within a user-specified survey area.  The simulation is assumed to be periodically repeated in all directions, and distance along the lightcone axis determines the redshift of the simulation snapshot from which galaxy and halo properties are taken.  The lightcones in this paper were allowed to pass through the same region of the simulation volume multiple times.  Because the lightcones are for pencil-beam surveys, this results in correlations only across large redshift ranges ($\Delta z > 1$).

We generate 8 lightcone realizations for each of the 5 CANDELS fields (COSMOS, EGS, GOODS-N, GOODS-S, and UDS; \citealt{Grogin11,Koekemoer11}) using the best-fit parameter set from the \textsc{UniverseMachine} DR1.  We repeat this process for nine additional parameter sets chosen randomly from the model posterior distribution.  Lightcone origins and viewing angles are maintained across different parameter sets so that the effects of sample variance and model variance can be evaluated independently.  The 400 lightcones thus generated ($8\times5\times10$) are publicly available.\footnote{\url{https://peterbehroozi.com/data.html}}  Of note, all figures shown in this paper use the full catalogues (with 1000 model parameter set realizations) instead of the reduced data available in the lightcones.

\section{Results}

\label{s:results}

\subsection{Mass and Luminosity Functions Visible with JWST}

\label{s:mfs}

Predicted stellar mass and UV luminosity functions (Fig.\ \ref{fig:nds}) from the \textsc{UniverseMachine} show rapidly decreasing number densities at $z>8$.  As at redshifts $4<z<8$ (Fig.\ \ref{fig:existing_data}), the redshift evolution in bright galaxy number counts is much larger than for faint galaxy counts.  This is mainly due to larger haloes having more efficient star formation than smaller haloes (Fig.\ \ref{fig:halo_sfrs}).  As the number density of larger haloes drops rapidly at higher redshifts, so too does the number density of bright galaxies.

The evolution of the UV luminosity function is less rapid than that of the stellar mass function.  Specific star formation rates increase at higher redshifts due to shorter halo assembly times \citep[e.g.,][]{BehrooziHighZ}, leading to higher light-to-mass ratios that partially offset lower number densities at a given stellar mass.  At $z>10$, dust is less significant (typically $<1$ mag) and metallicity uncertainties are $<$0.2 mag (Section \ref{s:effective}), so uncertainties in UV luminosity functions are similar to those for stellar mass functions.  Practically, this also means that UV luminosity function measurements are as useful as stellar mass function measurements for constraining the \textsc{UniverseMachine} at $z>10$.

To aid with planning future surveys, Fig.\ \ref{fig:cumul} shows the expected cumulative number densities for detected galaxies above specified mass and luminosity thresholds.  For comparison, this figure also shows the corresponding areas and redshift ranges for planned \textit{JWST} surveys (Table \ref{t:obs_summary}).  For example, the \mbox{JADES Deep} survey is expected to detect galaxies with $M_*>10^7\Msun$ or $M_\mathrm{1500} < -17$ out to $z\sim 13.5$ at a confidence level of $>85$\%.  Detecting $z\sim 15$ galaxies with similar confidence would, with current uncertainties, require a $\sim$2 magnitude deeper survey.

Fig.\ \ref{fig:survey_counts} summarizes the galaxy counts expected in planned Cycle 1 surveys.  The shallower, wider fields in Table \ref{t:obs_summary} are expected to find many galaxies with $M_*>10^8\Msun$ or $M_\mathrm{1500} < -18.5$ out to $z\sim 12$.  For redshifts $z>12$, the JADES Deep survey is likely to contain most of the galaxies in the combined fields.  Nonetheless, at $z\ge 12$, uncertainties in galaxy number densities rapidly increase, exceeding $\pm 1.5$ dex at $z\sim 15$ (Figs.\ \ref{fig:cumul} and \ref{fig:survey_counts}).  Hence, even non-detections in shallower \textit{JWST} surveys at these high redshifts will give valuable constraints on galaxy evolution models.

\subsection{Faint-End Slopes}

\label{s:faint}

The predicted mass and luminosity functions (Fig.\ \ref{fig:nds}) also show steep faint-end slopes at higher redshifts.  Constrained (at $z<8$) and predicted (at $z>8$) faint-end slopes for stellar mass and luminosity functions at observable thresholds ($M_*>10^7\Msun$ or $M_\mathrm{1500} < -17$) are shown in Fig.\ \ref{fig:slopes}.  We find generally excellent agreement with existing observations at $z\le 10$, and predict that observable faint-end power-law slopes will continue steepening past $-2$ at $z>10$.

Such steep slopes arise naturally from halo mass functions in $\Lambda$CDM.  Halo mass functions have exponential falloffs beyond the typical Press-Schechter collapse mass $M_C$ (i.e., the mass where typical density fluctuations are $\sigma(M_C)= \delta_c = 1.686$).  $M_C$ decreases rapidly as redshift increases.  For the cosmology we use here, $M_C$ is $10^{12.8}\Msun$ at $z=0$, but falls below $10^9\Msun$ by $z=3$, and is $\sim 10^5\Msun$ at $z=8$ \citep{RP16b}. This is far below the atomic cooling limit.  Fig.\ \ref{fig:hmfs} shows that, as a result, host haloes for all star-forming galaxies at $z\ge 8$ are in the exponentially falling region of the halo mass function.

For example, for the $10^{10}\Msun$ haloes expected to host $10^7\Msun$ galaxies \citep{BWHC19}, the slope of the halo mass function is steeper than its asymptotic value of $-2$ already by $z\sim 3$.  At $z\sim 8$, the slope steepens to $-2.6$, and at $z\sim 15$, it falls below $-4$.  To help understand how the halo mass function slopes in Fig.\ \ref{fig:hmfs} relate to the galaxy mass and luminosity function slopes in Fig.\ \ref{fig:slopes}, we can make a simple analytic calculation.  If the halo mass function has a power-law slope of $\alpha$ (so $\frac{\mathrm{d}N}{\mathrm{d}M_h} \propto M_h^\alpha$) and the stellar mass--halo mass relation has a power-law slope of $\beta$ (so $M_\ast \propto M_h^\beta$), we can derive the shape of the stellar mass function as:
\begin{equation}
    \frac{\mathrm{d}N}{\mathrm{d}M_*} =  \frac{\mathrm{d}N}{\mathrm{d}M_h}\frac{\mathrm{d}M_h}{\mathrm{d}M_*} \propto  M_*^{\frac{\alpha+1}{\beta}-1}.
\end{equation}
An identical calculation applies for luminosity functions.  For typical values of $\beta\sim2$ \citep{BWHC19}, the stellar mass and luminosity functions will have a shallower slope than the halo mass function.  For example, a halo mass function slope of $-2.6$ for $10^{10}\Msun$ haloes at $z=8$ corresponds to a stellar mass function slope of $\sim -1.8$ at $M_\ast = 10^{7}\Msun$, as shown in Fig.\ \ref{fig:slopes}.

As a result, stellar mass and luminosity function slopes much steeper than $-1.5$ imply a halo mass function slope that is much steeper than $-2$, which occurs at $z\ge6$ for both galaxy mass/luminosity functions (Fig.\ \ref{fig:slopes}) and halo mass functions (Fig.\ \ref{fig:hmfs}).  At the same time, because the halo mass function becomes shallower at lower halo masses, the galaxy mass and luminosity functions also become shallower for fainter galaxies.  A direct result is that the observed faint-end slopes for galaxy mass and luminosity functions that are steeper than $\sim -1.5$ cannot be constant, and must instead become shallower for galaxies below current observable limits.

Fig.\ \ref{fig:extrap} shows this effect at $z=10$, based on the infinite resolution computation from Section \ref{s:resolution}.  Because the \textsc{UniverseMachine} assumes a constant slope for the SFR--halo mass relation at low halo masses (Section \ref{s:effective}), the change in slope in the luminosity function in Fig.\ \ref{fig:extrap} is entirely due to the shallower slope of the halo mass function for low-mass haloes.  Any additional physics that reduces star formation efficiency in low-mass haloes would make this effect even stronger.  As a result, assuming a constant faint-end slope for the \textit{luminosity function} will overestimate the number density of faint galaxies as well as the total CSFR (Fig.\ \ref{fig:extrap}, right panel; see also Section \ref{s:csfr}).  The same effect occurs for all redshifts $z=8-15$, as shown in Appendix \ref{a:res_tests}.

The shape of the halo mass function also implies that luminosity functions likely do not have asymptotic faint-end slopes at $z\ge8$.  As above, $z=8$ is where the halo mass function achieves a steeper slope than $-2$ even for the lowest-mass haloes able to form galaxies ($\sim 10^8\Msun$).  Hence, the Universe runs out of haloes able to form galaxies before the halo mass function slope stops changing.  Fig.\ \ref{fig:extrap_slopes} shows that, as a result, the UV luminosity function at $z>8$ never has a region with a constant power-law slope.  At $z>8$, the power-law slope of the UV luminosity function gradually increases for fainter galaxies until the halo mass limit of $10^8 \Msun$ is reached, at which point the slope increases dramatically due to the rapidly falling number density in the luminosity function \citep[see also][]{Yung19b,Yung20}.  The location of this rapid increase in slope occurs at $M_{1500}=-10$ to $-12$, depending on redshift (see Appendix \ref{a:res_tests}, Fig.\ \ref{fig:uvlf_res}), corresponding to the characteristic luminosity of haloes near the atomic cooling limit.

\begin{figure*}
\vspace{-8ex}
\hspace{-14ex}\includegraphics[width=1.2\columnwidth]{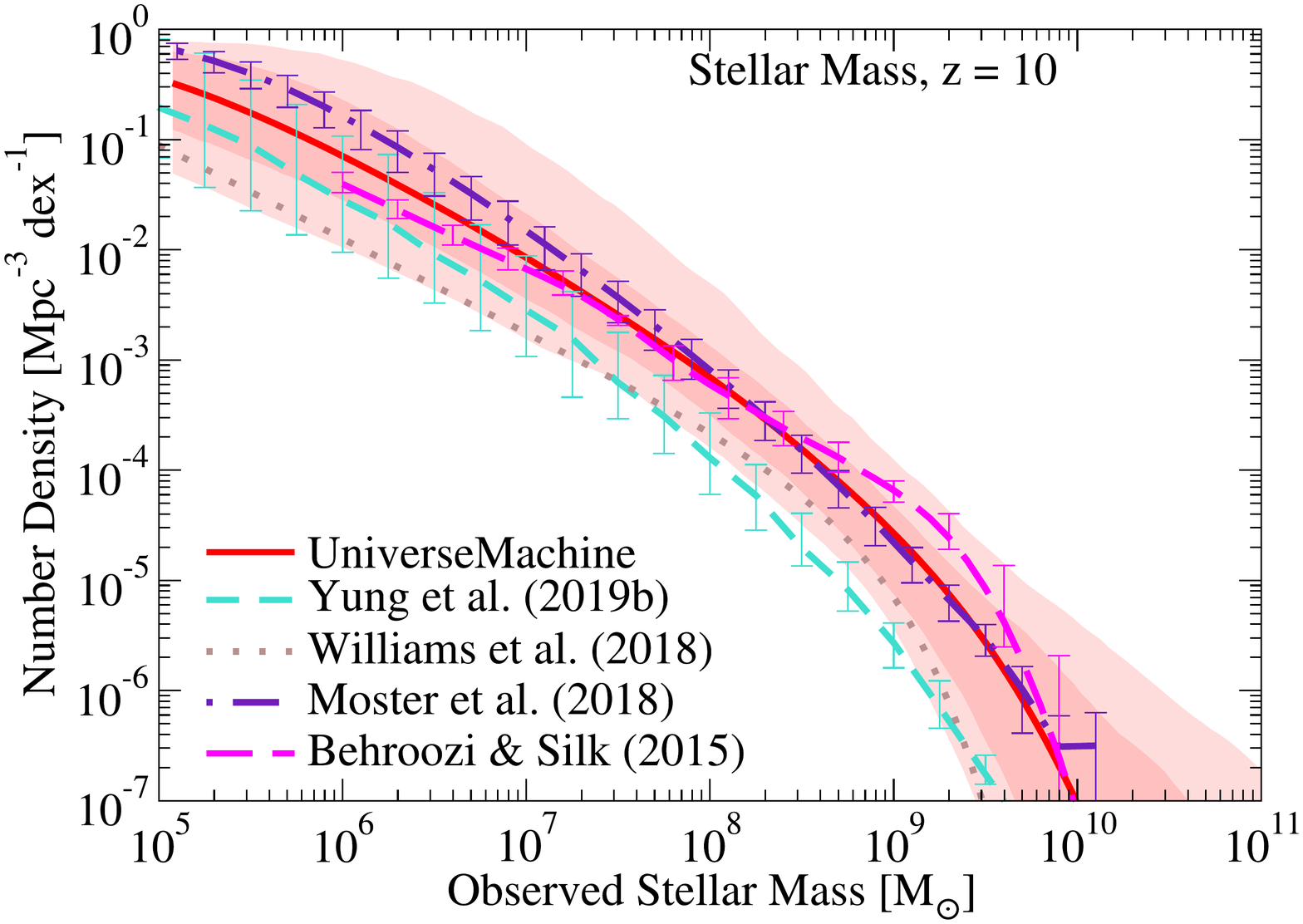}\hspace{-6ex}\includegraphics[width=1.2\columnwidth]{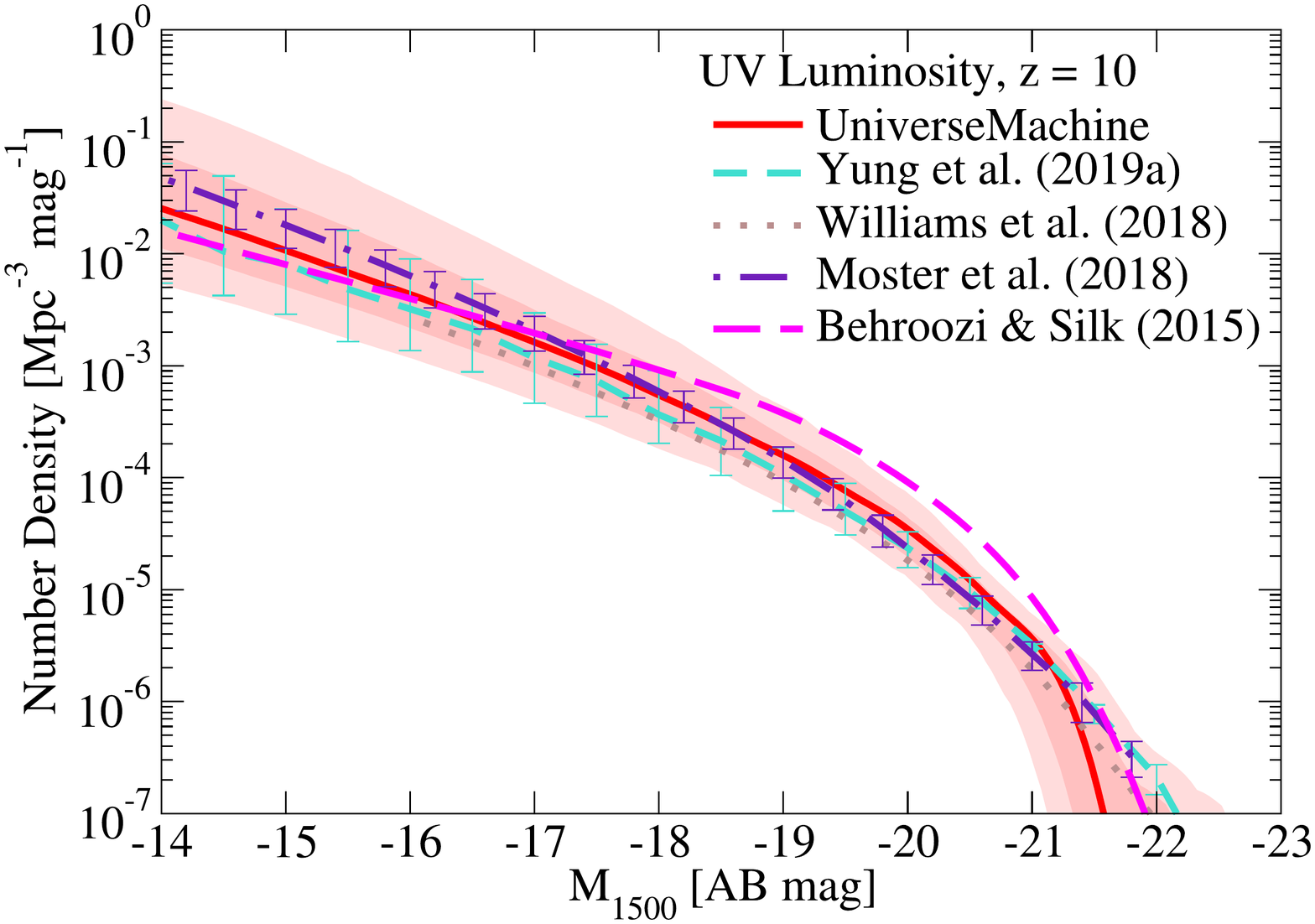}\hspace{-17ex}\\[-5ex]
\hspace{-14ex}\includegraphics[width=1.2\columnwidth]{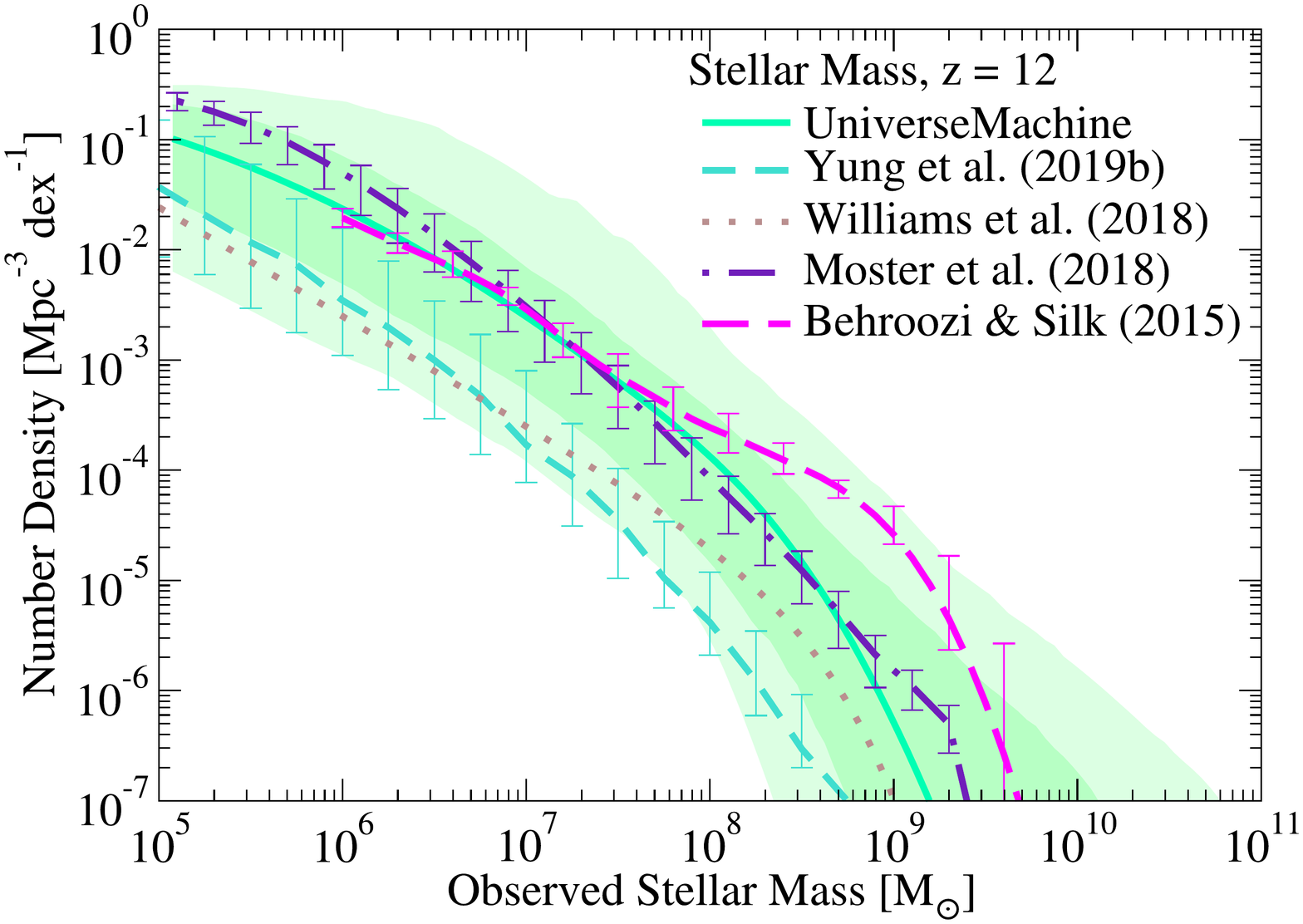}\hspace{-6ex}\includegraphics[width=1.2\columnwidth]{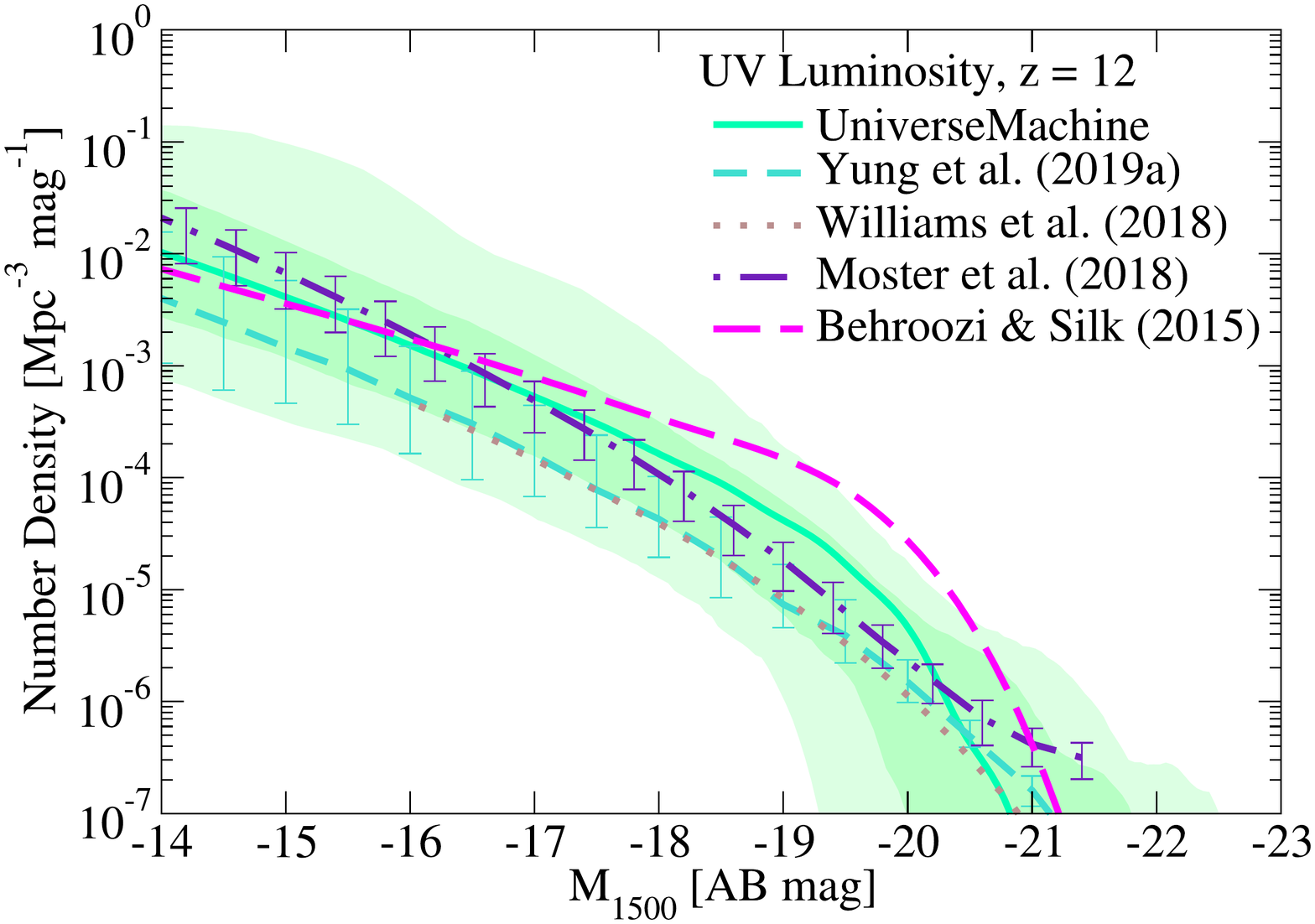}\hspace{-17ex}\\[-5ex]
\hspace{-14ex}\includegraphics[width=1.2\columnwidth]{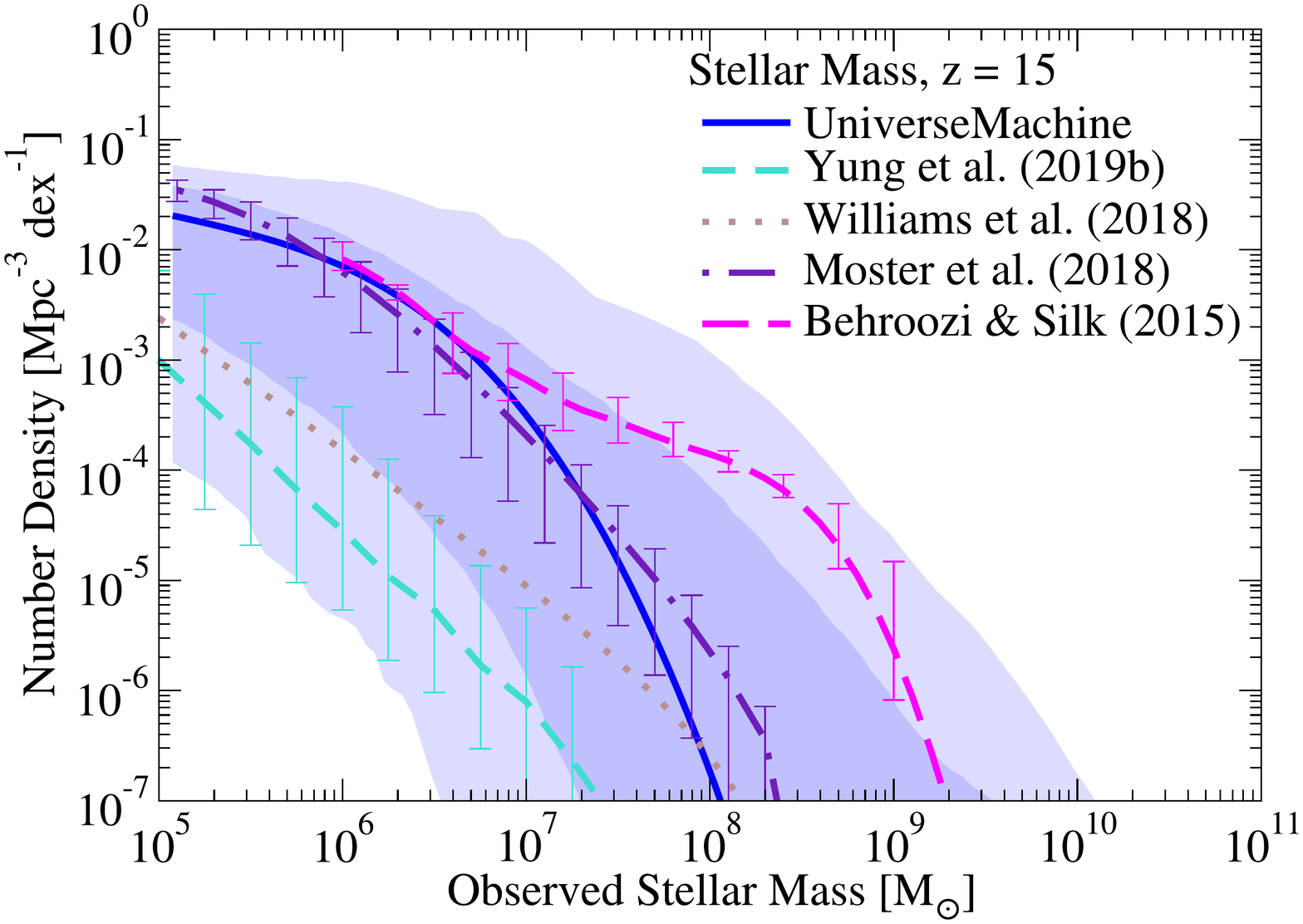}\hspace{-6ex}\includegraphics[width=1.2\columnwidth]{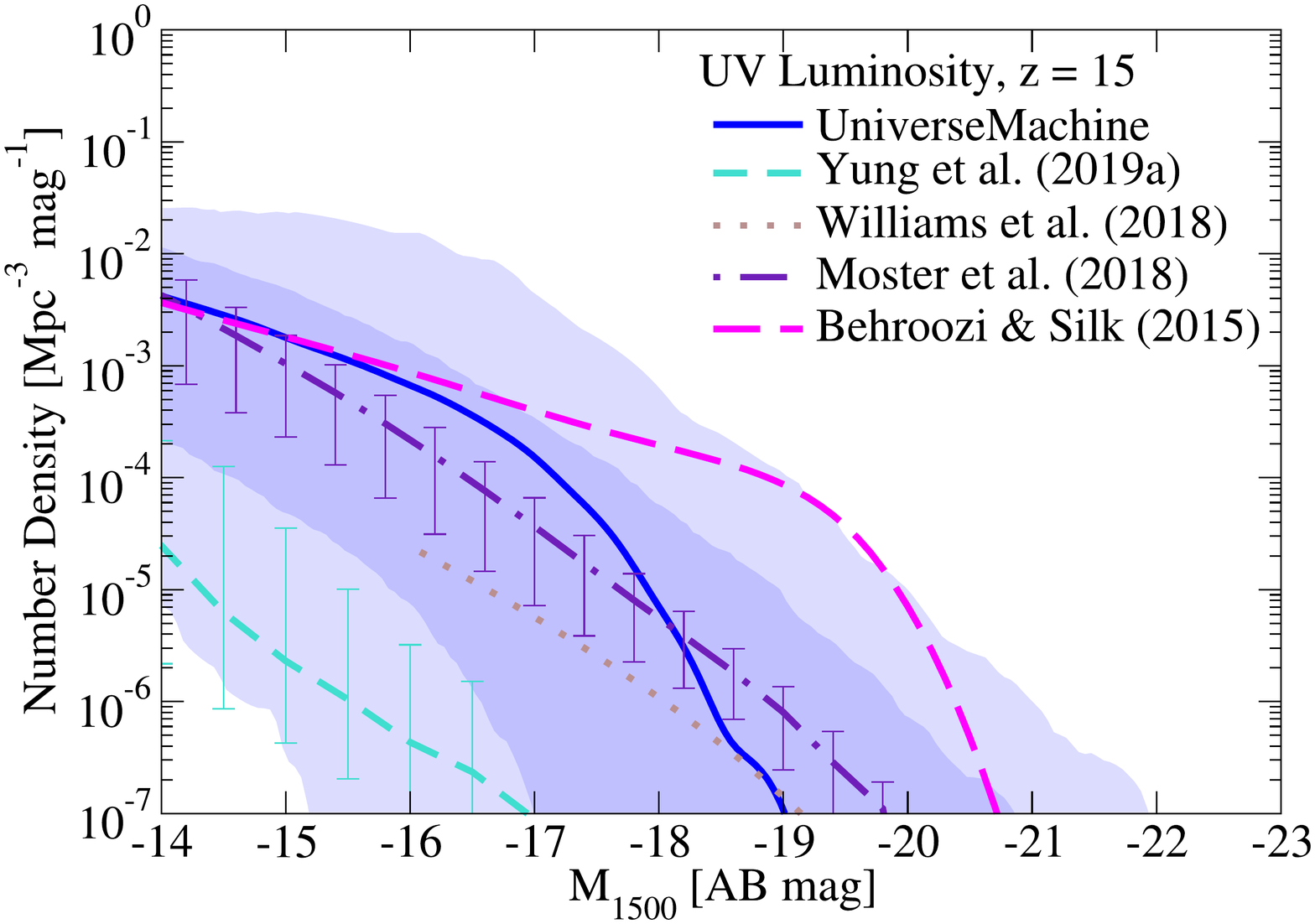}\hspace{-17ex}\\[-3ex]
\caption{Model comparisons for stellar mass functions (\textit{left panels}) and UV luminosity functions (\textit{right panels}) at $z=10$ (\textit{top panels}), $z=12$ (\textit{middle panels}), and $z=15$ (\textit{bottom panels}).  Bold and light shaded regions correspond to the  $16-84^\mathrm{th}$ and $3-97^\mathrm{th}$ percentile confidence intervals, respectively, for the \textsc{UniverseMachine}.  The predictions of these other models are within the \textsc{UniverseMachine} confidence intervals over this range of redshifts.}
\label{fig:comp}
\end{figure*}

\subsection{Stellar Mass--Halo Mass Relations}

\label{s:smhm}

Fig.\ \ref{fig:smhm} shows predicted stellar mass--halo mass relationships at \mbox{$z\ge8$}.  At all redshifts, the integrated star formation efficiency increases with halo mass at least up to $M_h \sim 10^{12}\Msun$.  The best-fit model also shows modestly increasing efficiency toward higher redshifts at fixed halo mass, by about a factor of 10 from $z=0$ to $z=12$ for $M_h = 10^{10}\Msun$.  For comparison, we also show constraints at lower masses from dwarf galaxies in the Milky Way \citep{Nadler20}, which are consistent with \textsc{UniverseMachine} results in the range of overlap ($10^{10}-10^{11}\Msun$).

Different empirical models currently suggest a wide range of redshift evolution (see discussion in \citealt{BWHC19}).  The need for redshift evolution is typically driven by a faster decline in halo cumulative number densities at fixed halo mass as compared to galaxy cumulative number densities at fixed stellar mass.  The redshift evolution expected here is less than that predicted in previous empirical models (e.g., \citealt{BehrooziHighZ}) because of adopted constraints from $z=9$ and $z=10$ that show accelerated UV luminosity function declines toward higher redshifts \citep{Bouwens15}.  Recent constraints with the \textit{Hubble Space Telescope} \citep[e.g.,][]{Oesch18,Ishigaki18,Bhatawdekar19, Bouwens19} continue to show this effect, albeit with some disagreement about its magnitude.

\textit{JWST} is very likely to settle questions about evolution in stellar mass--halo mass ratios.  At $z<10$, coverage of rest-frame optical colours will improve stellar masses, and spectroscopic clustering measurements will allow direct confirmation of halo masses \citep{Endsley20}.  At $z\ge 10$, improved constraints on galaxy number densities (by $\sim$1 dex or more) will dramatically shrink uncertainties on the stellar mass--halo mass relation implied by abundance matching.  This science will be accessible to currently planned Cycle 1 surveys at least to $z\sim 13.5$ (Fig.\ \ref{fig:survey_counts}).

\subsection{Cosmic Star Formation Rates}

\label{s:csfr}

Predicted observable ($M_{1500}<-17$) and total cosmic star formation rates (CSFRs) are shown in Fig.\ \ref{fig:csfrs}.  As expected, the \textsc{UniverseMachine} agrees well with data sets that were used as constraints (marked as ``Pre-2018 Observed Data'').  In the \textsc{UniverseMachine}, the observable CSFR evolution with redshift becomes steeper at $z>9$ than at lower redshifts, which is driven by rapid evolution in luminosity function constraints at $z\sim 10$ (Fig.\ \ref{fig:existing_data}).  More recent data (e.g., \citealt{Oesch18}) have suggested an even steeper redshift evolution at the two sigma lower bound of \textsc{UniverseMachine} predictions.  \textit{JWST} will extend these measurements to at least $z\sim 13$, resolving questions about the rate of observable CSFR evolution.  

The total CSFR has a shallower evolution with redshift than the observable CSFR, since an increasing fraction of the CSFR occurs in galaxies fainter than $M_\mathrm{1500}=-17$ at higher redshifts.  Indeed, observable CSFRs are predicted to be half as much as total CSFRs already by $z\sim 8$, and only a quarter as much by $z\sim 13$ (Fig.\ \ref{fig:csfr_completeness}, left panel).  While \textit{JWST} will not be able to resolve most of the total CSFR in blank fields at $z>8$,  lensed \textit{JWST} fields reaching to $M_\mathrm{1500}=-14$ may be able to detect up to $50-75$\% of the total CSFR (Fig.\ \ref{fig:csfr_completeness}, right panel).

The total CSFR is commonly estimated by assuming that the observed luminosity function slope is constant down to faint magnitudes.  However, from Fig.\ \ref{fig:extrap_slopes}, we expect that the slope of the UV luminosity function will continue to become shallower at fainter luminosities than observable with \textit{JWST}.  Although we expect many galaxies to exist at magnitudes fainter than $M_{1500}\sim -12$, Fig.\ \ref{fig:csfr_integrate} shows that integrating a constant slope from $M_{1500}= -17$ to $M_{1500}= -12$ achieves the lowest error in estimating the total CSFR in the \textsc{UniverseMachine}, at least to $z\sim 12$.  The \textsc{UniverseMachine} assumes a constant slope for the SFR--halo mass relationship (Eq.\ \ref{e:sfr}, Fig.\ \ref{fig:halo_sfrs}) down to the atomic cooling limit.  Other theoretical models have found that galaxy formation may become inefficient at higher halo masses due to, e.g., reduced H$_2$ formation in low-metallicity haloes \citep[e.g.,][]{Jaacks12,Xu16}.  Hence, integrating a constant luminosity function slope from $M_{1500}= -17$ to $M_{1500}= -12$ should be interpreted as giving a reasonable upper limit on the total CSFR.

This integration limit provides a simple rule of thumb for reionization modelling.  A similar rule of thumb can be estimated for other survey limits.  For example, integrating a constant slope from $M_{1500}= -18$ to $M_{1500}= -13.5$ again gives the total CSFR closest to the true value in the \textsc{UniverseMachine}; extrapolating from deeper surveys is of course preferred when possible.  Above $z\sim 12$, only the exponentially declining portion of the UV luminosity function is likely visible at $M_{1500}<-17$ (see Fig.\ \ref{fig:nds}), resulting in large uncertainties for extrapolated total CSFRs.

Because the faint-end slopes of the mass and luminosity functions become shallower than $-2$ for the faintest galaxies, such faint galaxies do not contribute much to the total CSFR.  The total CSFR is therefore not very sensitive to the threshold halo mass for star formation, as long as the threshold halo mass is less than about $10^{8.5}\Msun$ (Fig.\ \ref{fig:threshold_mass}).  For most \textsc{UniverseMachine} models, lowering the threshold halo mass from $10^8\Msun$ to $10^6\Msun$ with the same SFR--halo mass prescription results in less than a $30\%$ increase in the total CSFR out to $z\sim 15$.

\subsection{Model Comparisons}

We compare the \textsc{UniverseMachine} to three other empirical models and a semi-analytical model (SAM).  The semi-analytical model \citep[the Santa Cruz model;][]{Somerville15b,Yung19,Yung19b} employs analytic prescriptions for multiphase gas cooling, stellar and black hole feedback, metallicity enrichment, and dust-to-metal ratios; these prescriptions are integrated over Extended Press-Schechter dark matter halo merger trees to generate galaxy properties.  For the Santa Cruz SAM, error bars show the range of supernova feedback strengths explored in \cite{Yung19,Yung19b}.   The empirical models include EMERGE \citep{Moster17}, JAGUAR \citep{Williams18}, and the model of \cite{BehrooziHighZ}.  Each uses redshift-dependent scaling laws to describe star formation rates and stellar masses in dark matter haloes that are calibrated to match observations at $z\le 10$.  Of note, EMERGE has been recalibrated using a more flexible redshift scaling than in \cite{Moster17}, which yields lower stellar mass--halo mass ratios at $z\sim 3-6$ than previously published (B.\ Moster et al., in prep.).  Additionally, to generate UV luminosities, it uses the same approach described in \S \ref{s:effective}, with dust parameters taken from the best-fit \textsc{UniverseMachine} model.

Fig.\ \ref{fig:csfrs} compares CSFRs from the \textsc{UniverseMachine} to other data and models.  All models agree with all observations at $z<8$, with disagreements becoming more prominent at $z\sim 10$.  \cite{BehrooziHighZ} gives the most optimistic predictions at high redshifts, because the extrapolation technique used favours increasing stellar--halo mass ratios at higher redshifts.  JAGUAR and the Santa Cruz SAM give the most pessimistic predictions.  JAGUAR is driven by matching the rapidly-decreasing luminosity functions measured at $z>8$ in \cite{Oesch18}.  The Santa Cruz SAM requires star formation timescales for molecular ($\sim 10^2$K) gas that are increasingly significant compared to the age of the Universe at $z>10$.  EMERGE predictions are most similar (within one-sigma uncertainties of the \textsc{UniverseMachine}), likely due to the similar observational constraints used.  All models (and data) above are consistent within two-sigma uncertainty contours of the \textsc{UniverseMachine}.

Fig.\ \ref{fig:csfr_completeness} compares predicted ratios between total and observable ($M_\mathrm{1500} < -17$) CSFRs.  JAGUAR is not shown, as it integrates CSFRs only down to $M_\ast \sim 10^6\Msun$.  The remaining models are in excellent agreement from $z=5$ to $z=12$, at which point the predictions for the fraction of observable star formation diverge.  This is consistent with the divergence of uncertainties in the \textsc{UniverseMachine}.  At $z<5$, the Santa Cruz SAM has significantly more dust-obscured star formation at $M_\mathrm{1500}>-17$ than other models.

Lastly, we compare predicted stellar mass and luminosity functions in Fig.\ \ref{fig:comp}.  These show broad agreement with the CSFR trends in Fig.\ \ref{fig:csfrs}.  As with CSFRs, \cite{BehrooziHighZ} gives more optimistic predictions; JAGUAR as well as the Santa Cruz SAM give more pessimistic predictions; and EMERGE gives similar predictions.  Of note, predicted faint-end slopes for the luminosity function are similar up to $z\sim 12$, regardless of the approach, leading to similar predicted total to observed CSFR ratios in Fig.\ \ref{fig:csfr_completeness}.  As with CSFRs, the range of theoretical predictions generally falls within the two-sigma uncertainties of the \textsc{UniverseMachine}.

\section{Discussion}

\label{s:discussion}

In this paper, we present predictions from an empirical model at $z>10$.  As discussed in \cite{BehrooziHighZ}, such extrapolations can be valid as long as the dominant physical processes for galaxy formation remain the same and have no major discontinuities.  Confirmation or rejection of these predictions with \textit{JWST} will hence reveal whether 1) similar physics applies at $z>10$, or 2) new processes become important at these high redshifts.

We expect that \textit{JWST} will be able to observe galaxies with $M_*>10^7\Msun$ or $M_\mathrm{1500} < -17$ out to $z\sim 13.5$ with $>85\%$ confidence in planned Cycle 1 surveys (Fig.\ \ref{fig:cumul}).  Typical \textsc{UniverseMachine} models suggest that \textit{JWST} will also detect $z\sim 15$ galaxies.  However, the most pessimistic models suggest that $z=15$ galaxies will be inaccessible, even in a lensed survey, since greater depth will result in reduced effective volume (Fig.\ \ref{fig:comp}).  Given the more than $1.5$ dex one-sigma uncertainties in number density at $z=15$ (Fig.\ \ref{fig:comp}), even upper limits will be extremely useful to constrain galaxy evolution.

Lensed surveys may access the fainter galaxies that are believed to play important roles in reionization (e.g., \citealt{Bouwens12,Robertson15,Finkelstein19}; c.f., \citealt{Naidu20}).  With the \textit{Hubble Space Telescope}, observations of lensed galaxies in the Hubble Frontier Fields \citep{Lotz17} yielded UV luminosity functions $2-3$ magnitudes fainter than otherwise possible \citep{Livermore17,Bouwens17,Ishigaki18}.  Systematic uncertainties in lensing maps and galaxy intrinsic sizes prevent robust measurements at fainter magnitudes \citep{Bouwens17,Yue18}.  Hence, galaxies down to $M_{1500}\sim -14$ and $M_\ast \sim 10^{6}\Msun$ may be accessible in lensed fields with \textit{JWST}.

Consistent with past approaches, we find that faint galaxies ($M_\mathrm{1500}>-17$) should dominate cosmic star formation at $z\ge 8$ (Fig.\ \ref{fig:csfr_completeness}).  Nonetheless, we predict most cosmic star formation at $z<15$ to occur in galaxies brighter than $M_\mathrm{1500}=-14$ (Fig.\ \ref{fig:csfr_completeness}, right panel), which are accessible to \textit{JWST} in lensed fields.  We emphasize that the dominance of $M_\mathrm{1500}\lesssim-14$ galaxies in high-$z$ CSFRs does not require a change in SFR feedback in fainter galaxies that affects the slope of the stellar mass--halo mass relation.  Instead, this magnitude limit is a natural consequence of the changing slope of the halo mass function (Section \ref{s:mfs}), which becomes shallower for lower-mass haloes (Fig.\ \ref{fig:hmfs}). As a result, probes of the total cosmic CSFR at $z<12$ (e.g., short gamma-ray bursts) will not place significant constraints on the lower threshold for galaxy formation in haloes as long as it is $M_h\sim 10^{8.5}\Msun$ or below (Fig.\ \ref{fig:threshold_mass}).

UV luminosity and stellar mass functions similarly do not have constant faint-end slopes (Fig.\ \ref{fig:extrap}), again due to the shape of the halo mass function.  Extrapolating a constant faint-end slope to $M_\mathrm{1500}=-10$ (as done in \citealt{Bouwens12}) likely overestimates the CSFR by $\sim 20\%$ near reionization.  More recent reionization studies \citep[e.g.][]{Robertson15,Finkelstein19} extrapolated to $M_\mathrm{1500}\sim -13$, which more closely approximates the true CSFR.  We find that extrapolating to $M_\mathrm{1500}=-12$ results in the least expected error, at least up to $z\sim 12$ (Section \ref{s:csfr}; Fig.\ \ref{fig:csfr_completeness}).

Shallower UV luminosity functions below $M_\mathrm{1500}=-17$ at $z\ge 8$ also imply fewer ultrafaint dwarf galaxies at $z=0$.  As noted in \cite{Weisz17}, a turnover below $M_\mathrm{1500}=-13$ is likely necessary to reconcile ultrafaint dwarf galaxy counts with steep observed faint-end slopes at $M_\mathrm{1500}=-17$.  Quantitative comparison with expected luminosity functions at $z\ge 7$ is difficult due to uncertainties in the exact formation redshifts of ultrafaint dwarfs.  However, the \textsc{UniverseMachine} does give $z=0$ stellar mass--halo mass relations consistent with constraints from ultrafaint dwarf satellites of the Milky Way (Fig.\ \ref{fig:smhm}).  As more ultrafaint dwarf galaxies are observed (e.g., with the \textit{Vera Rubin Observatory} Legacy Survey of Space and Time), these will likely tighten constraints on the shape of high-redshift luminosity functions.  Our results here combined with constraints on the very low mass galaxy--halo connection from the dwarf galaxy analysis of \cite{Nadler20} indicate that for the foreseeable future, these very local measurements are likely to provide more insight into the low-mass threshold for galaxy formation than will very high redshift measurements on their own.

\section{Conclusions}

\label{s:conclusions}
In this paper, we apply the \textsc{UniverseMachine} to a high-resolution simulation (\textit{VSMDPL}) for redshifts from $z=0$ to $15$.  Using the posterior distribution of parameters for the \textsc{UniverseMachine} that match $z\le 10$ observations, we make predictions for what \textit{JWST} may observe at $10<z<15$.  Key results include:
\begin{enumerate}
    \item Planned \textit{JWST} Cycle 1 surveys will likely observe hundreds of $z>10$ galaxies, with a highest redshift of at least $z\sim 13.5$ (Table \ref{t:obs_summary} and Figs.\ \ref{fig:nds}-\ref{fig:survey_counts}).
    \item \textit{JWST} will likely be able to measure the evolution in the stellar mass--halo mass relation to at least $z\sim 13.5$ in planned Cycle 1 surveys (Section \ref{s:smhm}).
    \item Most star formation at $z>8$ is predicted to occur in galaxies brighter than $M_\mathrm{1500}=-14$, which would be accessible in lensed \textit{JWST} fields.
    \item The current uncertainty in galaxy number densities rises dramatically at $z\ge 12$ (Figs.\ \ref{fig:nds}--\ref{fig:survey_counts}, \ref{fig:csfrs}--\ref{fig:csfr_completeness}); both detections and non-detections at these redshifts will be extremely valuable to constrain galaxy formation models.
    \item Faint-end slopes ($\alpha$) for observed stellar mass and luminosity functions are expected to continue to steepen beyond $\alpha=-2$ with increasing redshift (Fig.\ \ref{fig:slopes}).  This is a natural consequence of $\Lambda$CDM halo mass functions, which have slopes much steeper than $-2$ at the halo masses which host observable galaxies at these redshifts (Fig.\ \ref{fig:hmfs}).
    \item Faint-end slopes for stellar mass and luminosity functions are expected to become shallower below observable thresholds (Figs.\ \ref{fig:extrap} and \ref{fig:extrap_slopes}; $M_\ast < 10^7\Msun$ or $M_\mathrm{1500}>-17$) because the haloes hosting these galaxies are in the exponentially falling region of the halo mass function.  For reionization models, a reasonable upper limit to the total CSFR can be obtained by extrapolating a constant faint-end slope from $M_\mathrm{1500}=-17$ to $M_\mathrm{1500}=-12$ (Fig.\ \ref{fig:csfr_completeness}), at least to $z\sim 12$.
    \item Other empirical and semi-analytic models of the high-redshift Universe 
    give predictions that are within the two-sigma uncertainties of the \textsc{UniverseMachine} (Figs.\ \ref{fig:csfrs}--\ref{fig:csfr_completeness}, \ref{fig:comp}).
    \item Mock catalogues and lightcones for CANDELS fields, including multiple realizations to evaluate sample variance and model variance, are available \href{https://peterbehroozi.com/data.html}{online}.
\end{enumerate}

\section*{Acknowledgements}

PB was partially funded by a Packard Fellowship, Grant \#2019-69646.
GY  acknowledges financial support  from  MICIU/FEDER (Spain) under research grant   PGC2018-094975-C21.
CCW acknowledges support from the National Science Foundation Astronomy and Astrophysics Fellowship grant AST-1701546.
Work done at Argonne National Laboratory was supported under the DOE contract DE-AC02-06CH11357.
This material is based upon High Performance Computing (HPC) resources supported by the University of Arizona TRIF, UITS, and RDI and maintained by the UA Research Technologies department.  The University of Arizona sits on the original homelands of Indigenous Peoples (including the \href{http://www.tonation-nsn.gov/history-culture/}{Tohono O'odham} and the Pascua Yaqui) who have stewarded the Land since time immemorial. 

The authors gratefully acknowledge the Gauss Centre for Supercomputing e.V. (www.gauss-centre.eu) and the Partnership for Advanced Supercomputing in Europe (PRACE, www.prace-ri.eu) for funding the MultiDark simulation project by providing computing time on the GCS Supercomputer SuperMUC at Leibniz Supercomputing Centre (LRZ, www.lrz.de).

\bibliography{master_bib} 

\appendix

\section{Key Parametrizations}

\label{a:full}

The \textsc{UniverseMachine} separately parametrizes the formation of star-forming and quiescent galaxies.  However, there are very few quiescent galaxies at high redshifts \citep[e.g.,][]{Muzzin13}, so the SFR--halo relationship for star-forming galaxies dominates.  The median SFR for haloes as a function of their $\vmp$ (i.e., $\vmax$ at the redshift of peak halo mass) is given by:
\begin{eqnarray}
\sfrsf & = & \epsilon\left[\left(v^\alpha + v^\beta\right)^{-1} + \gamma \exp\left(-\frac{\log_{10}(v)^2}{2\delta^2}\right)\right] \label{e:sfrsf}\\
v & = & \frac{\vmp}{V\cdot\mathrm{km}\;\mathrm{s}^{-1}}\\
\log_{10}(V) & = & V_0 + V_a (1-a) + V_{la} \ln(1+z) + V_z z  \label{e:v}\\
\log_{10}(\epsilon) & = & \epsilon_0 + \epsilon_a (1-a) + \epsilon_{la} \ln(1+z) + \epsilon_z z \label{e:sfr}\\
\alpha & = & \alpha_0 + \alpha_a (1-a) +  \alpha_{la} \ln(1+z) + \alpha_z z \label{e:alpha}\\
\beta & = & \beta_0 + \beta_a (1-a) + \beta_z z \label{e:beta}\\
\log_{10}(\gamma) & = & \gamma_0 + \gamma_a(1-a)+\gamma_z z \label{e:gamma}\\
\delta & = & \delta_0,\label{e:delta}
\end{eqnarray}
where $a$ is the scale factor.  Equation \ref{e:sfrsf} is a double power-law with an extra Gaussian bump near the transition between the two power laws.  Physically, this corresponds to the transition between two dominant modes of feedback, one for low-mass and one for high-mass haloes.  At high redshifts, however, the fraction of high-mass haloes declines exponentially (Fig.\ \ref{fig:hmfs}), so that Eq.\ \ref{e:sfrsf} reduces to  $SFR_\mathrm{SF} \sim  \epsilon v^{-\alpha}$.  Additionally, $\vmp$ is tightly correlated with halo mass, with $\vmp \propto M_\mathrm{peak}^\frac{1}{3}$.  Hence, typical star-forming behaviour is well-described by $SFR_\mathrm{SF} \propto  \epsilon \left(M_\mathrm{peak}\right)^{-\frac{\alpha}{3}}$ (Eq.\ \ref{e:sfr1}).  At high redshifts, the scalings of $\ln(1+z)$ and $a$ change only weakly with redshift, so the values of $V_z$, $\epsilon_z$, and $\alpha_z$ dominate the redshift scaling.  Of note, for a single power law, changes in $V_z$ are degenerate with changes in $\epsilon_z$, so the overall redshift scaling reduces to Eqs.\ \ref{e:sfr2}--\ref{e:sfr3}.

Scatter in SFRs is associated with halo accretion history, averaged over the past dynamical time ($1/\sqrt{G\rho_\mathrm{vir}}$).  In the \textsc{UniverseMachine}, this is limited to 0.3 dex for star-forming galaxies:
\begin{equation}
\sigma_\mathrm{SF} = \min(\sigma_\mathrm{SF,0} + (1-a)\sigma_\mathrm{SF,1}, 0.3) \;\mathrm{dex}.\label{e:sig_sf_a}
\end{equation}
At $z>4$, for all models in the \textsc{UniverseMachine} posterior distribution, a variation of 0.3 dex in SFR is very subdominant to the variation in SFR with halo mass and redshift.

\section{Resolution Tests}

\begin{figure}
\centering
    \vspace{-11ex}
    \includegraphics[height=\columnwidth,angle=90]{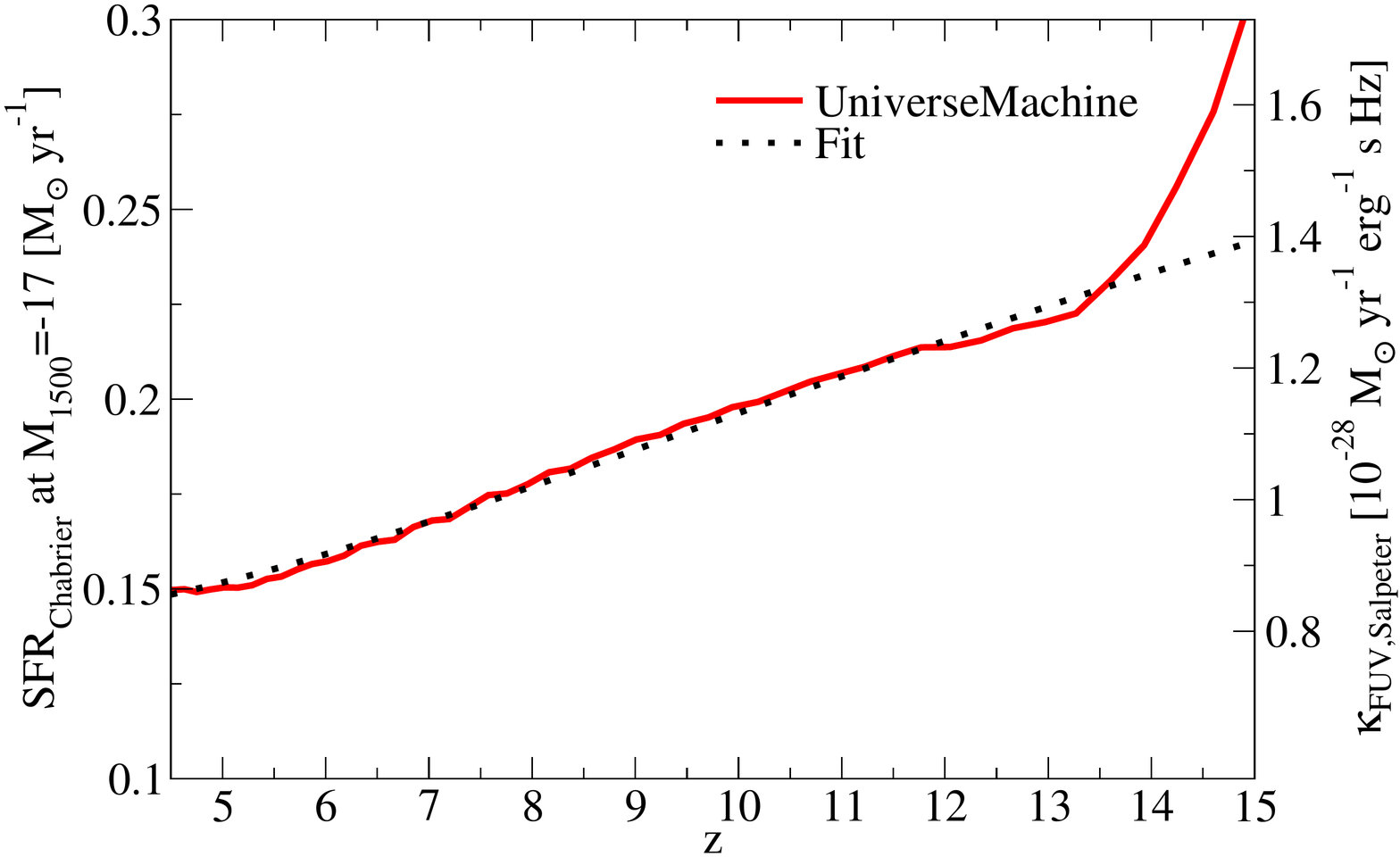} \\[-6ex]
    \caption{Median relation between SFR and UV luminosity in the \textsc{UniverseMachine}.  This is expressed as the SFR giving a UV luminosity of $M_\mathrm{1500}=-17$ on the left vertical axis, and as the SFR/UV luminosity ratio ($\kappa_\mathrm{FUV}$) on the right-hand axis.  Since the \citet{Salpeter55} IMF is typically used when reporting $\kappa_\mathrm{FUV}$ in the literature, we keep this convention here, but report SFRs for a \citet{Chabrier03} IMF.}
    \label{f:sfr_uv}
\end{figure}

\begin{figure*}
    \centering
    \vspace{-11ex}
    \hspace{-14ex}\includegraphics[width=1.2\columnwidth]{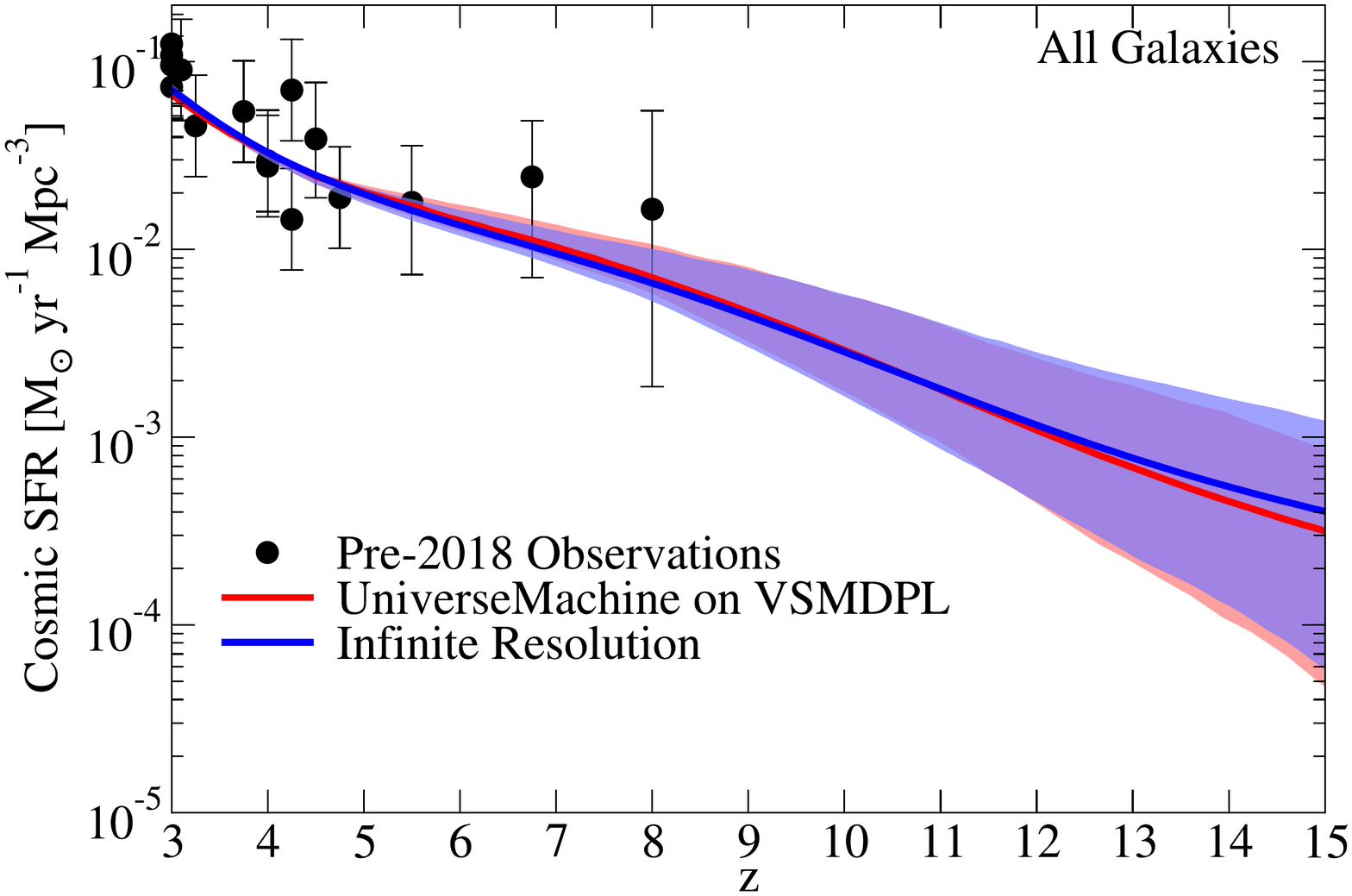} \hspace{-7ex}\includegraphics[width=1.2\columnwidth]{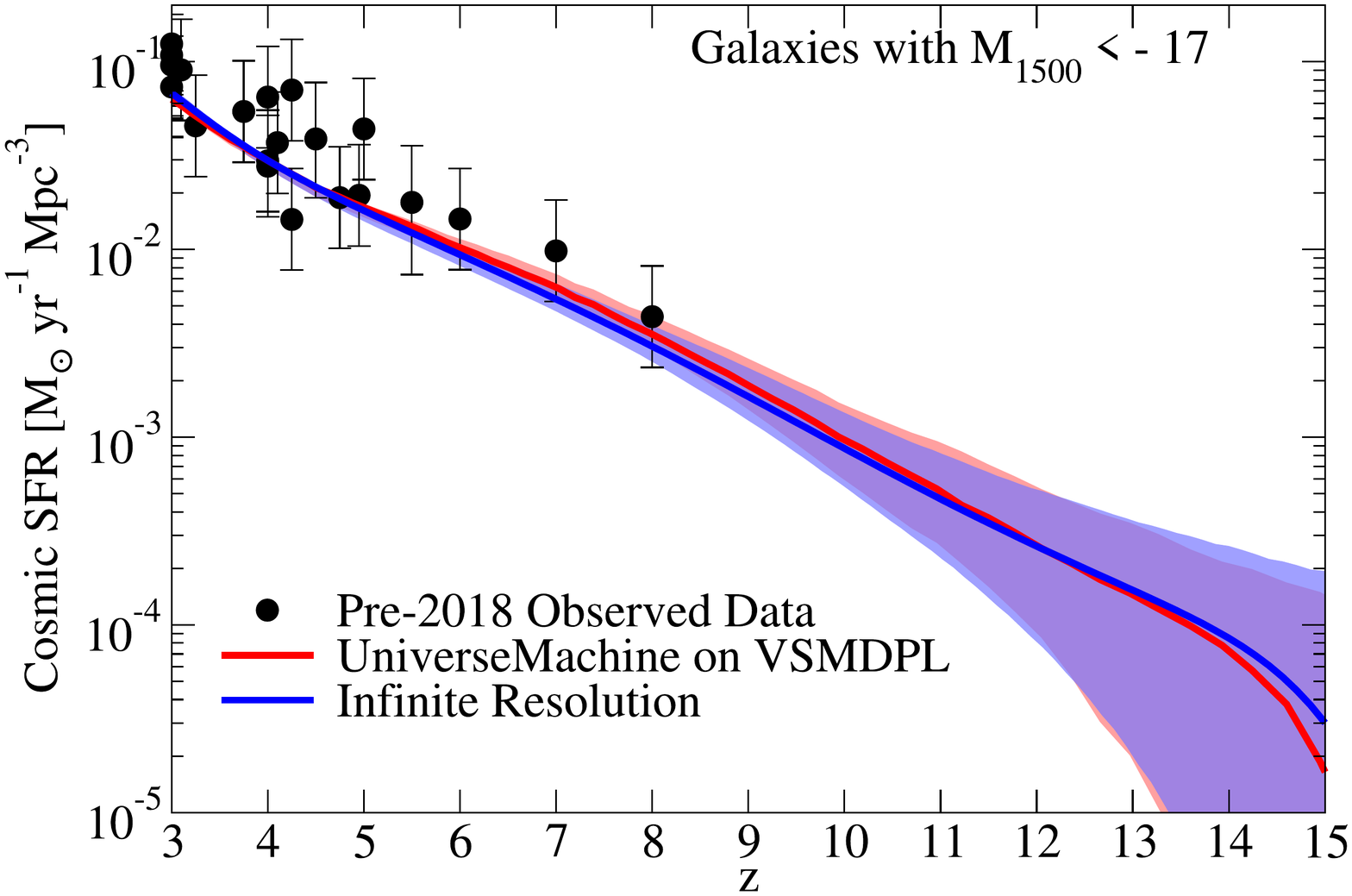}\hspace{-15ex}\\[-6ex]
    \caption{CSFRs compared between the \textsc{UniverseMachine} and the extrapolated ``Infinite Resolution'' test described in Appendix \ref{a:res_tests}.  The \textbf{left panel} shows CSFRs for all galaxies, and the \textbf{right panel} shows CSFRs for only bright galaxies ($M_{1500}<-17$).  In both cases, the \textsc{UniverseMachine} on \textit{VSMDPL} gives very similar results, even including the $16-84^\mathrm{th}$ percentile confidence intervals (shaded regions).  Observed data points are identical to those in Fig.\ \ref{fig:csfrs}.}
    \label{fig:csfr_res}
    \centering
    \vspace{-5ex}
    \hspace{-14ex}\includegraphics[width=1.2\columnwidth]{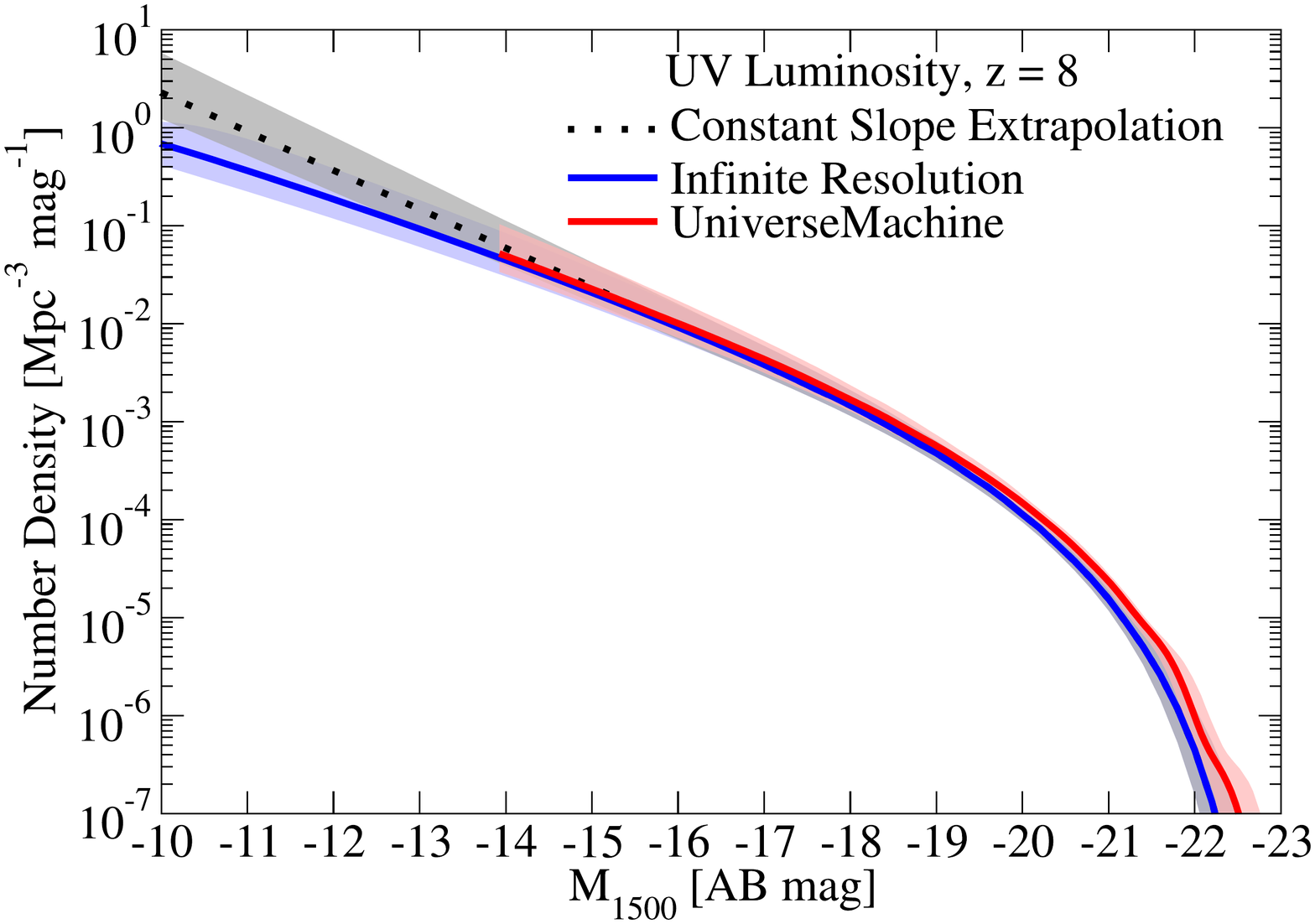}    \hspace{-7ex}\includegraphics[width=1.2\columnwidth]{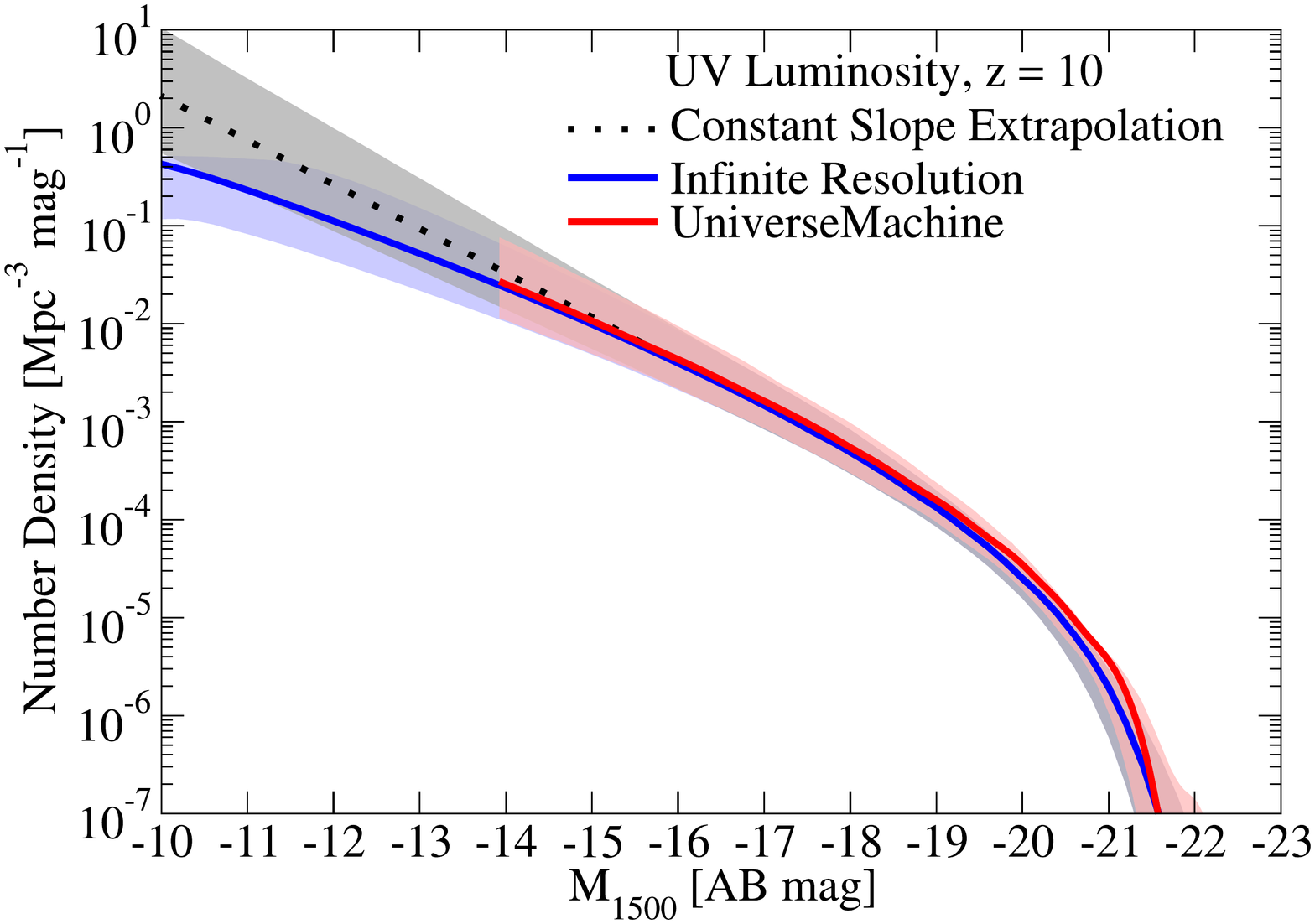}\hspace{-15ex}\\[-10ex]
    \hspace{-14ex}\includegraphics[width=1.2\columnwidth]{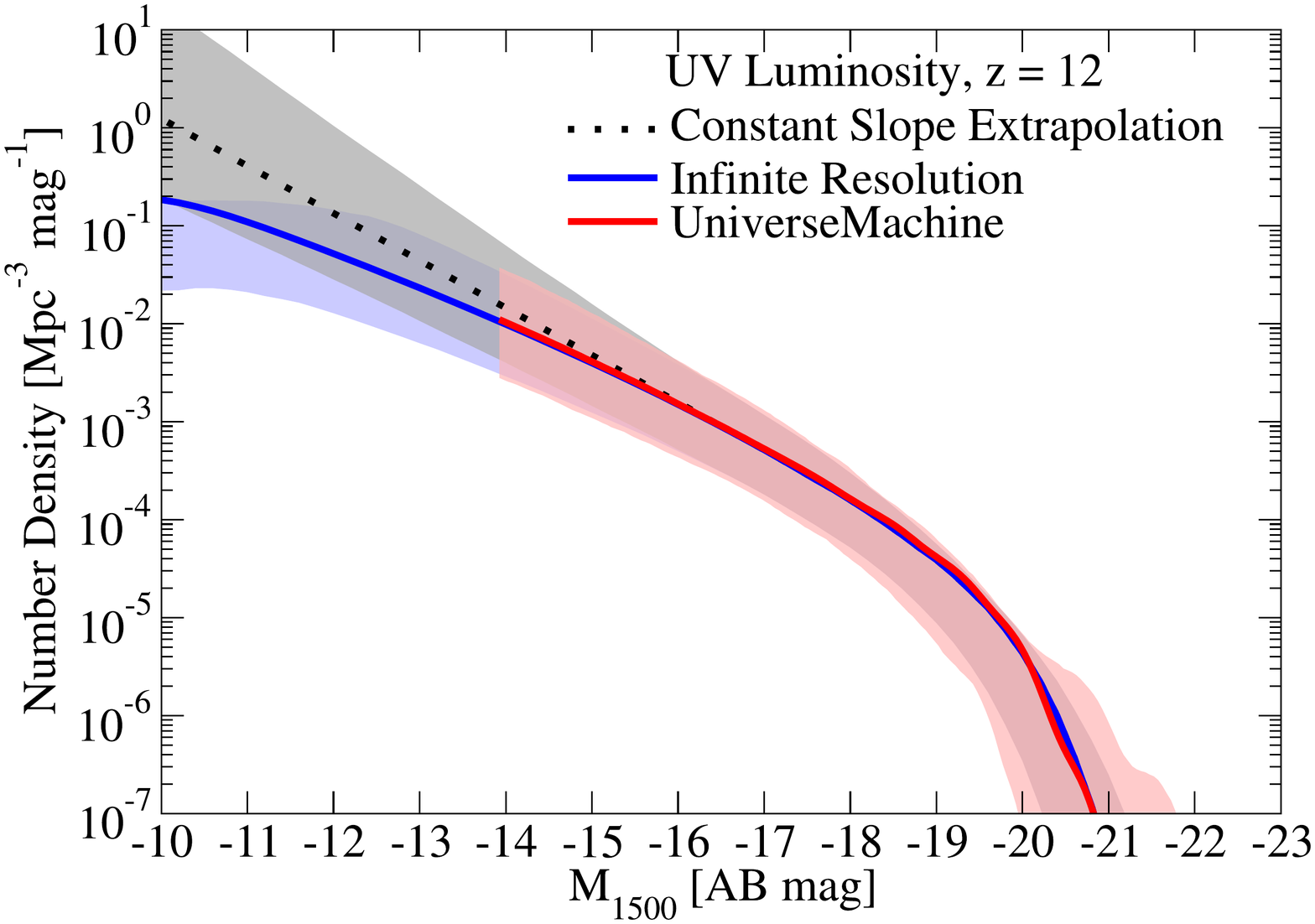}    \hspace{-7ex}\includegraphics[width=1.2\columnwidth]{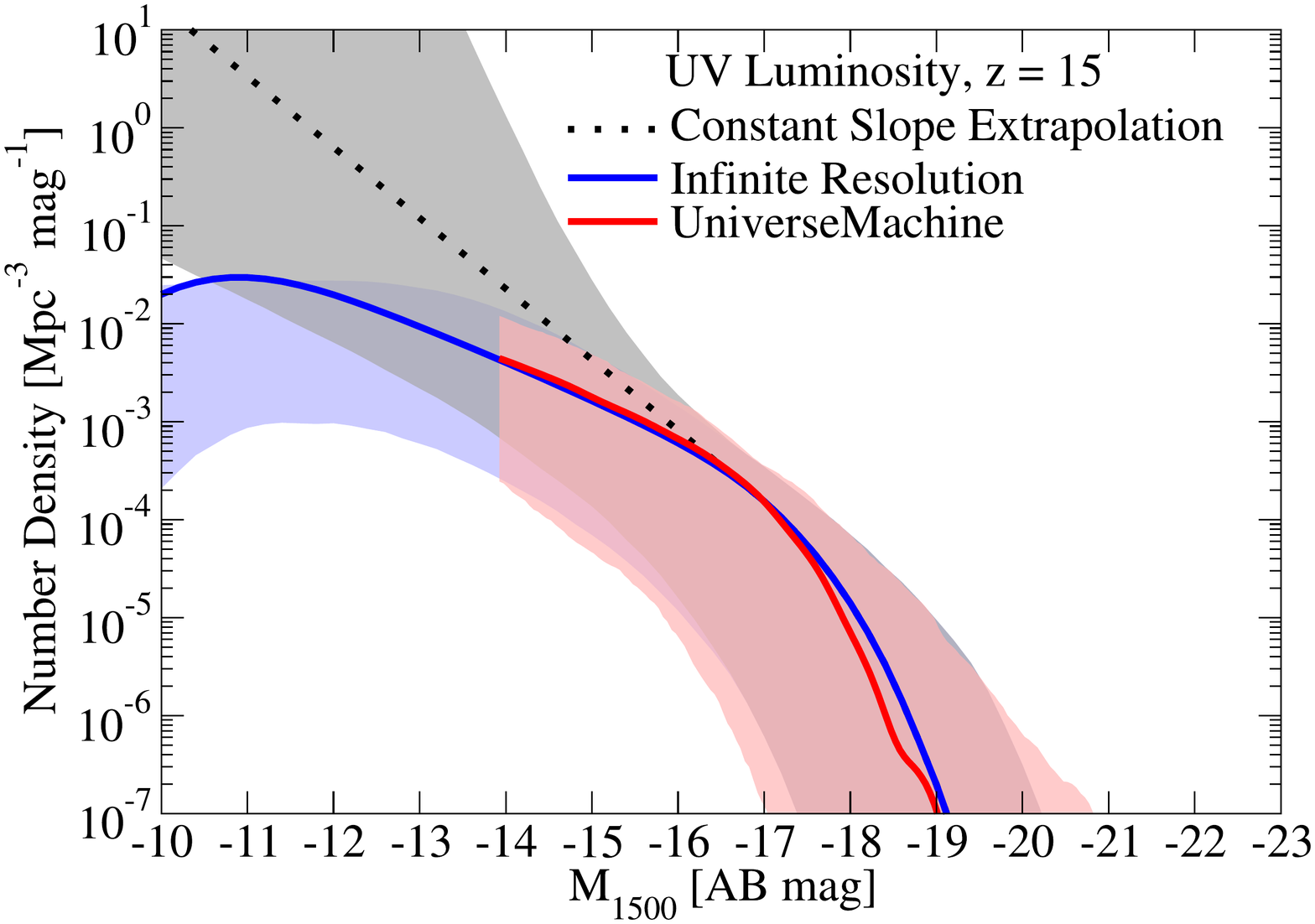}\hspace{-15ex}\\[-4ex]
    \caption{UV luminosity functions from the \textsc{UniverseMachine} on \textit{VSMDPL} compared to the extrapolated ``Infinite Resolution'' test in Appendix \ref{a:res_tests}.  In all cases, the reported results are very similar, indicating that \textit{VSMDPL} sufficiently resolves galaxy formation at $M_\mathrm{1500}<-14$.  Constant slope extrapolations below $M_{1500}=-17$ are shown for comparison with Fig. \ref{fig:extrap}.
    In all panels, error bars indicate the $16-84^\mathrm{th}$ percentile confidence interval.}
    \label{fig:uvlf_res}
\end{figure*}

\label{a:res_tests}

The parametrization of SFRs in Appendix \ref{a:full} can be applied to any halo mass function to closely approximate the SFR and UV luminosity distribution from the \textsc{UniverseMachine}.  Because halo mass function fits are available to very low masses, we can evaluate the behaviour of the \textsc{UniverseMachine} on an effectively infinite resolution simulation.

Here, we use the halo mass function fit in \cite{BWC13} (modified from \citealt{tinker-umf}) with the same cosmology as \textit{VSMDPL} (Fig.\ \ref{fig:hmfs}, left panel).  To convert between $\vmp$ and halo mass, we use the following relation from \citealt{BWHC19}:
\begin{eqnarray}
\vmp(M_h,a) & = & 200 \mathrm{km}\, \mathrm{s}^{-1} \left[\frac{M_h}{M_{200\mathrm{kms}}(a)}\right]^{1/3} \label{e:vmp}\\
M_{200\mathrm{kms}}(a) & = & \frac{1.64 \times 10^{12} \Msun}{\left(\frac{a}{0.378}\right)^{-0.142} + \left(\frac{a}{0.378}\right)^{-1.79}},\label{e:vmp2}
\end{eqnarray}
where $M_h$ is the peak virial halo mass \citep{mvir_conv}.  This relation was fit from the \textit{Bolshoi-Planck} simulation \citep{Klypin14,RP16}.

 Given Eqs.\ \ref{e:sfr}-\ref{e:delta} and \ref{e:vmp}-\ref{e:vmp2}, we can calculate the median SFR for any halo mass and redshift.  Subdividing the halo mass function (from the fit above) into 0.05 dex bins in halo mass, we compute the distribution of SFR in each mass bin using the log-normal scatter in Eq.\ \ref{e:sig_sf_a}.  For all results except for the mass threshold test in Fig.\ \ref{fig:threshold_mass}, we assume that star formation ceases to be efficient below a halo mass of $M_h=10^8\Msun$, corresponding to the atomic cooling limit \citep{OShea15,Xu16}.  Integrating across halo masses yields the SFR function (i.e., the number density of galaxies as a function of SFR), and integrating the SFR function yields the CSFR.

 To obtain UV luminosity functions, we need a scaling relation between SFR and UV luminosity.  For this, we evaluate the median unobscured UV luminosity as a function of SFR and redshift from the \textsc{UniverseMachine} best-fit model applied to \textit{VSMDPL}.  We find (as expected) that the unobscured UV luminosities are linear functions of SFR.  However, the normalization depends on redshift, because higher-redshift galaxies have higher specific growth rates \citep{BehrooziHighZ}, leading to more rapidly-rising star formation histories.  Specifically, we obtain a median ratio $\kappa_\mathrm{FUV}$ between SFR and UV luminosity of:
\begin{eqnarray}
    \kappa_\mathrm{FUV,Chabrier}(a) & = & 5.1\times 10^{-29} \left(1+\exp(-20.79a + 0.98)\right) \nonumber \\
    && \times \Msun\,\mathrm{yr}^{-1}\,\mathrm{erg}^{-1}\,\mathrm{s}\,\mathrm{Hz}, \label{e:kappa_fuv}
\end{eqnarray}
where $a$ is the scale factor.  The corresponding $\kappa_\mathrm{FUV,Salpeter}$ for a \cite{Salpeter55} IMF is a factor 1.58 larger \citep{Salim07}; the fit is shown in Fig.\ \ref{f:sfr_uv}.  Due to scatter in star formation histories, the typical scatter in $\kappa_\mathrm{FUV}$ across galaxies is 0.12 dex.  We convolve the SFR function with this scatter and divide by the median $\kappa_\mathrm{FUV}$ from Eq.\ \ref{e:kappa_fuv} to obtain the unobscured luminosity function.  Finally, we apply Eqs.\ \ref{e:dust1}--\ref{e:dust2} to obtain the observed (attenuated) UV luminosity function.

CSFR comparisons are shown in Fig.\ \ref{fig:csfr_res}, and UV luminosity function comparisons are shown in Fig.\ \ref{fig:uvlf_res}.  We find excellent agreement between the ``Infinite Resolution'' calculation above and \textit{VSMDPL}, even though \textit{VSMDPL} is only formally complete down to $M_\mathrm{h}=10^9\Msun$.  This is due to the fact that the SFR function becomes shallower for faint galaxies, so the contribution from low-mass haloes becomes less important than would be expected assuming a  steep, constant faint-end slope (Fig.\ \ref{fig:extrap}).  At the highest redshifts, more star formation occurs in low-mass haloes, but \textit{VSMDPL} is still at least 80\% complete at $z=15$ (Fig.\ \ref{fig:csfr_res}).

We note in passing that the extrapolated UV luminosity functions have a turnover at low luminosities.  This is not due to any change in the slope of the stellar mass--halo mass relation, but is instead due to the lower halo mass limit of $M_h = 10^8\Msun$.  This turnover moves to brighter magnitudes at higher redshifts, due to expected higher star formation rates at fixed halo mass as redshift increases (Figs.\ \ref{fig:halo_sfrs} and \ref{fig:halo_uv}).

\section{Cosmology Uncertainties}

\label{a:cosmo}

\begin{figure}
 \centering
    \vspace{-11ex}
    \hspace{-10ex}\includegraphics[width=1.15\columnwidth]{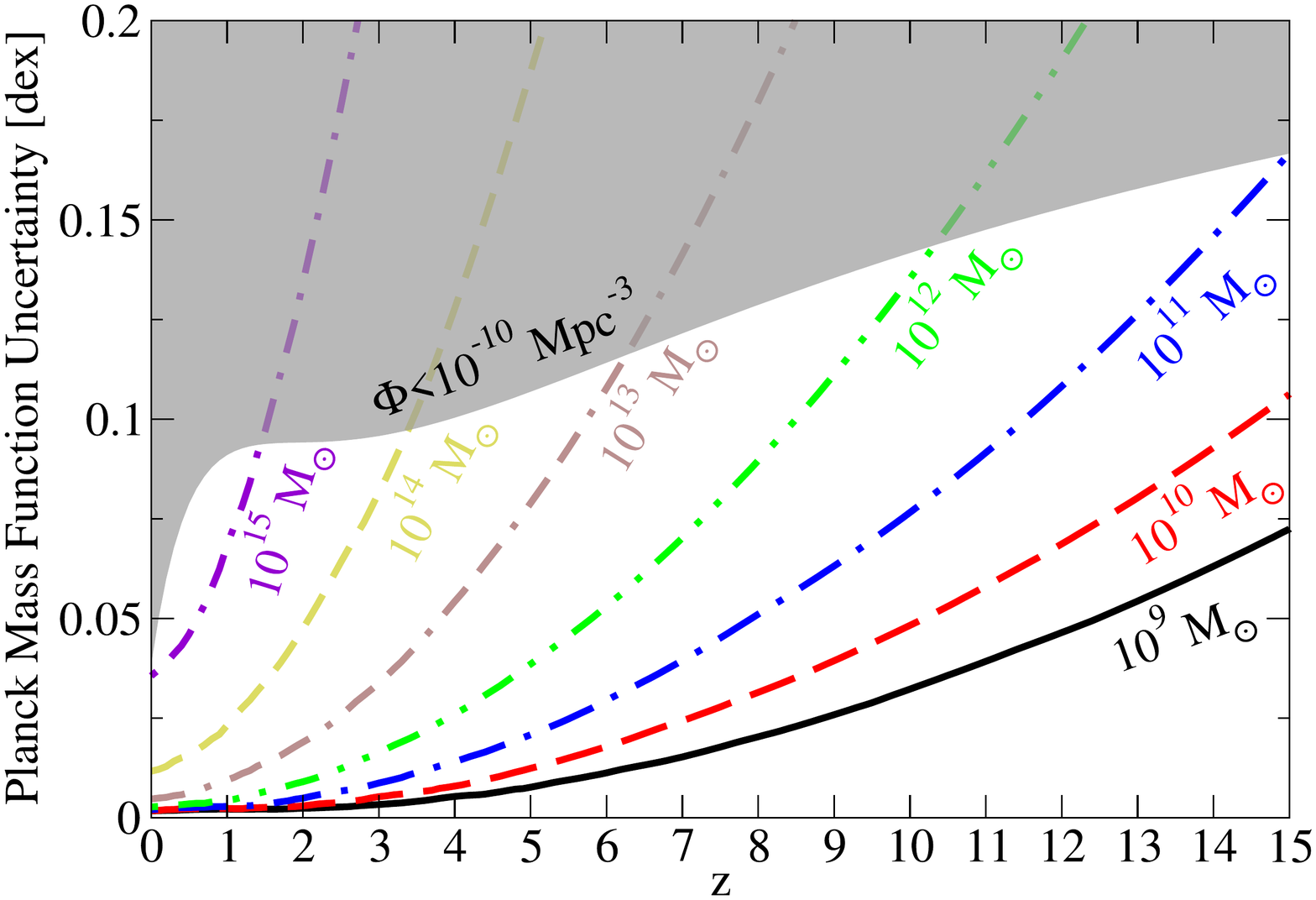}\\[-4ex]
    \caption{\textit{Planck} 2018 uncertainties in the differential number densities of haloes as a function of redshift and virial halo mass.  These represent half of the $16-84^\mathrm{th}$ percentile range. Uncertainties for halo masses less than $10^9\Msun$ (not shown) are less than those for $10^9\Msun$ haloes. The \textit{grey shaded region} shows uncertainties corresponding to haloes with cumulative number densities ($\Phi$) less than $10^{-10}$ Mpc$^{-3}$.  Uncertainties for haloes relevant to \textit{JWST} ($\Phi>10^{-6}$ Mpc$^{-3}$) are always $<0.2$ dex.}
    \label{fig:planck_uncertainties}
\end{figure}

For this analysis, cosmology uncertainties are subdominant to galaxy formation uncertainties.  We validate this by selecting 200 points at random from the posterior distribution of the baseline \textit{Planck} 2018 cosmological results (\texttt{plikHM\_TTTEEE\_lowl\_lowE\_lensing}; \citealt{Planck18}) and computing halo mass functions according to \cite{tinker-umf}.  Uncertainties (half the $16-84^\mathrm{th}$ percentile range) are shown in Fig.\ \ref{fig:planck_uncertainties}.  For the halo masses forming most stars at $z>8$ ($M_h<10^{11}\Msun$), relative uncertainties in number densities are at the $<0.2$ dex level even at $z=15$, well below uncertainties from existing constraints on galaxy number densities (Fig.\ \ref{fig:comp}).  Of note, systematic disagreements for $h$ at the present 0.036 dex level \citep{Riess19} would result in number density differences at the $\sim 0.1$ dex level.

\bsp	
\label{lastpage}
\end{document}